%% file: 0Main.tex
\documentclass[12pts]{article}

\usepackage{scicite}
\usepackage[style=nature, isbn=false, date=year, doi=true, natbib=true]{biblatex}
\BiblatexSplitbibDefernumbersWarningOff
\AtEveryBibitem{\ifentrytype{article}{\togglefalse{bbx:url}}{}}
\usepackage{times}
\usepackage{textcomp}

\usepackage{amsmath}
\usepackage{graphicx}
\usepackage{subfig}
\usepackage{todonotes}
\usepackage[utf8]{inputenc}
\usepackage{nameref}
\usepackage{float}
\usepackage{lineno}
\linenumbers

\newcommand{\lt}{<}
\newcommand{\gt}{>}

\usepackage[hidelinks]{hyperref}

\newcommand{\beginsupplement}{%
  \setcounter{table}{0}
  \setcounter{figure}{0}
  \renewcommand{\thetable}{S\arabic{table}}%
  \renewcommand{\thefigure}{S\arabic{figure}}%
  \renewcommand{\theHtable}{Supplement.\thetable}%
  \renewcommand{\theHfigure}{Supplement.\thefigure}
}

\newenvironment{sciabstract}{%
\begin{quote} \bf}
{\end{quote}}

\title{Dynamic restrengthening and fault heterogeneity explain megathrust earthquake complexity}

\begin{document}
\baselineskip16pt

\author
{Jeremy Wing Ching Wong,$^{1\ast}$ Alice-Agnes\ Gabriel,$^{1,2}$ Wenyuan\ Fan$^{1}$\\
\\
\normalsize{$^{1}$Institute of Geophysics and Planetary Physics, Scripps Institution of Oceanography}\\
\normalsize{University of California, San Diego, CA, USA}\\
\normalsize{$^{2}$Department of Earth and Environmental Sciences,}\\ \normalsize{Ludwig-Maximilians-Universit\"at M\"unchen, Munich, Germany}\\
\\
\normalsize{$^\ast$E-mail: jeremywong@ucsd.edu}
}

\let\leqslant=\leq
\label{firstpage}
\maketitle

\textbf{This manuscript is an arXiv preprint and has been accepted for publication in a peer-reviewed journal. Please note that this version may differ slightly from the final published version.}

\clearpage
\newpage

\begin{sciabstract}
\section*{Abstract}
\input{1Abstract}

\end{sciabstract}

\begin{refsegment}
\section*{Introduction}
\input{2Introduction.tex}

\section*{Results}
\input{3Results.tex}
\input{4Discussion.tex}

\clearpage
\section*{Methods}

\input{5Methods.tex}

\clearpage
\newpage
\input{6Acknowledgments.tex}

\printbibliography[segment=1]
\end{refsegment}

\begin{refsegment}
\begin{titlepage}
\centering

\vspace*{2.5cm}

{\LARGE \bfseries Supplementary Information \par}
\vspace{1.2cm}

{\Large Dynamic restrengthening and fault heterogeneity explain megathrust earthquake complexity \par}

\vspace{2cm}

{\large
Jeremy Wing Ching Wong$^{1\ast}$, Alice-Agnes Gabriel$^{1,2}$, Wenyuan Fan$^{1}$ \par
}

\vspace{1.2cm}

{\normalsize
$^{1}$ Institute of Geophysics and Planetary Physics, Scripps Institution of Oceanography,\\
University of California San Diego, La Jolla, CA, USA \par
\vspace{0.5em}
$^{2}$ Department of Earth and Environmental Sciences,\\
Ludwig-Maximilians-Universit\"at M\"unchen, Munich, Germany \par
}

\vfill

{\normalsize
$^\ast$ Correspondence to: jeremywong@ucsd.edu \par
\vspace{0.5em}
\today
}

\end{titlepage}
\beginsupplement
\input{8Supplementary}

\clearpage
\newpage
\printbibliography[title={Supplementary References}]
\end{refsegment}
\end{document}

%% file: 1Abstract.tex
Megathrusts host Earth's largest earthquakes. 
Understanding the physical conditions controlling their rupture dynamics is critical for assessing seismic and tsunami hazards. 
These earthquakes often display complex rupture dynamics, exemplified by the 2011 Tohoku-Oki earthquake, which exhibited multiple rupture episodes, depth-dependent seismic radiation, and substantial tsunamigenic slip near the trench. 
However, how such complexity arises from preexisting physical conditions remains uncertain. 
Here, we demonstrate that the observed rupture complexity of the Tohoku-Oki earthquake can spontaneously and self-consistently emerge, driven by rapid coseismic frictional restrengthening and data-informed fault heterogeneity.
We use an ensemble of 3D dynamic rupture simulations to identify that mixed downdip pulse-like and updip crack-like rupture are driven by dynamic stress redistribution with episodic rupture reactivation. 
By featuring low fault strength compared to its dynamic stress drop, a preferred model can consistently reproduce the observed complex depth-dependent propagation speeds, multiple rupture fronts as imaged by back-projection, and large tsunamigenic slip at the trench. 
Our findings demonstrate that preexisting fault heterogeneity conjointly with dynamic frictional weakening and restrengthening drives seemingly unexpected megathrust rupture complexity, highlighting the need to include dynamic effects into physics-based seismic and tsunami hazard assessments of future earthquakes.

%% file: 2Introduction.tex


Large megathrust earthquakes propagate rapidly, rupture over hundreds of kilometers within minutes, generate strong ground shaking, and, in certain instances, cause devastating tsunamis. 
The resulting seismic and tsunami hazards are directly controlled by rupture dynamics and shallow fault slip behavior (e.g., \cite{Wirth2022Occurrence}). However, the physical mechanisms controlling devastating earthquake rupture dynamics remain poorly understood. 
The 2011 $M_W$9.0 Tohoku-Oki, Japan, earthquake, one of the most destructive earthquakes of the 21st century, exhibited unexpected complexities throughout its rupture process: possible reactivation at the hypocenter \cite{Lee2011Evidence, Ide2011Shallow, Melgar2015Kinematic}, depth-dependent seismic radiation \cite{Meng2011Window, Lay2012Depthvarying}, large slip to the trench exceeding 50--60~m \cite{Fujiwara20112011, Kodaira2020Large}, and an unusually limited along-strike rupture extent for its magnitude \cite{Uchida2021Decade}.
Although the Tohoku-Oki earthquake is among the best-recorded megathrust events, the physical mechanisms underlying its complexity remain debated \cite{Uchida2021Decade}, and its slip models show significant variability \cite{Wong2024Quantitative}. 
Previous studies attribute some of this event's complexities to preexisting stress or frictional-strength fault asperities \cite{Duan2012Dynamic, Ide2013Historical, Kozdon2013Rupture, Huang2014SlipWeakening, Sallares2019Upperplate, Galvez2020Dynamic, Ma2023Wedge}. 
Here, we use 3D dynamic rupture simulations to show that the surprising characteristics of the Tohoku-Oki earthquake arise dynamically during rupture evolution. 
We analyze the interplay between preexisting fault heterogeneity and dynamically evolving rupture processes as drivers of earthquake complexity, as well as their distinct observational signatures.

The frictional properties of fault rocks and gouges govern fault strength and slip, and thus are a fundamental component of dynamic rupture models \cite{Ramos2022Working}. 
A key factor is the frictional response during fault slip, which controls earthquake nucleation, propagation, and arrest \cite{Ampuero2008Earthquake, Ke2018Rupture, Lambert2021Propagation}. 
Rate-and-state friction laws effectively describe fault friction at interseismic to slow slip rates, capturing the dependence of friction on sliding velocity and state evolution \cite{Dieterich1979Modeling, Ruina1983Slip}.
Laboratory and theoretical studies show that at coseismic slip rates, fault friction likely exhibits even stronger velocity-weakening, followed by equally rapid frictional healing as slip rate decreases \cite{Noda2009Earthquakea, DiToro2011Fault, Ujiie2013Low}. 
Such rapid dynamic friction evolution generates complex faulting behavior, facilitating diverse earthquake rupture styles and speeds, slip reactivation, and multi-fault interaction in laboratory experiments and numerical simulations \cite{Gabriel2012Transition, Ulrich2019Dynamic, Rubino2022Intermittent}. 
However, currently, no fully dynamic rupture model of the Tohoku-Oki earthquake accounted for fast-velocity weakening rate-and-state friction, and none in combination with data-constrained fault heterogeneity. 

The temporal evolution of frictional strength governs the local stability of slip, whereas the spatial distribution of fault stress and strength is equally important (e.g., \cite{Kammer2024Earthquake}). 
Fault heterogeneity is expected along the megathrust interface, as evidenced by the broad range of faulting behavior observed along the Japan Trench \cite{Nishikawa2019Slow} and in other subduction zones \cite{Bassett2025Variation}. 
Such variability may arise from differences in prestress, fault strength, fluid content, geometry, and material properties \cite{Nishikawa2023Review}. 
While some fault heterogeneity reflects relatively static factors such as depth, temperature, and lithology, these alone cannot explain the full range of observed slip behavior. 
In addition, fault heterogeneity can evolve dynamically during slow and fast slip. 
For example, stress heterogeneity can spontaneously emerge even under spatially uniform frictional conditions \cite{Lapusta2003Nucleation, Cattania2019Complex, Barbot2019Slowslip}. 
Therefore, understanding megathrust rupture dynamics requires studying both pre-existing fault heterogeneity and dynamically evolving stresses and friction.

Here, we present fully dynamic 3D rupture simulations of the Tohoku-Oki earthquake incorporating fast velocity-weakening rate-and-state friction and coseismic restrengthening, constrained by observationally informed fault heterogeneity.
Our simulations spontaneously reproduce the event's reported complex rupture behavior, including repeated rupture reactivation, depth-dependent rupture styles, and realistic trench slip.
The physical processes identified here, driven by rapid coseismic frictional restrengthening and fault heterogeneity, are likely fundamental controls on rupture behavior in other megathrust settings, carrying important implications for earthquake dynamics, tsunami generation, and hazard assessment globally. 

\clearpage
\newpage

%% file: 3Results.tex
\begin{figure}
\noindent\includegraphics[width=\textwidth]{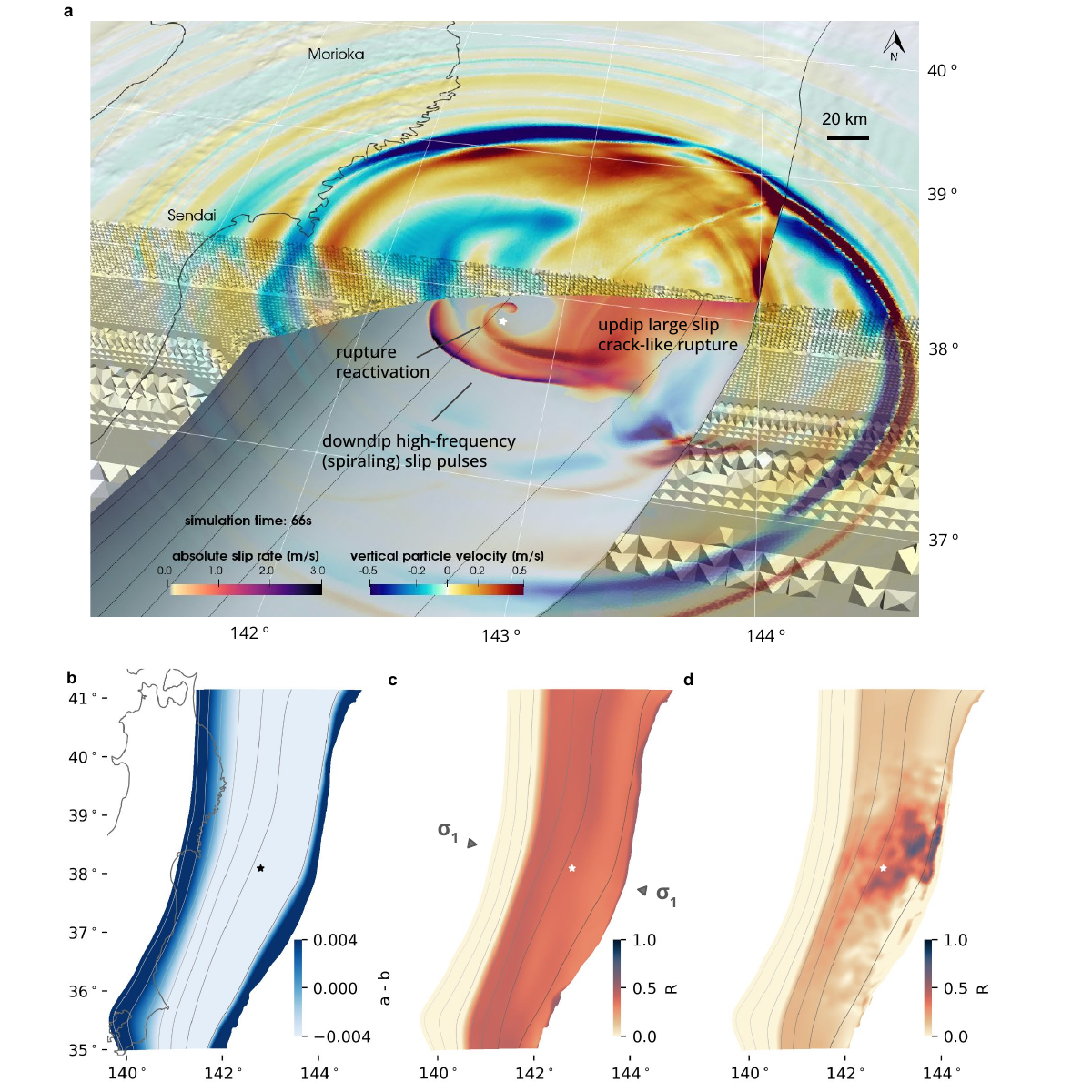}
\caption{\textbf{Overview of 3D dynamic rupture simulations of the 2011 Tohoku-Oki earthquake and their initial conditions.} Depth contours (gray, 10 km intervals) and hypocenter location (star) are shown in all panels. (a) Snapshot of the simulated absolute slip rate and seismic wavefield evolution (vertical particle velocity) at 66~s, highlighting multiple reactivated slip pulses propagating downdip and crack-like rupture accumulating large slip near the trench.
The model incorporates realistic slab geometry and high-resolution topobathymetry within an unstructured tetrahedral mesh refined near the slab and onshore region. 
(b) Depth-dependent frictional properties ($a-b$, see Methods Sec.~``\nameref{Fault friction}'') with velocity-strengthening behavior in the shallow ($\lt$9 km) and deep ($\gt$45 km) regions, transitioning to velocity-weakening in the seismogenic zone.
(c) Laterally homogeneous, depth-dependent ambient stress and frictional strength initial conditions informed by regional tectonics (see Methods Sec.~``\nameref{Prestress}''), showing the relative prestress ratio $R$ (maximum possible stress drop over frictional strength drop, Eq.\ref{eq:R}, \cite{Aochi20031999}). The principal stress direction ($\sigma_1$ at an azimuth of 100$^\circ$ and a plunge angle of 8$^\circ$, \cite{Heidbach2018World}) is indicated with arrows.
(d) Heterogeneous stress initial conditions combining the ambient background stress shown in (c) and heterogeneous initial stress inferred from the median slip distribution of 32 data-constrained slip models (Supplementary Figs.~\ref{SFig:Kinematic_med_slip_model},~\ref{SFig:Traction_cij}, \cite{Wong2024Quantitative}). 
Along-dip initial shear stresses are shown in Supplementary Fig.~\ref{FigE1:Initial stress conditions}.
}
\label{Fig1:Initial_conditions}
\end{figure}

\clearpage
\newpage

\label{Results}
3D dynamic rupture simulations can capture the nonlinear interactions between seismic wave propagation, fault friction, stress heterogeneity, and fault geometry, leveraging high-performance computing, reaching megathrust earthquake spatial and temporal scales at high resolution \cite{Uphoff2017Extreme}.
Dynamic rupture simulations have been applied to subduction zones worldwide (Methods). 
However, many studies have been restricted to 2D cases with imposed ad-hoc fault friction or stress heterogeneities \cite{Duan2012Dynamic, Huang2014SlipWeakening, Galvez2020Dynamic} or simplified friction laws \cite{Ulrich2022Stress}, restricting the direct integration of observational constraints and verification. 
Here, for the first time, we present fully dynamic 3D rupture simulations of the Tohoku-Oki earthquake incorporating fast velocity-weakening and coseismic frictional restrengthening, constrained by observationally informed fault heterogeneity. 
Unlike previous dynamic rupture models that prescribed frictional or stress asperities to control slip distribution and steer the depth-dependent rupture behavior, our simulations spontaneously reproduce the observed complex rupture behavior, including repeated rupture reactivation, depth-dependent rupture styles, and realistic trench slip, solely through dynamic friction evolution and stress conditions.

Our 3D dynamic rupture models (Fig. \ref{Fig1:Initial_conditions}), resolving up to 1.5~Hz of the seismic wavefield (Methods Sec.~``\nameref{Model geometry and mesh}'' and Supplementary Sec. ~`\nameref{SM: Model domain and resolution}''). ), investigate physical controls on complex, spontaneous rupture processes by incorporating regional-tectonic constraints (\cite{Heidbach2018World}, Fig.~\ref{Fig1:Initial_conditions}c), and initial stress heterogeneity (Fig.~\ref{Fig1:Initial_conditions}d). 
We construct an initial stress state (Methods Sec. ``\nameref{Prestress}'') that combines the regional maximum principal stress orientation with stress heterogeneity from a median slip model that captures common slip features among 32 finite-fault slip models (\cite{Wong2024Quantitative}, Supplementary Fig.~\ref{SFig:Kinematic_med_slip_model}). 
The dynamic models use a realistic slab geometry and high-resolution topobathymetry (Fig. \ref{Fig1:Initial_conditions}a), along with fast velocity-weakening rate-and-state friction \cite{Noda2009Earthquakea} (Fig.~\ref{Fig1:Initial_conditions}b, Extended table~\ref{T:Friction}) and off-fault plasticity (Supplementary Fig.~\ref{SFig:CF_cohesion_profile}). 

Dynamic rupture is governed by a fast velocity-weakening rate-and-state friction law (Methods Sec. ``\nameref{Fault friction}''), in which frictional strength is inversely proportional to slip rate at high slip rates. 
This formulation represents thermal weakening processes that can operate on natural faults at co-seismic slip rates, such as thermal pressurization and flash-heating \cite{Rice2006Heating, Beeler2008Constitutive}. 
We use the full rate-and-state friction formulation, including state evolution (Methods Sec. ``\nameref{Fault friction}'', Equations 5-6).
In this formulation, the steady-state friction coefficient is given as
\begin{equation}
    f_{ss} = f_{w}+\frac{f_{LV, ss}(V)-f_w}{(1+(V/V_w)^4)^{1/4}}.
\end{equation}
It depends on slip rate $V$ and transitions from a low-velocity steady-state friction coefficient $f_{LV, ss}(V)$ to a fully weakened level $f_w$ once $V$ exceeds the onset of weakening velocity $V_w$. The low-velocity steady-state friction coefficient is defined as 
\begin{equation}
    f_{LV,ss} = f_0 - (b-a)\ln(V/V_0),
\end{equation}
where $f_0$ is the reference friction coefficient, $V_0$ is the reference velocity, $a$ is the direct-effect parameter, and $b$ is the state-evolution parameter.
At low slip-rates ($V<<V_w$), friction follows the classical rate-and-state friction law, whereas for high slip rates ($V>=V_w$), an additional fast velocity-weakening term is activated, producing a rapid reduction in frictional strength.

When the fault local slip rate decelerates below $V_w$, the inverse slip rate dependence of $f_{ss}(V)$ recovers towards $f_{LV, ss}(V)$, producing rapid frictional healing in the tail of slip pulses \cite{Nielsen2003SelfHealing, Beeler2008Constitutive}. 
This fast velocity-weakening law is capable of generating a wide spectrum of rupture styles, including sub- and super-shear, crack-like, and pulse-like ruptures \cite{Gabriel2012Transition}.

In our simulations, the frictional parameters ($a$, $b$, $f_w$, and $V_w$) are fixed across the ensemble (Table~\ref{T:Friction}), following previous studies \cite{Gabriel2012Transition, Ulrich2019Dynamic}. 
We prescribe a depth-dependent profile of the rate-and-state friction parameters $a-b$ (Figure~\ref{Fig1:Initial_conditions}b).  The shallow megathrust is assigned velocity-strengthening friction ($a-b>0$), consistent with laboratory constraints for clay-rich lithified rock and rock gouges at low slip velocities \cite{Saffer2003Comparison}.  The seismogenic zone (10--45 km depth) is assigned velocity-weakening friction ($a-b<0$). At the downdip limit of the seismogenic zone ($>$ 45 km), we revert to velocity-strengthening friction, consistent with inferences from repeating earthquakes and slow-slip events \cite{Uchida2021Decade}. For more details, see the Methods Sec.~``\nameref{Fault friction}''.
Aside from the depth-dependent variation in $a-b$, we do not prescribe any additional frictional asperities. 
All other frictional parameters remain spatially uniform and are held constant across the model ensembles (Table~\ref{T:Friction}).

\subsection*{Grid search for preferred rupture model}
\label{Initial stress condition grid search for the preferred rupture model}
We use a systematic grid-search approach to identify a preferred 3D dynamic rupture model that minimizes the misfit with respect to the onshore and offshore geodetic observations and SCARDEC seismic moment-rate \cite{VallA2016New}, with each dataset weighted equally. 
The preferred model is selected from an ensemble of simulations that systematically vary (i) the amplitude of initial stress heterogeneity informed by finite-fault slip models ($\alpha$) and (ii) the regionally constrained ambient stress, expressed as the ratio of maximum possible stress drop over fault strength on an optimally oriented fault plane ($R_0$). 
Larger $\alpha$ represents higher prestress heterogeneity informed by finite-fault slip models, whereas larger $R_0$ reflects an ambient stress state closer to failure. 
Figures~\ref{Fig1:Initial_conditions} and ~\ref{FigE1:Initial stress conditions} show the resulting initial shear stress, relative prestress ratio $R$ across the megathrust. 
The along-strike and downdip variation of $R$ and initial shear stress $\tau_0$ along the megathrust reflects the resolved initial shear traction on the 3D fault geometry. 

Together, $\alpha$ and $R_0$ govern the earthquake energy balance between the available energy release rate that sustains rupture propagation and the fracture energy required for continued rupture growth \cite{Kammer2024Earthquake}. 
We vary $\alpha$ from 1.0 to 1.2 in increments of 0.01 and $R_0$ from 0.1 to 0.2 in increments of 0.005, generating 33 dynamic parameter sets. 
Each corresponding spontaneous dynamic rupture scenario is quantitatively evaluated by the variance reduction of observed onshore and offshore static displacements and seismic moment release rate (Supplementary Fig.~\ref{SFig:MR_grid_search} and Fig.~\ref{SFig:grid_search_obs_VR}). 
The ensemble results suggest that $\alpha$ primarily controls peak slip amplitude, whereas $R_0$ controls the overall rupture area. 
Variations in crack- or pulse-like rupture style and the occurrence of healing fronts arise from the nonlinear interaction with the initial stress conditions parameterized by $\alpha$ and $R_0$, which control the amplitude and heterogeneity of the prestress but do not modify the frictional healing law itself.

\begin{figure}
\noindent\includegraphics[width=\textwidth]{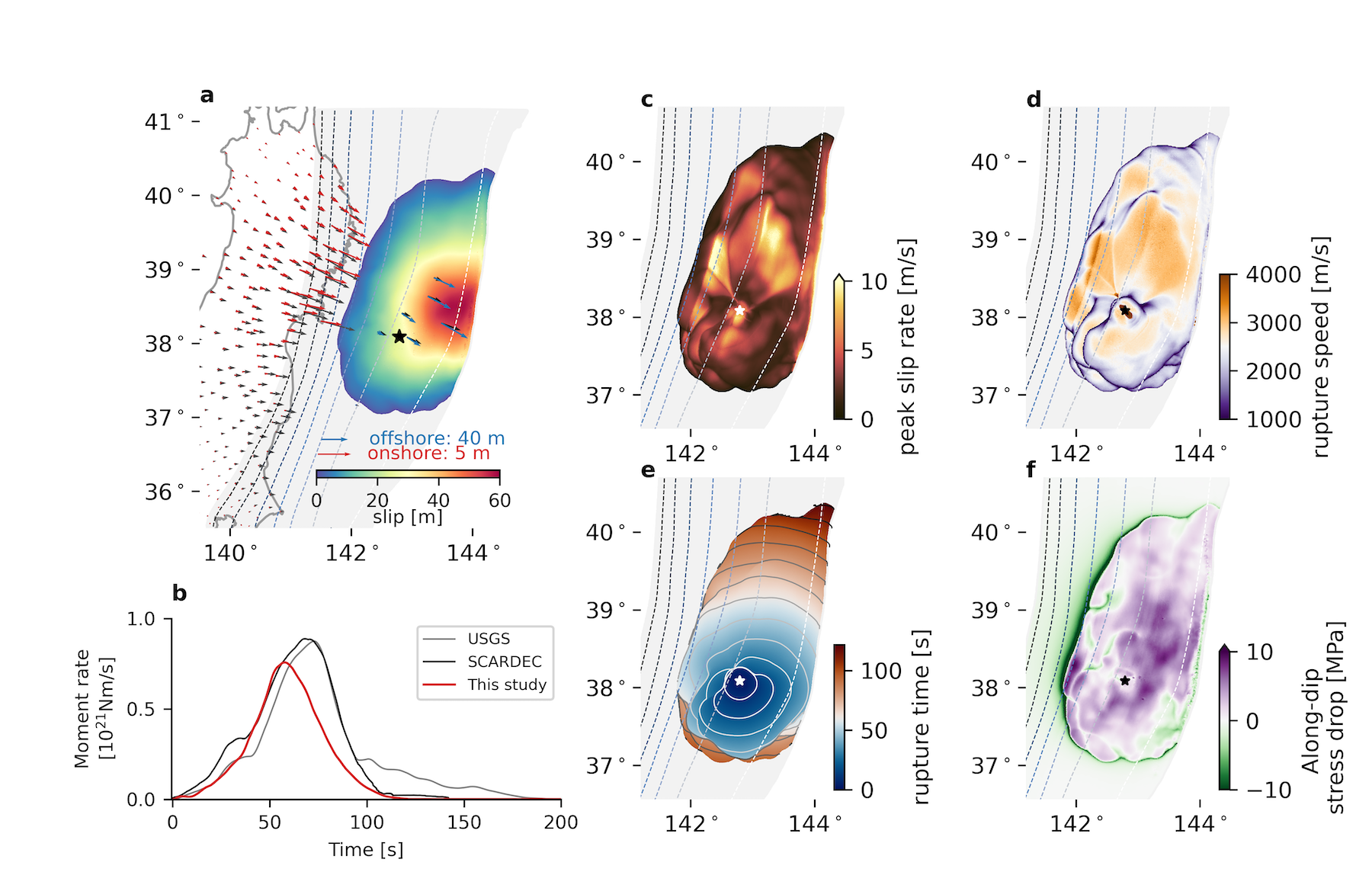}
\caption{\textbf{Preferred 3D dynamic rupture scenario of the Tohoku-Oki earthquake constrained by geodetic observations and seismic moment release rate.}
Gray contours indicate depth (10 km intervals), and the star is the hypocenter location \cite{Hayes2011Rapid}. 
Rupture extends 200~km along-dip and 360~km along-strike, producing a moment magnitude of $M_W~8.97$ and a duration of 120~s. The total radiated seismic energy is  $\approx 7.7\times 10^{17}J$, within observational estimates of $4.2-9.1 \times 10^{17}J$ for the Tohoku-Oki earthquake \cite{Ide2011Shallow, Lay2012Depthvarying}. 
(a) Fault slip distribution with comparison between observed and simulated geodetic displacements onshore and offshore. 
Black arrows denote observed horizontal displacements from offshore and onshore stations. 
Blue and red arrows represent simulated horizontal displacements offshore and onshore, respectively, achieving variance reductions of 77\% (onshore) and 55\% (offshore). 
(b) Synthetic moment rate release compared with observational inferences from teleseismic by the USGS model \cite{Hayes2011Rapid} and SCARDEC inversion results \cite{VallA2016New}.
Heterogeneous spatial distributions of
(c) peak slip rate, (d) rupture speed, (e) rupture front timing (10~s intervals, gray contours), and (f) along-dip stress drop.}
\label{Fig2:preferred_overview}
\end{figure}

The preferred model reproduces the overall slip pattern of the Tohoku-Oki earthquake.
Despite not being a full-scale inversion, the model achieves variance reductions of 77\% and 55\% for onshore and offshore geodetic observations, respectively (Fig. \ref{Fig2:preferred_overview}a).
It features a smooth slip distribution, with major slip concentrated updip of the hypocenter, and a triangular moment-rate function consistent with observations (Fig.~\ref{Fig2:preferred_overview}b). 
We capture the gradual seismic moment release during the earthquake's initiation phase, in contrast to previous dynamic models (e.g., \cite{Kozdon2013Rupture, Galvez2020Dynamic}, Supplementary Sec. ~`\nameref{SM2: Nucleation}''). 
Despite the simple slip distribution and moment rate release and the absence of imposed frictional heterogeneity, the model produces substantial rupture complexity, including variations in peak slip rate, rupture speed, and stress drop (Fig.~\ref {Fig2:preferred_overview}c-f), as reported in kinematic slip models (e.g., \cite{Yagi2011Rupture, Melgar2015Kinematic}).
The model yields an average stress drop of 2.4~MPa, matching the average reported stress drop estimates \cite{Brown2015Static} (Supplementary Sec. ~`\nameref{SM3: Dynamic stress drop}''). 
The coseismic stress drop distribution from our model aligns with the major afterslip pattern of the Tohoku-Oki earthquake between 40-60 km depth \cite{Uchida2021Decade}, where negative stress drop is prominent at the downdip edge of the simulated rupture area.
In the following, we examine four key dynamic rupture characteristics of the preferred model and their corresponding observational signatures. 
\clearpage
\newpage

\clearpage
\newpage

\begin{figure}
\noindent\includegraphics[width=\textwidth]{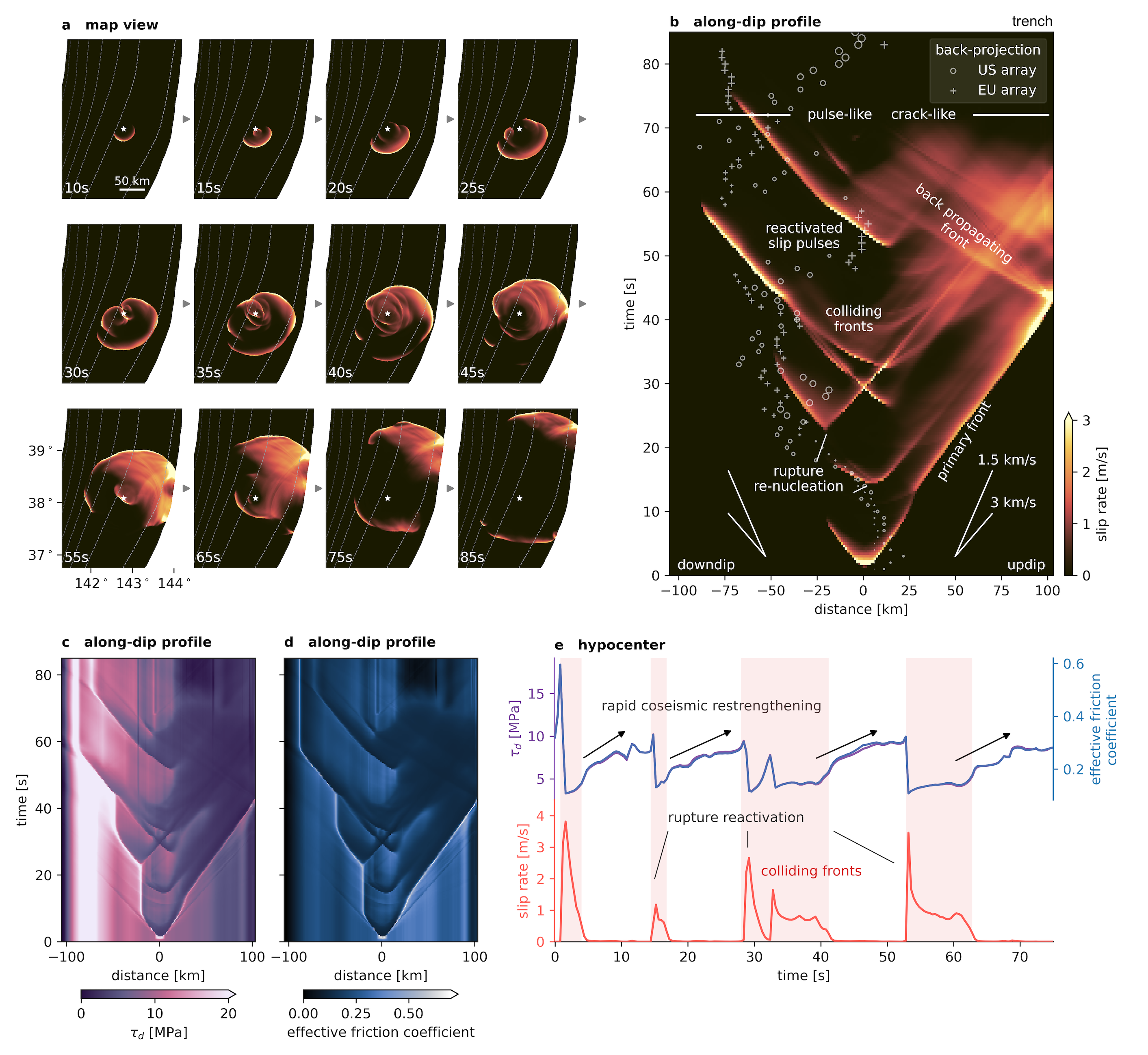}
\caption{
\textbf{Repeated dynamic rupture reactivation enabled by rapid coseismic weakening and restrengthening during the preferred Tohoku-Oki earthquake dynamic rupture model}.
(a) Map-view snapshots of rupture evolution from 10~s to 85~s simulation time, showing three main re-nucleation episodes at 15~s, between 25--40~s, and at 50~s. The white star indicates the hypocenter. Similarly ``spiraling'' rupture fronts have been observed in recent laboratory experiments \cite{Cochard2024Propagation}. 
Supplementary Fig.~\ref{FigE2:All_slip_rate_5s} and Supplementary Video~S1 show the complete rupture evolution. 
Supplementary Fig.~\ref{FigE3:Spiral_rupture} shows the detailed evolution of ``spiraling'' rupture fronts.
(b) Slip rate evolution along a dip profile through the hypocenter, highlighting multiple episodes of rupture reactivation.
Crosses and circles indicate the locations of high-frequency radiation from back-projection using the US and European arrays, respectively \cite{Meng2011Window}.
Rupture propagates faster updip ($\approx$2.5~km/s) compared to downdip ($\approx$1.7~km/s), matching the observational results from the back-projection analyses \cite{Meng2011Window}.
(c)-(d) Temporal evolution of along-dip shear stress (purple) and effective friction coefficient (blue, Methods ``~\nameref{Fault friction}'') along the hypocentral dip profile, highlighting rapid variations coincident with dynamic rupture reactivation; lighter colors indicate higher values. 
(e) Time series at the hypocenter of slip rate (red), along-dip shear stress (purple), and effective friction coefficient (blue), showing repeated rupture reactivation (slip rate $\ge$ 0.05 m/s, shaded red) and rapid frictional restrengthening.
Supplementary Fig.~\ref{FigE3:Profile_comp} shows updip and downdip time series evolution of slip rate, along-dip shear stress, and effective friction coefficient. 
}
\label{Fig3:reactivation}
\end{figure}
\clearpage
\newpage

\subsection*{Frictional restrengthening drives rupture reactivation}
Our preferred dynamic rupture model spontaneously produces episodic pulse- and crack-like slip reactivation originating near the hypocenter 
(Fig.~\ref{Fig3:reactivation}, Supplementary Fig.~\ref{FigE2:All_slip_rate_5s}, and Supplementary Video~S1). 
Crack-like rupture involves prolonged local slip durations (rise time) comparable to the total event duration. In contrast, self-healing pulse-like rupture is characterized by short rise times and a "healing" front following rupture front driven by rapid coseismic fault strength recovery \cite{Heaton1990Evidence, Perrin1995Selfhealing, Nielsen2003SelfHealing}.  
Our preferred model has complex rupture dynamics, including multiple spiraling \cite{Cochard2024Propagation}, back-propagating \cite{Gabriel2012Transition, sun2025back}, and colliding rupture fronts driven by rapid coseismic frictional weakening and restrengthening. 
Figure~\ref{Fig3:reactivation}a shows the complexity of slip rate evolution through multiple rupture reactivation episodes. 
The rupture initiates as a primary slip pulse with short local slip duration, followed by a secondary slip pulse nucleating at its healing front. 
Interaction between updip and downdip propagating rupture and healing fronts leads to a successive second, third, and (unsustained) fourth episodes of slip reactivation in the hypocentral region.
When the updip propagating rupture fronts reach the seafloor, strong dynamic interactions with the free surface generate trench-reflected, back-propagating phases, which coalesce with secondary arriving rupture fronts to form a sustained updip crack-like rupture with prolonged slip duration.
In the later stage, rupture simplifies into a bilateral slip pulse propagating along-strike, saturating the seismogenic zone width before spontaneously arresting (Supplementary Fig.~\ref{FigE2:All_slip_rate_5s}; Video~S1).

Our model explains the observed contrast between the slow downdip rupture propagation speed \cite{Meng2011Window} and the faster updip speed \cite{Lee2011Evidence, Yagi2011Rupture, Melgar2015Kinematic}. 
Figure~\ref{Fig3:reactivation}b shows how episodic pulse-like rupture reactivation successively extends the rupture duration downdip, producing a slower apparent rupture speed ($\sim$1.5~km/s) despite each reactivated front propagating at a regular rupture speed ($\sim$2.5~km/s). 
In contrast, the updip rupture front propagates steadily as a primary crack-like rupture front at $\sim$2.5 km/s.

Rapid coseismic restrengthening emerges as the principal mechanism controlling downdip repeated reactivation, causing a fault portion to slip and stop more than once during the same earthquake. 
The mechanisms driving rupture reactivation are illustrated by the along-dip evolution of shear stress and frictional strength  (Figs.~\ref{Fig3:reactivation}c--d), and by the corresponding time series at the hypocenter (Fig.~\ref{Fig3:reactivation}e).

Initially, as slip rate increases, both along-dip shear stress and effective friction coefficient sharply rise due to the instantaneous response of rate-and-state friction to slip rate changes, quickly followed by dynamic weakening (see Methods), causing dynamic stress drops of up to 10~MPa (Fig.~\ref{Fig3:reactivation}e). 
As slip rate subsequently ceases, a healing front follows.
The growing slip pulse gradually concentrates shear stress in its hypocentral region eventually overcoming local fault strength and reactivating slip, consistent with theoretical predictions for singular, self-similar pulse-like rupture \cite{Nielsen2003SelfHealing} and simpler 2D numerical simulations \cite{Gabriel2012Transition}. 
This process repeats multiple times, resulting in six distinct slip episodes at the hypocenter in our preferred rupture model.
Notably, all secondary ruptures are pulse-like in the downdip direction and coalesce with the free-surface reflected front into crack-like ruptures in the updip region.
The sustained crack-like slip updip and continued dynamic stressing limit the formation of healing fronts, and pulse-like rupture does not develop there. Thus, a key difference is not whether coseismic restrengthening occurs, but whether it is sufficient to arrest dynamic slip. Downdip, restrengthening more readily arrests slip, allowing self-healing pulses and repeated reactivation.
 
\clearpage
\newpage

\begin{figure}
\noindent\includegraphics[width=\textwidth]{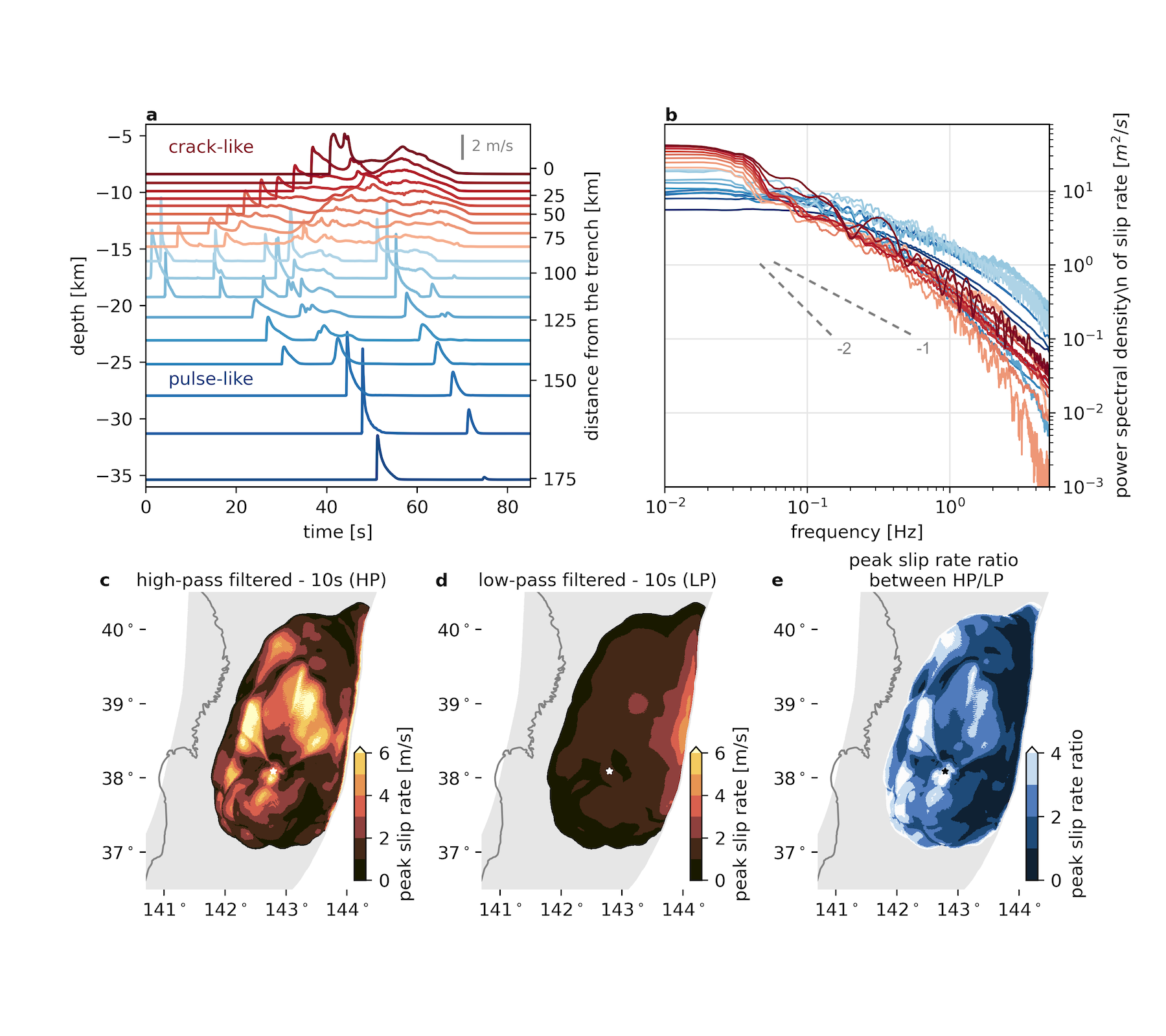}
\caption{\textbf{Depth-varying rupture styles featuring downdip short-duration slip pulses and updip large-slip crack-like ruptures}. 
(a) Slip-rate evolution with depth along the same hypocenteral dip profile in Fig.~\ref{Fig3:reactivation}, highlighting crack-like ruptures shallower than 15~km depth (red) and pulse-like ruptures at greater depths (blue).
(b) Power spectra of the slip rates, illustrating systematically higher-frequency content in downdip pulses compared to the shallower crack-like ruptures. The shallow crack-like rupture spectra follow a -1 slope over the 0.02-1~Hz range, whereas the pulse-like rupture exhibits a shallower -0.7 slope within the same frequency band. 
(c--d) Spatial distribution of high-pass filtered (HP) and low-pass filtered (LP) peak slip rates, respectively. 
(e) Spatial variation in the ratio of high-frequency to low-frequency peak slip rates (HP/LP). 
Shallow regions ($<$15 km depth) exhibit predominantly crack-like rupture with low HP/LP ratios, while downdip and hypocentral areas show pulse-like ruptures enriched in high-frequency content, consistent with observations from back-projection and regional strong-ground motion analyses \cite{Meng2011Window, Kurahashi2013ShortPeriod}. Supplementary Fig.~\ref{FigE4:Depth_dependent_freq_ranges} shows additional comparisons in three frequency ranges: 10--2~s, 2--0.5~s, and $>$ 0.5~s.
}
\label{Fig4:depth_dependent_rupture_style}
\end{figure}
\clearpage
\newpage

\begin{figure}
\noindent\includegraphics[width=\textwidth]{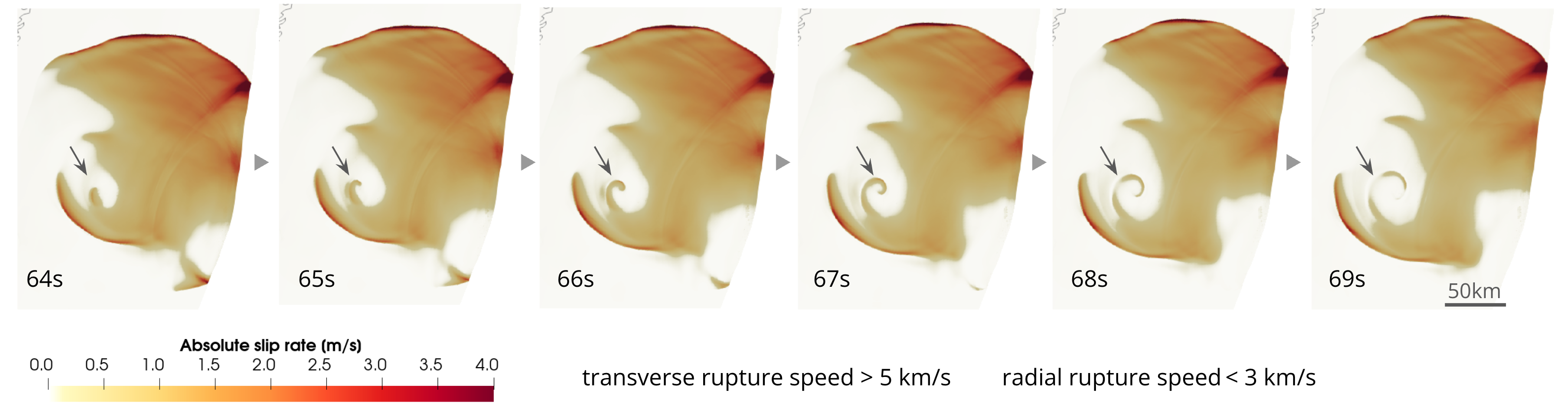}
\caption{\textbf{Rapid transverse expansion of a circular rupture front, resembling recent laboratory observations \cite{Cochard2024Propagation}}. Snapshots show slip rate evolution of a circular, spiraling rupture front from 64~s to 69~s. The black arrows mark the onset location where the circular rupture front forms. The spiraling rupture reaches a radius of approximately 12~km from the hypocenter and sweeps a 180$^{\circ}$ arc within 5~s. Its transverse rupture front expansion speed exceeds 7~km/s, which is higher than the local shear wave speed at the corresponding depth of $\approx$20~km (Supplementary Table~S1).}
\label{FigE3:Spiral_rupture}
\end{figure}

\subsection*{Downdip slip pulses and updip crack-like rupture}
Observational studies, including teleseismic back-projection \cite{Meng2011Window}, regional strong-ground motion analyses \cite{Kurahashi2013ShortPeriod}, and finite-fault inversions \cite{Yagi2011Rupture, Melgar2015Kinematic}, have consistently documented depth-dependent seismic radiation characteristics in the Tohoku-Oki earthquake. 
In our preferred model, we observe depth-varying dynamic rupture styles, characterized by short-duration slip pulses \cite{Heaton1990Evidence} radiating high-frequency seismic waves downdip and prolonged crack-like rupture accumulating large slip updip (Fig.~\ref{Fig4:depth_dependent_rupture_style}). 
Short slip pulses at depth are associated with relatively small total slip but strong high-frequency radiation, whereas the shallow crack-like rupture produces large slip with relatively reduced high-frequency content. Thus, along the fault, high-frequency radiation is inversely correlated with local total slip.
Figure~\ref{Fig4:depth_dependent_rupture_style}a highlights contrasting slip-rate functions across different depths. At depths shallower than 15~km, rupture propagation is crack-like, characterized by continuous slip and prolonged rise times exceeding 50~s.
The prolonged rupture is sustained by successive reactivation of slip triggered by free-surface reflections that generate back-propagating rupture fronts \cite{dunham2005dissipative, Vallee2023Selfreactivated}. 
In contrast, rupture transitions to multiple reactivated sharp slip pulses as it propagates deeper than 15 km, each with rise times less than 10~s. 
These depth-dependent rupture styles become particularly evident during 30--80~s rupture time, when multiple reactivated rupture fronts are present, and the deeper ruptures manifest discrete short-duration pulses (Fig.~\ref{Fig4:depth_dependent_rupture_style}a, Supplementary Fig.~\ref{FigE3:Profile_comp}).
Additional profiles from north to south show consistent behavior, with rupture durations systematically increasing towards shallower depth (Supplementary Fig.~\ref{FigS:Depth_dependent_profile_comparison}). 
This rupture style variability influences the associated seismic radiation.
Figure~\ref{Fig4:depth_dependent_rupture_style}b shows slip-rate amplitude spectra, revealing that shallow crack-like rupture episodes are relatively depleted in high-frequency energy than the deeper pulse-like rupture portions. 
However, northern profiles exhibit less contrast in seismic radiation due to a strong peak at the rupture front across all depths (Supplementary Fig.~\ref{FigS:Depth_dependent_profile_comparison}). 
This behavior reflects an intrinsic limitation of our model setup, which does not prescribe any depth-dependent frictional variations that control rupture-tip behavior \cite{Huang2014SlipWeakening}.

Observational studies indicate that high-frequency radiation inversely correlates with the total slip (e.g., \cite{Lay2012Depthvarying, Melgar2015Kinematic}). 
Our preferred dynamic rupture model reproduces these observations, showing the amplitude ratio of high-pass to low-pass filtered peak slip-rate function at 10~s period reaching approximately 400\% in the downdip and hypocentral regions (Fig.~\ref{Fig4:depth_dependent_rupture_style}e). 
Supplementary Fig.~\ref{FigE4:Depth_dependent_freq_ranges} further compares peak-slip rate distribution across multiple frequency bands and shows that the high-frequency radiations are dominated within the 0.5--2~Hz frequency range used in back-projection studies \cite{Meng2011Window}.
Furthermore, the multiple reactivated downdip slip pulses can explain migrating downdip and hypocentral high-frequency seismic radiation imaged through back-projection methods (Fig.~\ref{Fig3:reactivation}b; \cite{Meng2011Window}).
Downdip reactivated slip pulses also exhibit spiraling rupture.
These spiraling rupture fronts are localized and expand rapidly in the transverse direction with rupture speed higher than shear wave speed ($\gt$~7~km/s), while the radial rupture front expansion speed remains sub-shear ($<$~3~km/s) (Fig.~\ref{FigE3:Spiral_rupture}). 
Such rapid expansion of the rupture area can further enhance the short-period radiation burst at depths \cite{Kurahashi2013ShortPeriod}.
In contrast, shallow regions are dominated by large slip occurring primarily at low frequencies (below 10~s), consistent with crack-like behavior suggested in finite-fault models (e.g., \cite{Yagi2011Rupture, Melgar2015Kinematic}, Fig.~\ref{Fig4:depth_dependent_rupture_style}c–e).

\clearpage
\newpage

\begin{figure}
\noindent\includegraphics[width=\textwidth]{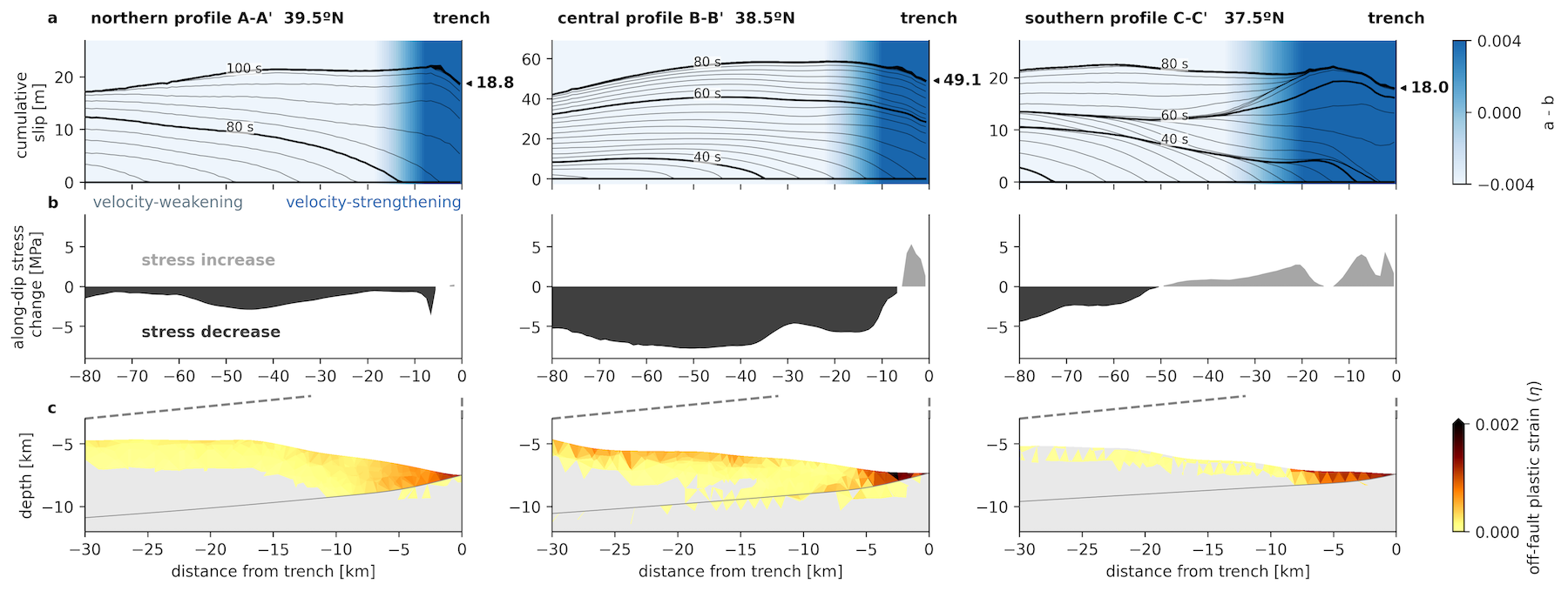}
\caption{
\textbf{Large near-trench slip occurs despite shallow velocity-strengthening frictional behavior, shallow off-fault plastic deformation, and negative stress drop.}
Along-dip profiles of slip, stress drop, and off-fault deformation at northern (39.5$^\circ$, first column), central (38.5$^\circ$, second column), and southern (37.5$^\circ$, third column) cross-sections.
(a) Evolution of cumulative slip at 2~s (thin lines) and 20~s intervals (thick lines). Background color shading indicates frictional behavior, transitioning from velocity-weakening (light blue) at depth to velocity-strengthening (dark blue) near the trench. Bold values at the right indicate the fault slip amplitude at the trench.
(b) Along-dip shear stress change ($\Delta\tau$), with dark gray representing stress decrease and light gray representing stress increase.
(c) Close-up view of the distribution of off-fault plastic strain (color shading, quantified as $\eta$, Methods Sec.~``\nameref{Off-fault plasticity}'') and the megathrust interface geometry (gray line).
}
\label{Fig5:trench_slip}
\end{figure}
\clearpage
\newpage

\subsection*{Large slip to the trench}
Shallow slip reaching the trench is critical for assessing tsunami hazards associated with large megathrust earthquakes. Our simulation demonstrates substantial near-trench slip, driven by coseismic frictional weakening, despite competing effects from shallow off-fault plastic deformation,
velocity-strengthening frictional behavior, and the associated negative stress drop.  
Figure~\ref{Fig5:trench_slip}a presents three along-dip profiles of slip from south to north. The modeled trench slip magnitudes of 18.8~m, 49.1~m, and 18.0~m at the southern, central, and northern cross-sections are comparable to differential bathymetry measurements, which indicate horizontal displacements of approximately 50-70~m at the central region and up to 20~m to the northern and southern extents (Supplementary Fig.~\ref{FigE5:disp_field}, \cite{Kodaira2020Large, Ueda2023Submarine}). 

In our model, dynamic frictional weakening effectively sustains shallow rupture propagation, enabling up to 50~m of slip near the trench.
Wave-mediated dynamic stressing, including both free-surface-reflected rupture fronts and updip-propagating fronts reactivated downdip, increases shear traction and slip rate on the shallow interface. As shallow slip rates approach and eventually exceed $V_w$,  fast-velocity weakening friction is activated, producing a second acceleration phase  (Fig.~\ref{Fig3:reactivation}b, Supplementary Figure~S5a).
The velocity-strengthening friction ($\le$9~km depth) and fast velocity-weakening friction adopted in our model are motivated by laboratory friction measurements of borehole-recovered fault gouge samples from the shallow high-slip region of the Tohoku-Oki earthquake \cite{Ujiie2013Low} and by laboratory constraints indicating that clay-rich lithified rock and rock gouges commonly exhibit velocity-strengthening to transitional behavior at low effective normal stress \cite{Saffer2003Comparison}.
This assumption is intended as a generalized representation of shallow fault materials and does not imply that the entire shallow megathrust segment behaves as unlithified sediment.
Additionally, off-fault plastic deformation dissipates seismic energy in the uppermost 10 km (Fig~\ref{Fig5:trench_slip}, \cite{Ma2023Wedge}), though its overall contribution remains limited, accounting for only 2.9\% of the total on-fault seismic moment.
Lastly, our assumed initial stress state (Supplementary Fig.~\ref{FigE1:Initial stress conditions}b) features low to negative shear stress near the trench, implying limited near-trench strain accumulation prior to the Tohoku-Oki earthquake \cite{Loveless2016Two}.
Nonetheless, the collective effects of velocity-strengthening friction, off-fault plasticity, and low prestress conditions can only modestly reduce trench slip, by about 10\% relative to the maximum slip further down-dip (Fig.~\ref{Fig5:trench_slip}).
This reduction in trench slip aligns with near-trench bathymetric evidence of inelastic deformation and a decrease in horizontal displacement at the trench \cite{Zhang2023Complex}.

\subsection*{Spontaneous along-strike rupture arrest}
\begin{figure}
\noindent\includegraphics[width=\textwidth]{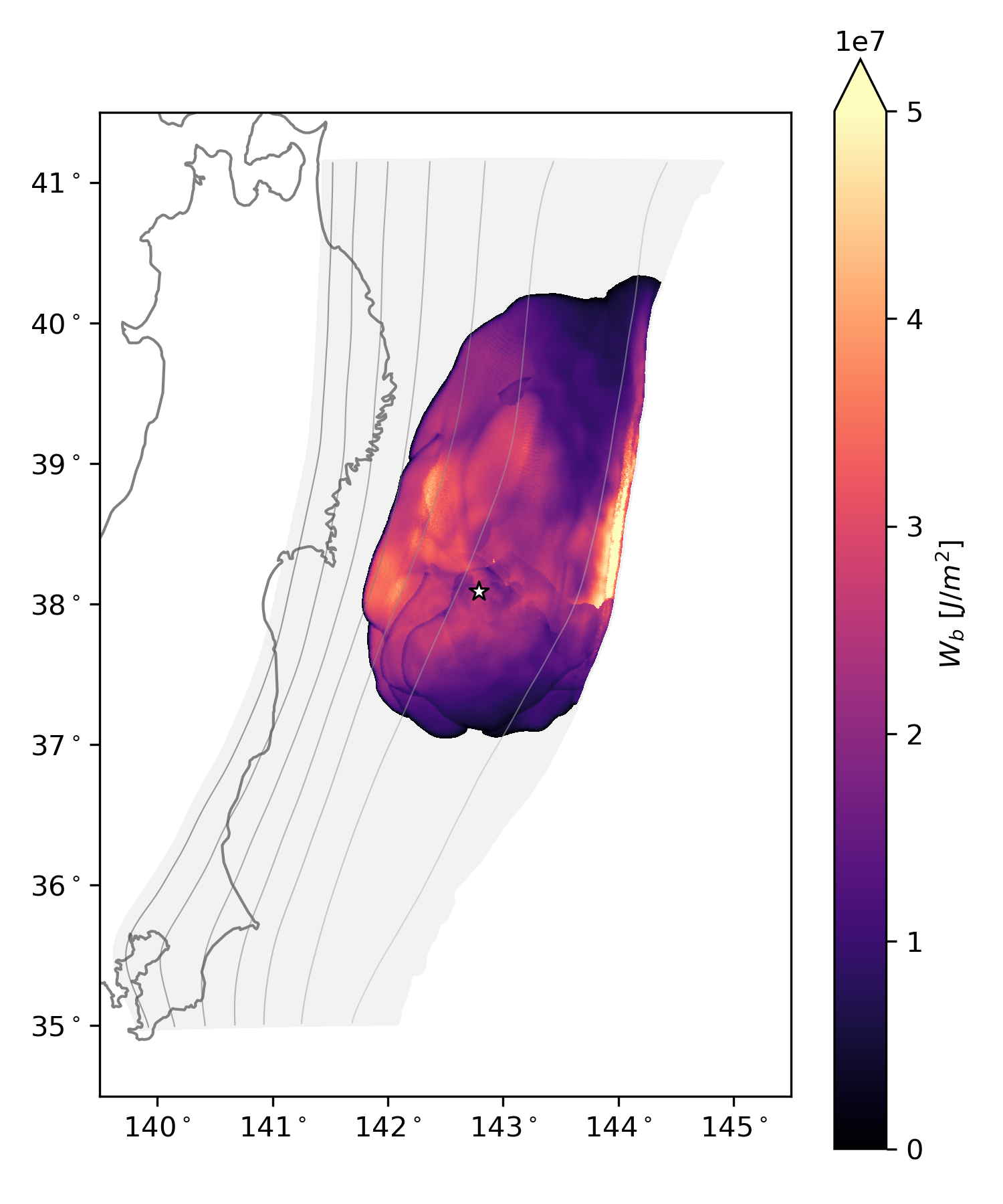}
\caption{\textbf{Breakdown work density ($W_b'$, Methods Sec.~``\nameref{Breakdown work density}'') distribution for the preferred dynamic rupture model.} The preferred model shows pronounced spatial variation in breakdown work density, where downdip slip pulses increase the breakdown work density, compared to the updip region.}
\label{Fig7:Breakdownenergy}
\end{figure}
Our preferred model successfully reproduces the spontaneous along-strike rupture arrest of the Tohoku-Oki earthquake by incorporating data-informed stress heterogeneity rather than prescribing ad hoc frictional or structural barriers.   
In linear elastic fracture mechanics, dynamic rupture arrest occurs where available strain energy becomes insufficient to exceed local fracture energy, halting further rupture propagation \cite{Ke2018Rupture, Barras2023How}.
In our simulation, rupture arrest results from strain energy depletion, as evidenced by a reduction in breakdown energy near the fault areas where rupture terminates (Fig.~\ref{Fig7:Breakdownenergy}, Methods Sec.~``\nameref{Breakdown work density}'').
During the later stages of rupture (70--120~s), bilateral rupture pulses propagate into regions with lower relative prestress levels ($R$) (Fig.~\ref{Fig1:Initial_conditions}~d), progressively decreasing in slip rate amplitude, rupture speed, and pulse width (Fig.~\ref{Fig2:preferred_overview}d,e). 

An exception occurs in the shallow northern slab section, where rupture arrest is delayed.
Here, a localized region of elevated relative prestress, located at the northeast and updip of the hypocenter (Fig~\ref{Fig1:Initial_conditions}d), generates a high slip rate rupture pulse propagating toward the northern shallow margin (Fig.~\ref{Fig3:reactivation}a). 
This interaction facilitates extended shallow rupture in the northern slab region, producing uplift patterns consistent with those derived from tsunami waveform inversions (Supplementary Fig.~\ref{FigE5:disp_field}, e.g., \cite{Satake2013Time}). 

In contrast, alternative simulations that use solely depth-dependent prestress conditions informed by regional principal stress orientations without stress heterogeneities fail to spontaneously arrest rupture (Fig.~\ref{Fig1:Initial_conditions}b, \cite{Heidbach2018World}). Under comparable average prestress levels, these laterally homogeneous prestress models result in rupture of the entire megathrust, yielding an unrealistic moment magnitude of $M_W 9.61$ and a prolonged rupture duration of 220~s (Supplementary Fig.~\ref{FigE7:Compare_arrest}).

%% file: 4Discussion.tex


\clearpage
\newpage
\begin{figure}
\noindent\includegraphics[width=\textwidth]{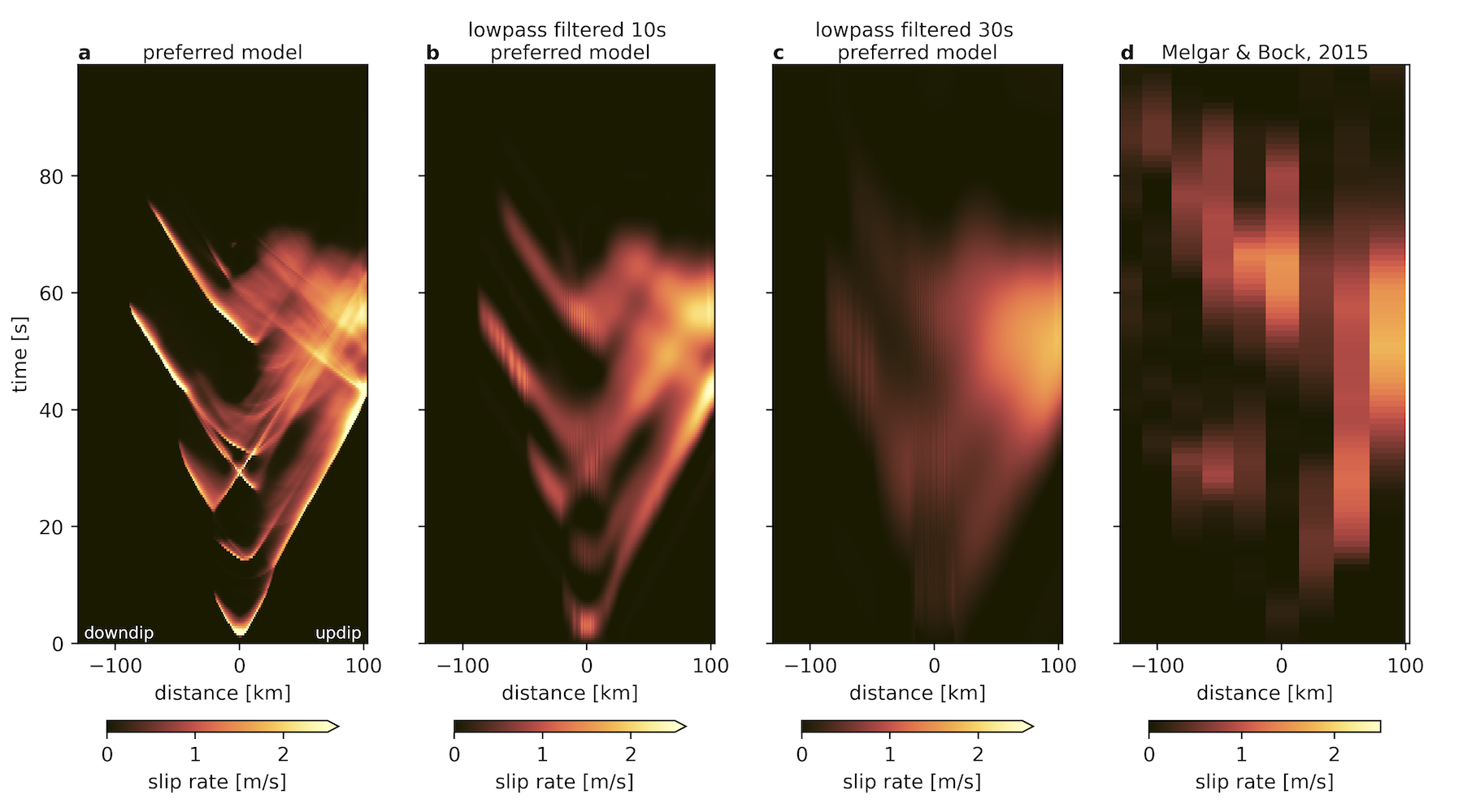}
\caption{\textbf{Comparison of slip-rate profile along a hypocentral dip profile between dynamic rupture models and the finite-fault model of \citet{Melgar2015Kinematic}.} (a) Along-dip slip-rate evolution of the preferred dynamic rupture model with heterogeneous prestress. (b) The same modeled slip-rate evolution low-pass-filtered at 10~s. (c) Modelled slip-rate evolution low-pass-filtered at 30~s period. (d) Along-dip slip-rate evolution of the \citet{Melgar2015Kinematic} finite-fault slip model. The profile is constructed from nine subfaults. Each subfault slip-rate function is represented by 20 triangular source time functions that are 10s-long and 50\%-overlapping .}
\label{Fig9:SR_FFM_comp}
\end{figure}

\clearpage
\begin{figure}
\noindent\includegraphics[width=\textwidth]{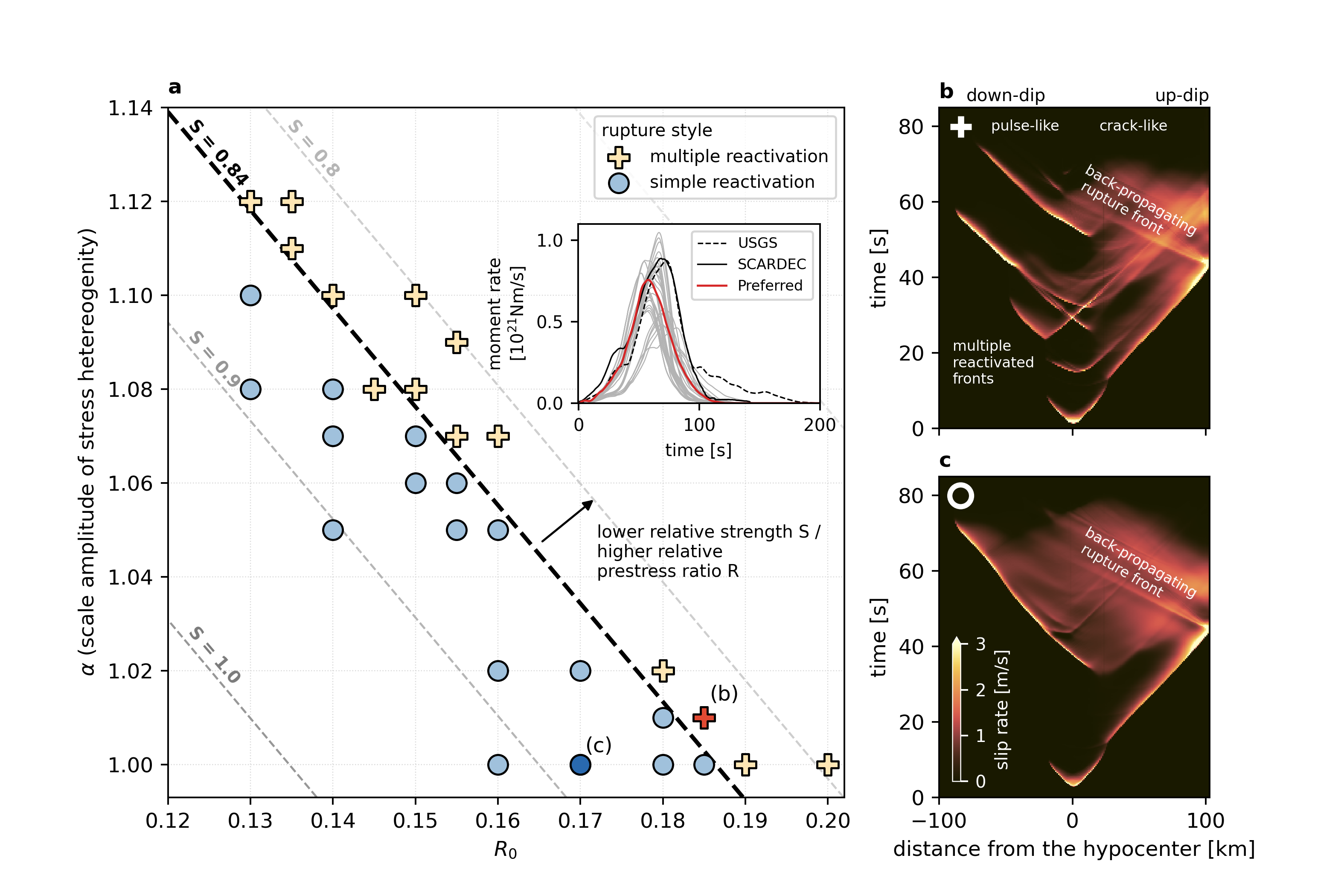}
\caption{\textbf{Distinct rupture styles under hypocentral friction and prestress.} (a) Phase diagram showing how variations in stress heterogeneity amplitude ($\alpha$) and regional ambient stress level ($R_{0}$) control rupture style. 
Blue circles represent a family of dynamic rupture models dominated by prominent pulse-like rupture with free-surface reflection and back-propagating crack-like rupture, while yellow crosses denote models exhibiting repeated rupture reactivation near the hypocenter resembling the preferred model.
The preferred model and an exemplary simpler model are indicated by the red cross and dark blue circle, respectively, with their corresponding along-dip slip-rate evolutions shown in (b) and (c), respectively.
Dashed contours illustrate the variation of hypocentral seismic $S$ ratio (initial strength excess over dynamic stress drop, Equation \ref{eq:S}), with the thick dashed line marking the transition boundary between rupture styles.
We observe that the amplitude of stress heterogeneity ($\alpha$) primarily controls peak slip magnitude, whereas the regional ambient stress level ($R_0$) largely determines rupture extent in both families of dynamic rupture models.
The inset compares the moment-rate functions of all models (gray lines), the preferred model (red), the USGS inversion \cite{Hayes2011Rapid} (black dashed), and the SCARDEC inversion result \cite{VallA2016New} (solid black). 
(b--c) Along-dip slip-rate evolution of the preferred model and exemplary simpler model, representing the rupture reactivation family and simpler rupture family, respectively. 
}
\label{Fig6:Grid_search}
\end{figure}
\clearpage

\begin{figure}
\noindent\includegraphics[width=\textwidth]{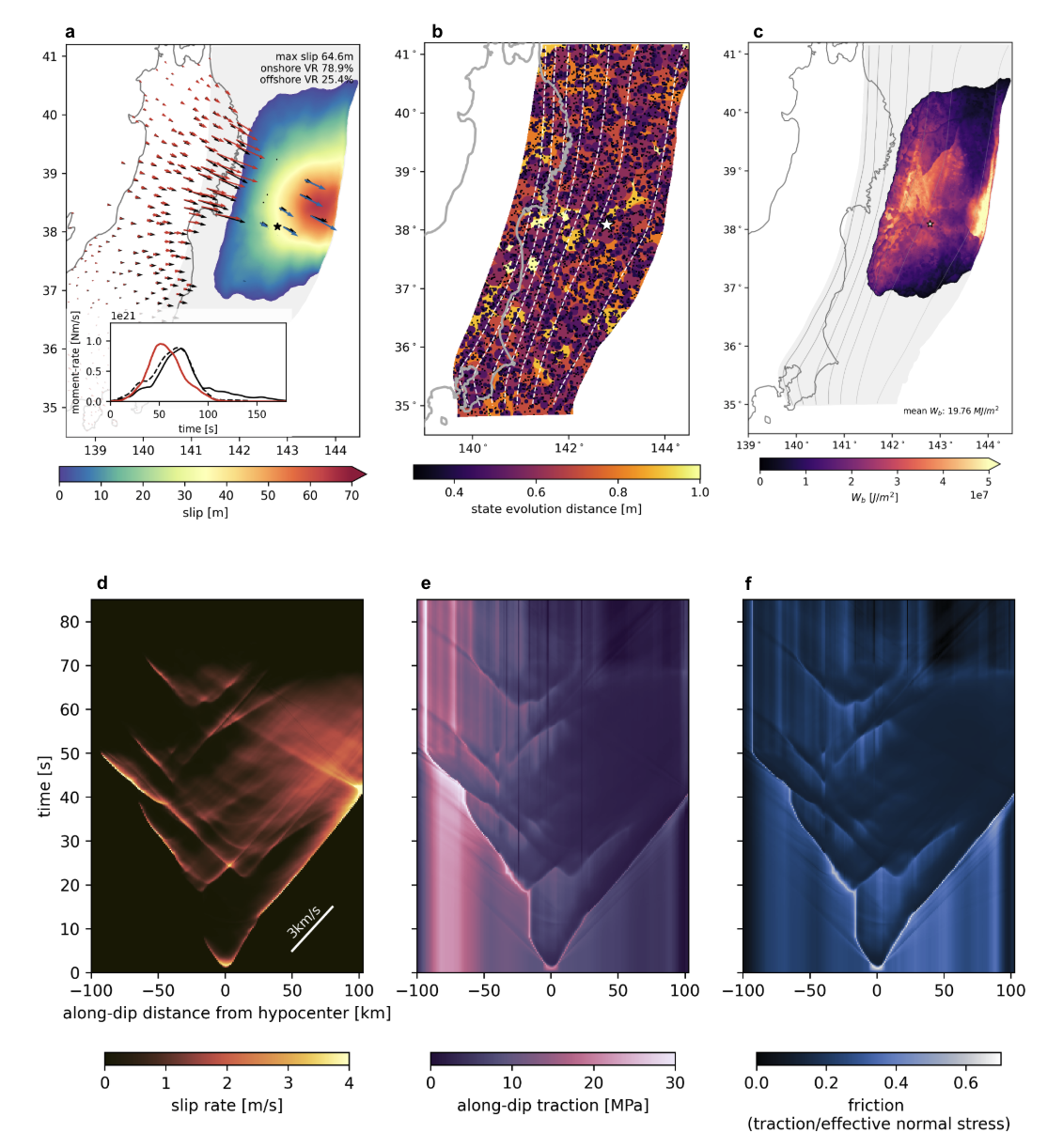}
\caption{\textbf{Dynamic rupture scenario incorporating heterogeneous state-evolution distance.} (a) Simulated slip distribution and corresponding geodetic deformation. Red and blue arrows indicate synthetic onshore and offshore displacements, respectively. Black arrows show the observations. The inset compares the simulated moment-rate function (red), with the USGS (solid black), and SCARDEC (dashed black) source model moment-rate functions \cite{Hayes2011Rapid, Vallee2013Source}. (b) Spatial distribution of the state-evolution distance $L$ with multiscale heterogeneity ranging from 0.3 to 1.0 m. (c) Distribution of breakdown work density. (d-f) Along-dip profiles of slip-rate, along-dip traction, and effective friction coefficient evolution, from left to right, respectively. See also Supplementary Figure~\ref{SFig:Sl0_snapshot} and Video S4.}
\label{Fig10:Sl0_het_model}
\end{figure}

\section*{Discussion}
\subsection*{Comparison of modeled rupture reactivation with observations}

Strong-ground motion move-out patterns and finite-fault slip models of the Tohoku-Oki earthquake provide evidence for multiple rupture episodes \cite{Ide2011Shallow, Lee2011Evidence, Yue2011Inversion, Melgar2015Kinematic}. 
Similar rupture complexities have been documented in other large megathrust earthquakes, including the 2010 $M_W$ 8.8 Chile earthquake \cite{Okuwaki2014Relationship}, and the 2019 $M_W$ 8.0 Northern Peru earthquake \cite{Vallee2023Selfreactivated}.

Our simulations reproduce this first-order behavior, showing spontaneous rupture reactivation in the hypocentral region. 
However, our preferred model exhibits more reactivation episodes than the two to three inferred from most finite-fault models. 
These discrepancies likely reflect resolution limits in finite-fault inversions. The spectra of the slip-rate functions of the finite-fault models exhibit fast spectral decay starting at a period of 30~s \cite{Melgar2015Kinematic}, whereas the slip-rate functions of our models show a more gradual spectral decay that starts between 30 and 10~s.
Therefore, finite-fault models are likely to miss episodic, high-frequency slip pulses at depth.
When low-pass filtered at 30~s, our simulated slip-rate evolution reproduces the reported patterns in \cite{Melgar2015Kinematic}, including the updip rupture propagation during the first 50~s followed by a back-propagating rupture front between 50 and 90~s  (Fig.~\ref{Fig9:SR_FFM_comp}). 
In addition, the preferred model reproduces key qualitative features of strong-motion records near the main rupture area, particularly the relative timing and spectral content up to ~0.5 Hz, although long-period trends and absolute amplitudes are not fully captured (Supplementary Figs. S9–S10).

\subsection*{Impact of prestress on dynamic rupture complexity}
Our preferred model indicates that the interplay of dynamically evolving friction and stresses gives rise to mixed rupture styles and depth-dependent rupture propagation. We next investigate how rupture evolution depends on the amplitude of stress heterogeneity and the relative prestress level by systematically varying $\alpha$ and $R_0$, which represent the stress heterogeneity amplitude and regional ambient stress level. 
We find two distinct dynamic rupture styles (Fig.~\ref{Fig6:Grid_search}): one resembling the preferred model, and another dominated by a prominent updip-propagating pulse that reflects at the free surface, back-propagates, and transitions to crack-like rupture.

We find that our preferred rupture style requires low fault strength compared to the dynamic stress drop, corresponding to a low seismic $S$ ratio \cite{Das1977Numerical} (Equation~\ref{eq:S}, ratio of initial strength excess over dynamic stress drop, See Methods Sec.``\nameref{Prestress}''). The $S$ ratio measures the balance between fault proximity to failure and the available stress release during dynamic weakening.
Lower effective normal stress and fault strength or higher initial shear stress favor reactivation-dominated scenarios. 
The abrupt transition in rupture behavior with small incremental changes in prestress is consistent with prior work on the interplay between evolving friction and stress (e.g., \cite{Zheng1998Conditionsa, Gabriel2012Transition}). 
However, theoretical and numerical models generally predict that higher prestress downdip favors crack-like rupture, whereas lower prestress promotes updip pulse-like propagation.
Consistent with the expectation, models without stress heterogeneity (Method Sec.~``\nameref{Regional stress rupture model without stress heterogeneity}''), exhibit a pulse-to-crack rupture transition, characterized by a distinguishable rupture reactivation episode (Supplementary Figs.~\ref{FigE8:Regional_model}, \ref{FigE9:Reg_depth-varying},\ref{SFig:Regional_model_5s_snapshot}, and Supplementary Video~S2).

In contrast, heterogeneous stress conditions in our models cause locally low prestress regions, particularly near the hypocenter and downdip, which facilitate the formation of self-healing pulse-like ruptures \cite{Zheng1998Conditionsa}. 
Subsequent slip arrest at depth reaccumulates shear stress along previously ruptured segments \cite{Nielsen2003SelfHealing}, thereby enabling reactivation.
All reactivated ruptures are pulse-like down-dip, due to the locally heterogeneous nature of the slip left behind by the primary rupture front, which correlates with the distribution of the primary residual stresses \cite{Gabriel2012Transition}.
Updip, reactivated fronts coalesce with the free-surface reflected primary rupture to form a sustained crack-like rupture (Fig.~\ref{Fig4:depth_dependent_rupture_style}a, Supplementary Fig.~\ref{FigE3:Profile_comp}).
Comparison with observed regional strong ground motion records shows that the preferred model with multiple reactivation shows multiple move-out branches, whereas the single reactivation model produces a single dominant pulse that is inconsistent with the observations (Supplementary Fig.~\ref{SFig:SMA_comp}).

\subsection*{Origin and role of prestress heterogeneity}
Defining initial stress conditions is a key challenge in dynamic rupture modeling. 
Prestress heterogeneity has been constrained in several ways, including using kinematic geodetic coupling models (e.g., \cite{Hok2011Dynamic}), Coulomb stress changes from preceding earthquakes or slow slip events  (e.g., \cite{Oral2022Method}), and coseismic slip distributions (e.g., \cite{Tinti2021Constraining}). 
Importantly, prestress heterogeneity does not uniquely determine dynamic rupture evolution, which is governed by the nonlinear interaction between frictional evolution, stress redistribution, and seismic wave propagation (e.g., \cite{Kammer2024Earthquake}), rather than by static initial conditions alone, including prestress. 
The stress patterns inferred from finite-fault slip models capture large-scale stress variations associated with the preceding strain accumulation and coseismic stress release, but identical stress heterogeneities can lead to different rupture evolution, e.g., depending on frictional properties \cite{Tinti2021Constraining}. 

\subsection*{Rupture scenarios from alternative finite-fault slip models}
To evaluate the effects of different prestress heterogeneity patterns, we explore dynamic rupture scenarios using three additional sets of prestress distributions derived from three finite-fault slip models: \cite{Melgar2015Kinematic, Yamazaki2018Self, Kubota2022New}. 
\citet{Melgar2015Kinematic} and \citet{Yamazaki2018Self} jointly invert seismic, geodetic, and tsunami data, whereas \citet{Kubota2022New} inverts using tsunami data only. 
In contrast to the median slip model, which features a single smooth slip patch, these slip models exhibit more heterogeneous slip. 
The \citet{Melgar2015Kinematic} and \citet{Yamazaki2018Self}models contain small-scale slip patches updip of the hypocenter, while the \citet{Kubota2022New} model shows slip asperities in the southern region (Supplementary Figs.~\ref{SFig:Kubota_sim}, ~\ref{SFig:Melgar_sim}, \&~\ref{SFig:Yamazaki_sim}). 
The corresponding stress-change patterns display greater heterogeneity and distinct spatial patterns. 

Despite the contrasting prestress patterns, all alternative dynamic rupture models yield comparable rupture complexity, including multiple reactivation, large shallow slip near the trench, and spontaneous rupture arrest (Supplementary Figs.~\ref{SFig:Kubota_sim}-~\ref{SFig:Yamazaki_SR_snapshot}. Stronger prestress heterogeneity in the \citet{Melgar2015Kinematic} and \citet{Yamazaki2018Self} models produces more complex rupture evolution with multiple slip pulses across depth. 
For example, the \citet{Melgar2015Kinematic} model initiates with an updip slip pulse followed by reactivation updip of the hypocenter. Although these alternative models reproduce surface displacements well, they generate multiple peaks in moment-rate functions that are inconsistent with the observationally inferred triangular moment-rate function.
In contrast, the preferred rupture model based on the median model of \cite{Wong2024Quantitative} can dynamically reproduce the triangular moment-rate function. 

Taken together, these alternative simulations also indicate that the key dynamic rupture complexity identified in our preferred model arises robustly from the underlying dynamic processes rather than from a particular choice of prestress distribution.

\subsection*{Frictional heterogeneity}
In addition to prestress heterogeneity, frictional heterogeneity is also an inherent part of the subduction megathrust heterogeneity. 
To compare the effects of these two types of heterogeneity, we conduct two complementary sets of dynamic rupture simulations introducing explicit frictional heterogeneity \cite{Ide2013Historical, Tinti2021Constraining}, totaling ten models. 
In the following, we discuss alternative `preferred' scenarios for each ensemble, defined as the models that yield the highest variance reduction relative to the geodetic observations and the SCARDEC seismic moment-rate function \cite{VallA2016New}. 
These alternative models qualitatively agree with the preferred heterogeneous prestress model presented in the Results Section. 
This highlights that similar dynamic rupture behavior can emerge from different forms of pre-existing heterogeneity in stress or frictional strength, when coupled with rapid frictional weakening and restrengthening.

In the first set of four simulations, we impose multiscale spatial variations in the state-evolution distance ($L$, Equation 5, and Methods: ``\nameref{Fault friction}''), following the fractal parameterization of \citet{Ide2005Earthquakes} (Supplementary Fig.~\ref{Fig10:Sl0_het_model}b).
We superimpose heterogeneity in the $L$ (0.3--1.0~m) onto the preferred model, which includes prestress heterogeneity.
Despite this additional frictional heterogeneity, dynamic rupture evolution remains qualitatively consistent with the preferred model. 
We observe similar multiple reactivation episodes, mixed downdip pulse-like and updip crack-like rupture, and large shallow slip toward the trench (Fig.~\ref{Fig10:Sl0_het_model}d). 
This heterogeneous friction model re-nucleates even more frequently than the depth-dependent friction case, exceeding the six reactivation episodes observed in the preferred model (Supplementary Fig.~\ref{SFig:Sl0_snapshot}; Video~S4). 
It yields a breakdown work density comparable to the preferred model (19.8 MJ/m$^2$ versus 19.6 MJ/m$^2$), but a higher total radiated energy ($10.1 \times 10^{17} J$ versus $7.7 \times 10^{17}~J$). 
The patches of shorter local nucleation length promote this frequent reactivation in the downdip and hypocentral regions.

The second set of six simulations maps heterogeneity inferred from the median slip distribution onto the fully-weakened frictional strength ($f_w \sigma’$, Equation~1, and Methods section:``\nameref{Fault friction}''), while maintaining a homogeneous, depth-dependent initial shear stress, following \cite{Tinti2021Constraining} (Supplementary Fig.~\ref{SFig:Fw_compare}). 
Despite the contrasting assumptions about frictional heterogeneity, the heterogeneous-$f_w \sigma’$ preferred model reproduces key rupture complexities of the preferred model, including multiple reactivation, mixed rupture styles, and spontaneous arrest. 
However, unlike the preferred heterogeneous prestress model, depth-dependent rupture style variability is less clear, and rupture transitions into a pure crack-like style after 50~s simulation time (Supplementary Fig.~\ref{SFig:Fw_sim}) and arrests prematurely (Supplementary Figs.~\ref{SFig:Fw_sim}–\ref{SFig:Fw_snapshot}; Video~S5). 
Because mapping heterogeneity into the fully-weakened friction coefficient decreases the local strength excess relative to the dynamic stress drop (Supplementary Fig.~\ref{SFig:Fw_slip_strength_illustration}), the heterogeneous dynamic-friction model has a lower seismic $S$ ratio and therefore tends toward crack-like rupture \cite{Zheng1998Conditionsa, Nielsen2000Rupture, Gabriel2012Transition}. 
In contrast, the heterogeneous prestress model has a higher $S$ ratio, which promotes self-healing pulse-like rupture and a more gradual arrest through decaying slip pulses (e.g., \cite{Nielsen2003SelfHealing, Barras2023How}).

\subsection*{Comparison to previous models of the Tohoku-Oki earthquake}

The observed penetration of slip into the velocity-strengthening near-trench region is consistent with the 2D dynamic simulations of \cite{Kozdon2013Rupture}, which used classical rate-and-state friction and a realistic elastic structure including bimaterial contrasts. In their models, interaction with the free surface and enhanced shear loading from deeper slip sustain trench-breaching rupture despite velocity-strengthening friction. In our 3D simulations, enhanced frictional weakening and inelastic deformation reduce the near-trench peak slip by only about 10\%, reproducing the large near-trench slip required by seafloor displacement measurements \cite{Fujiwara20112011, Ito2011Frontal, Kodaira2020Large,Zhang2023Complex}. The agreement corroborates the potential for large coseismic displacements on shallow faults driven by strong coseismic weakening, despite the presence of velocity-strengthening friction and off-fault plastic deformation.

Our simulations differ from previous studies in the physical conditions required to obtain rupture complexity.
Most previous Tohoku dynamic rupture models have relied on linear slip-weakening friction laws and prescribed asperities in stress or strength to promote multiple subevents and along-depth variability \cite{Huang2014SlipWeakening, Galvez2020Dynamic}. 
In contrast, our simulations employ a fast velocity-weakening rate-and-state friction law that incorporates rapid coseismic restrengthening and heterogeneous prestress constrained by observation-derived slip heterogeneity. 
Within this framework, multiple rupture reactivation and mixed pulse-like and crack-like behavior arise spontaneously from the co-evolution of friction, stress, and 3D fault geometry, without prescribing depth-dependent frictional patches or time-dependent reactivation schemes.

While our simulations reproduce pronounced downdip complexity and a systematic reduction of rise time at depth, back-projection studies of the 2011 Tohoku-Oki earthquake show comparatively weak high-frequency radiation from the shallow rupture \cite{Meng2011Window, Lay2012Depthvarying}. Reconciling the discrepancy with our models may require additional mechanisms that may damp shallow peak slip-rates and high-frequency radiation, such as inelastic yielding of the accretionary wedge \cite{Ma2023Wedge, Ulrich2022Stress} and depth-dependent frictional variations \cite{Huang2014SlipWeakening}.

\subsection*{Limitations} 
We do not evaluate the effects of geometrical complexity and 3D velocity structure on rupture dynamics.
Our simulations use a 1D velocity model and a smoothed megathrust geometry and therefore do not capture bimaterial contrasts across the plate interface or detailed upper-plate and sedimentary structure. 
Whereas these simplifications enable systematic exploration of how frictional and prestress parameters control dynamic rupture, additional processes proposed to influence rupture complexity in large megathrust earthquakes are not represented here, including along-strike variations in shallow material properties and sediment heterogeneity \cite{Moore2015Sediment}, accretionary wedge geometry \cite{Tsuru2002Alongarc}, interface roughness and upper-plate structural heterogeneity \cite{Bassett2015Gravity}, bimaterial effects across the plate interface \cite{Shi2006Dynamic, scholz2014rupture, Kozdon2013Rupture}, and spatial variations in pore-fluid pressure \cite{Noda2013Stable}. 
Incorporating these 3D effects is an important direction for future work.

The goal of this study is to identify dynamically viable rupture scenarios that reproduce the key source characteristics and illuminate the governing physical mechanisms, rather than to perform an inversion of all available observations. 
However, characterizing dynamic parameters, including shear stress and frictional properties, remains challenging because of strong trade-offs (e.g., \cite{schmedes2010correlation}). 
Whereas this work focuses on coseismic dynamic rupture, future work may integrate observations across time-scales to help constrain the governing faulting conditions \cite{Premus2022Bridging, Schliwa2024Linked}.

Using 3D dynamic rupture simulations, we demonstrate that the complex rupture behavior of the Tohoku-Oki earthquake can spontaneously arise from dynamic rupture processes.
We highlight that fault heterogeneity, although challenging to quantify, can be informed by existing observational data and fundamentally controls the complexity and scale of dynamic rupture. 
Our models capture episodic re-nucleation in the hypocentral region, the co-existence of short-duration, spiraling slip pulses at depth and shallow crack-like rupture with large slip near the trench, and spontaneous rupture arrest.
To this end, this work provides a robust, self-consistent framework applicable to other megathrust settings towards physics-based earthquake and tsunami hazard assessment worldwide.

%% file: 5Methods.tex
We perform 3D dynamic rupture simulations of the 2011 $M_w$ 9.0 Tohoku-Oki earthquake that simultaneously solve for seismic wave propagation,  on-fault frictional failure, and off-fault inelastic deformation.
2D and 3D dynamic rupture simulations have been applied to subduction zones worldwide 
 (e.g., \cite{Hok2011Dynamic, Yao2020Rupture, prada2021influence, Ramos2021Assessing, Ulrich2022Stress, Chan2023Impact, Wirp2024Dynamic, Li2024Linking}), 
including the Tohoku-Oki earthquake \cite{Duan2012Dynamic,Ide2013Historical, Kozdon2013Rupture, Huang2014SlipWeakening, Galvez2020Dynamic, Ma2023Wedge}.
We use the open-source software SeisSol (https://seissol.org) for all dynamic rupture simulations on two supercomputers, SuperMUC-NG, at the Leibniz Supercomputing Center, Germany, and Frontera, at the Texas Advanced Computing Center, United States.
SeisSol employs the Arbitrary High-order Derivative (ADER) Discontinuous Galerkin (DG) method \cite{Dumbser2006Arbitrary}, which enables higher-order accuracy in space and time on unstructured tetrahedral meshes, which are well-suited to capture geometric complexities, including shallowly dipping megathrust interfaces in subduction zones \citep[e.g.,][]{Ramos2021Assessing, Ulrich2022Stress}. 
SeisSol is optimized for high-performance computing \citep[e.g.,][]{Heinecke_et_al_2014, Uphoff2017Extreme, Krenz20213D} and verified in dynamic rupture community benchmarks \citep{Pelties2012Threedimensional, Pelties2014Verification, Harris2018Suite, Taufiqurrahman2022Broadband}.
We employ SeisSol with sixth-order accuracy in time and space, i.e., the polynomial order of the basis functions is $p$~=~5.
 
\subsection*{Model geometry and mesh}
\label{Model geometry and mesh}
Our 3D dynamic rupture models incorporate realistic megathrust geometry, high-resolution topobathymetry, and velocity-aware adaptive mesh refinement to accurately capture rupture processes and seismic wave propagation up to 2~Hz \cite{Breuer2022NextGeneration}. 
Our megathrust geometry is adapted from the 3D Japan Integrated Velocity Structure Model geometry (JIVSM) \cite{Koketsu2009Proposal, Koketsu2012Japan}, which is based on seismic imaging, waveform inversion, and seismicity studies.
We extract the top layer of the oceanic plate as the megathrust interface. 
To ensure the interface connects to the trench, we extend and smooth the interface to the USGS trench with a 12$^\circ$ extension from the surface following \citet{Wong2024Quantitative}. 
The constructed megathrust interface spans a region from 35$^\circ$N to 41$^\circ$N and from 139.5$^\circ$E to 145$^\circ$E, extending approximately 700~km along strike, 250~km along dip, and reaching a depth of 80~km (Fig.~\ref{Fig1:Initial_conditions}). 
This large extent of the megathrust geometry can prevent model boundary effects on dynamic rupture arrest. 

Our model incorporates realistic topobathymetry using the Geobco dataset at 15 arc~s (380~m) resolution \cite{Geobco2024}. 
We use a 1D velocity structure consisting of five layers, which we modify from \cite{Fukuyama1998Automated} (Supplementary table~\ref{ST:Velocity_structure}) by prescribing a 20\% shear modulus reduction in the two uppermost layers. 
This adjustment accounts for the presence of lower-rigidity materials in shallow subduction zone regions and closely matches the average shear modulus derived from the 3D JIVSM velocity model at equivalent depths.

Our structure model is refined near the fault interface to accurately resolve the process zone at the rupture tip, near the free surface to capture topobathymetry, and near Honshu Island to resolve the seismic wavefield up to 1.5~Hz (Supplementary Sec.:~``\nameref{SM: Model domain and resolution}'', Fig. \ref{SFig:Knet_vel_spec}). 
The process zone width $\Lambda$ \cite{Day2005Comparison} is defined as the area behind the rupture front in which the shear stress decreases from the static value to the dynamic value. 
The resulting unstructured tetrahedral mesh consists of 50.3 million elements, and one simulation requires 58,000~CPU hours on Frontera and 64,000~CPU hours on SuperMUC-NG. 
The mesh uses the following Cartesian projection: WGS84/UTM transverse mercator centered at (143$^\circ$E, 39$^\circ$N).

\subsection*{Fault friction}
\label{Fault friction}
We use a fast velocity-weakening rate-and-state friction law that replicates the severe coseismic friction reduction observed in high slip-rate laboratory experiments \cite{DiToro2011Fault, Goldsby2011Flash}, including studies of using drilled samples from the Japan subduction zone \cite{Ujiie2013Low, Brodsky2020State}. 
Such pronounced weakening at elevated slip rates can result from flash heating of highly stressed, short-lived contact asperities and thermal pressurization due to shear heating of pore fluids \citep[e.g.,][]{Rice2006Heating, Beeler2008Constitutive, Ujiie2010Highvelocity, Viesca2015Ubiquitous}.

This friction law allows the megathrust fault interface to operate under low average shear stress while producing realistic fault slip and stress drop during dynamic rupture \cite{Noda2009Earthquakea, Ulrich2019Dynamic}, and promoting complex rupture styles including cascading ruptures across multi-fault systems \citep[e.g.,][]{Bizzarri2006Thermal, Noda2013Stable, Gabriel2012Transition, Schmitt2015Nucleation, Wang2017Seismic, Perry2020Nearly, Palgunadi2024Rupture}. 
Low average shear stress conditions align with the limited thermal signature of the Tohoku-Oki earthquake, which may imply rupture under low ambient stress levels \cite{Fulton2013Low}. 
Although this friction law has been extensively used in 2D simulations to examine its control on rupture dynamics \citep[e.g.,][]{Noda2009Earthquakea, Dunham2011Earthquake, Lambert2023Absolute} and in 3D models of crustal earthquakes \citep[e.g.,][]{Ulrich2019Dynamic, Premus2022Bridging}, it has not yet been explored in a 3D full-scale model of a large megathrust earthquake. 

We use the formulation suggested in the community benchmark problem TPV104 of the Southern California Earthquake Center \cite{Harris2018Suite}, which is similar to the friction law introduced by \citet{Dunham2011Earthquake}.
All frictional parameters are listed in Supplementary Table~\ref{T:Friction}. 

In the rate-and-state friction framework, frictional strength depends on both the state of the slipping surface and the current slip rate \cite{Dieterich1994Direct, Dieterich1979Modeling, Ruina1983Slip}. 
The shear traction $\tau$, is assumed to equal fault strength, and is given by
\begin{linenomath*}
\begin{equation}
\tau = f(V,\theta)\sigma'_n\,.
\end{equation}
\end{linenomath*}
$f$ is the effective friction coefficient, $V$ is the slip rate, $\theta$ is the state variable and $\sigma'_n$ is the effective normal stress. 

The frictional coefficient $f$ depends on $V$ and $\theta$, as 
\begin{linenomath*}
\begin{equation}
f(V,\theta) = a \sinh^{-1} \left[\frac{V}{2V_0} \exp\left(\frac{\theta}{a}\right)\right]\,,
\end{equation}
\end{linenomath*}
where $a$ is the direct-effect parameter and $V_0$ is the reference velocity. 
The evolution of $\theta$ is governed by 
\begin{linenomath*}
\begin{equation}
\frac{d\theta}{dt} = - \frac{V}{L}(\theta-\theta_{ss})\,,
\end{equation}
\label{eq:L}
\end{linenomath*}
where $L$ is the characteristic slip distance, $t$ is time, and $\theta_{ss}$ is the steady-state value of the state variable, which is given by
\begin{linenomath*}
\begin{equation}
\theta_{ss}(V) = a \ln\left[\frac{2V_0}{V}\sinh\left(\frac{f_{ss}V}{a}\right)\right]\,.
\end{equation}
\end{linenomath*}
The steady-state friction coefficient $f_{ss}$ is given by
\begin{linenomath*}
\begin{equation}
f_{ss}(V) = f_w + \frac{f_{LV_{ss}}(V) - f_w}{(1+(V/V_w)^4)^{1/4}}\,,
\end{equation}
\label{eq:F_ss}
\end{linenomath*}
where $V_w$ is the onset of the weakening velocity, $f_w$ is the fully weakened friction coefficient, and the steady-state low-velocity friction coefficient is:
  \begin{linenomath*}
 \begin{equation}
 f_{LV_{ss}} = f_0 - (b-a)\ln(V/V_0)\,,
\end{equation}
\end{linenomath*}
with $b$ as the state-evolution parameter, and $f_0$ as the reference friction coefficient.
The steady-state friction behavior is asymptotic, such that $f_{ss}(V)\approx f_{LV_{ss}}(V) \text{ for } V \ll V_w$ and $f_{ss}(V) \approx f_w \text{ for } V \gg V_{w}$. 
This behavior aligns with laboratory observations, capturing classic rate-and-state frictional behavior at low sliding velocities and pronounced frictional weakening at high sliding velocities.

Velocity-strengthening, $(a-b)>0$, friction describes materials whose frictional strength increases with rising slip rate, thus stabilizing fault slip. 
Conversely, velocity-weakening, $(a-b)<0$, friction characterizes materials that decrease in strength with increasing slip rate, facilitating the nucleation and propagation of unstable slip \cite{Dieterich1992Earthquake, Ampuero2008Earthquake}.
 
In our models, we prescribe a depth-dependent distribution of $(a-b)$ to represent realistic frictional behavior along the megathrust interface.  
The shallow portion ($<$10~km depth) of the Japan subduction zone is characterized by velocity-strengthening friction, consistent with laboratory measurements of frictional behavior of clay-rich accretionary wedge material at low slip velocities \cite{Saffer2003Comparison, Ikari2011Cohesive, Ujiie2013Low, Ikari2015Strength}. 
This velocity-strengthening region is constrained by the transition depth from accretionary wedge to bedrock, as defined by the JIVSM \cite{Koketsu2009Proposal, Koketsu2012Japan}. 
Between 10–45 km depth, we define the seismogenic zone by parameterizing velocity-weakening friction.  
Further downdip, between 40--50~km depth, friction gradually transitions back to velocity-strengthening, consistent with observed downdip limits of seismicity and diverse faulting behavior \cite{Nishikawa2019Slow, Nishikawa2023Review}. 
We assign uniform frictional parameters along strike within each depth interval, except near the hypocenter, where a modified state evolution distance is imposed for smooth rupture nucleation (Supplementary Section ~`\nameref{SM2: Nucleation}''). 
We emphasize that our assumptions of depth-dependent frictional properties do not imply a frictionally homogeneous megathrust. Our assumptions on depth-dependent frictional properties are sought to explore rupture dynamics driven solely by dynamic processes and without introducing ad-hoc frictional asperities and barriers. We acknowledge that the Japan trench subduction zone hosts diverse faulting behavior, including slow-slip events, tremors, low-frequency earthquakes, and moderate-to-large thrusting earthquakes \cite{Nishikawa2023Review}.

\subsection*{Depth-dependent effective normal stress}
Pore fluid pressure plays an important role in controlling the effective normal stress and thus the shear stress conditions governing earthquake rupture dynamics \citep[e.g.,][]{Madden2022State}. 
Drilling observations and seismic reflection studies \citep[e.g.,][]{Saffer2011Hydrogeology, JamaliHondori2022Connection} as well as stress orientation analyses \cite{Hardebeck2018Temporal} suggest elevated pore-fluid pressure in the Japan subduction zone, with measurements within the accretionary wedge and along the shallow fault interface reaching 80-95\% of lithostatic stress. 
Drilling observations and seismic reflection studies in the Japan subduction zone have documented elevated pore fluid pressure within the accretionary wedge and along the shallow fault interface, reaching 80-95\% of lithostatic stress \cite{Saffer2011Hydrogeology, JamaliHondori2022Connection}. 
Stress orientation analyses support elevated ambient pore fluid pressure at seismogenic depths \cite{Hardebeck2018Creeping}. 
We assume that pore fluid pressure reaches 90\% of lithostatic stress in all layers.
The lithostatic stress is defined as $P_{litho}(z) = \int_0^z(\rho_i g h_i) dz$, where the subscript $i$ refers to the respective layer in the velocity model (Supplementary Table~\ref{ST:Velocity_structure}) and $g = 9.81 m/s^2$ is the gravitational acceleration. 
This assumption results in depth-dependent effective normal stress and relatively low fault strength everywhere (Supplementary Fig.~\ref{FigE1:Initial stress conditions}c).

\subsection*{Off-fault plasticity}
\label{Off-fault plasticity}
We account for off-fault inelastic energy dissipation using a Drucker-Prager visco-elasto-plastic rheology \cite{Andrews2005Rupture, Wollherr2018Fault}. 
Models incorporating off-fault plasticity require specifying initial stress, bulk friction, and cohesion throughout the entire simulation domain.
We employ a depth-dependent cohesion model following \cite{Ulrich2022Stress} and motivated by laboratory-inferred shallow low  cohesion \cite{Saffer2003Comparison, Ikari2011Cohesive}, where bulk cohesion $C(z)$ varies linearly with effective confining pressure:
\begin{linenomath*}
\begin{equation}
C(z) = C_0 + C_1(z)\sigma'_c \, ,
\end{equation}
\end{linenomath*}
where $C_1(z)$ represents rock hardening with depth and $\sigma'_c = \sigma_{litho}-P_f$ is the effective confining stress. 
We set $C_0$ to 1.0~MPa to represent partially consolidated sediments, while $C1$ linearly reduces from 1 to 0 at depths shallower than 18~km.

The Drucker-Prager yield criterion is given by
\begin{linenomath*}
\begin{equation}
\tau_c = C(z) \cos(\Phi) - \sigma_m\sin(\Phi)\,,
\end{equation}
\end{linenomath*} 
where $\Phi = \arctan(f')$ is the internal angle of friction and $\sigma_m = \Sigma^3_{n=1}\sigma_{ii}/3$ as the mean stress. 

The closeness-to-failure (CF) metric \cite{Templeton2008Fault} is defined as the ratio between the magnitude of the deviatoric shear stress ($J_2$) and $\tau_c$:
\begin{linenomath*}
\begin{equation}
CF = \frac{\sqrt{J_2}}{\tau_c}\,.\label{Eq:CF}
\end{equation}
\end{linenomath*}
This parameterization results in shallow regions (above 10 km depth) being close to yielding ($CF~\approx 0.8$) under both preferred and regional stress conditions (Supplementary Fig.~\ref{SFig:CF_cohesion_profile}).

The total seismic moment $M_{0,t}$ is the sum of the moment due to the slip on fault, $M_{0,e}$, and the moment due to off-fault plastic strain, $M_{0,p}$ \cite{Gabriel2013Source,Ulrich2022Stress}, as:
\begin{linenomath*}
 \begin{equation}
M_{0,p} = \sum^N_{i=1}\mu V\eta\,,
 \end{equation}
 \end{linenomath*}
where $\mu$ is the rigidity, $V$ is the volume of each tetrahedral element $i$, and $\eta$ is a scalar quantity measuring the accumulated off-fault plastic strain at the end of the dynamic rupture simulation. 
Following \cite{Ma2008Physical}, $\eta$ is defined as: 
\begin{linenomath*}
\begin{equation}
\eta(t) =\int^t_0 \sqrt{\frac{1}{2}\dot{\epsilon}^p_{ij}\dot{\epsilon}^p_{ij}} dt\,,
\end{equation}
\end{linenomath*}
with $\dot{\epsilon}^p_{ij}$ as the 3D inelastic strain rate tensor. 

The contribution of plastic strain to the total moment is small for our rupture models. 
Ratios of $M_{0,p}/M_{0,e}$ for both the heterogeneous and regional relative prestress rupture scenarios are on the order of a few percent (preferred model: 2.9\%), consistent with 2D dynamic-rupture simulations and large scale megathrust simulation \cite{Gabriel2013Source, Ma2019Dynamic, Ulrich2022Stress} at comparable relative prestress levels. 

\subsection*{Prestress} 
\label{Prestress}

Variations in prestress and fault strength significantly influence rupture style and complexity \cite{Dunham2011Earthquake, Gabriel2012Transition, Nielsen2000Rupture, Zheng1998Conditionsa}. 
However, the prestress and strength conditions that govern earthquake rupture are challenging to directly constrain by observations \cite{Brown2015Static, Gallovic2019Bayesian}. 
Multiple approaches have been proposed to parameterize friction and/or stress conditions from locking models \cite{Yang2019Hypocenter, Yao2020Rupture, Chan2023Impact, Glehman2024Partial}, finite-fault slip distributions \cite{guatteri2000can, Weng2018Constraining, Weng2018Constraining, Tinti2021Constraining, Jia2023Complex, Hayek2024}, and stress change from prior events \cite{Taufiqurrahman2023Dynamics, Li2024Linking}.  
Previous dynamic rupture models of the Tohoku-Oki earthquake often assume rupture complexity from frictional properties and have relied on prescribed frictional asperities \citep[e.g.,][]{Ide2013Historical, Noda2013Stable, Ma2023Wedge} and, in some cases, additional stress asperities \citep[e.g.,][]{Duan2012Dynamic, Galvez2014Dynamic, Galvez2016Rupture, Galvez2020Dynamic, Huang2012Dynamic, Huang2014SlipWeakening}, both of which require ad-hoc assumptions. 

Here, we use a data-informed framework to explore the prestress conditions.
We define the initial stress tensor $s_{ij}$ as a linear combination of the regional-tectonically constrained stress tensor $b_{ij}$ and the stress changes inferred from finite-fault slip models $c_{ij}$, following \cite{Jia2023Complex, Hayek2024NonTypical}.
The initial full stress tensor $s_{ij}$ is defined as:
\begin{linenomath*}
 \begin{equation}
 s_{ij}(x,y,z) = \Omega(z)(b_{ij}(x,y,z) + \alpha c_{ij}(x,y,z)) + (1-\Omega(z)) \sigma'_n(x,y,z) \delta_{ij}\,,\label{Eq_initial_stress}
 \end{equation}
 \end{linenomath*}
with $\Omega(z)$ as a depth-dependent modulation function smoothly tapering deviatoric stresses below the seismogenic zone (45~km), $\alpha$ as a scaling factor controlling the amplitude of stress heterogeneity, and $\delta_{ij}$ is the Kronecker Delta.

Dynamic rupture simulations often exhibit strong trade-offs between friction and initial stress conditions \citep[e.g.,][]{Tinti2021Constraining}), which can be characterized by the relative prestress ratio $R$ between the maximum potential stress drop and frictional strength drop \cite{Aochi20031999}.
Following \citet{Ulrich2019Dynamic}, to define $R$ in our velocity-weakening rate-and-state friction framework, 
we approximate peak shear strength as $f_0 \sigma'_n$ and residual strength as the fully weakened frictional state, $f_w \sigma'_n$. 
During rupture, the shear stress level typically approaches this fully weakened frictional state (Supplementary Fig.~\ref{FigE3:Profile_comp}).
$R$ is then defined as
\begin{linenomath*}
\begin{equation}
R = \frac{\tau_0 - 
\mu_d\sigma'_n}{(\mu_s - \mu_d)\sigma'_n} \approx \frac{\tau_0 - f_w\sigma'_n}{(f_0 - f_w)\sigma's}\,,
\label{eq:R}
\end{equation}
\end{linenomath*}
where $\tau_0$ is the initial shear traction projected from $s_{ij}$ on the 3D megathrust interface.

Alternatively, initial stress and fault strength can be characterized by the seismic ratio $S$ \cite{Das1977Numerical}, which represents the ratio of initial strength excess to maximum dynamic stress drop:
\begin{linenomath*}
\begin{equation}
S = \frac{\mu_s\sigma'_n -\tau_0 }{\tau_0 - \mu_d\sigma'_n} \approx \frac{f_0 \sigma'_n - \tau_0}{\tau_0 - f_w\sigma'_n}\,,
\label{eq:S}
\end{equation}
\end{linenomath*}
 with a direct relationship between $R$ and $S$:
\begin{linenomath*}
\begin{equation}
R = \frac{1}{1+S}\,.
\end{equation}
\end{linenomath*}

The $R$ and $S$ ratios capture different aspects of the balance between available strain energy and fracture energy, thus influencing dynamic stress drop and acceleration or deceleration of the rupture front. For non-planar fault geometries and spatially variable prestress and initial fault strength, these ratios vary across the fault interface(s). 
As detailed below (Sec.~\nameref{subsubsec:bij}), prescribing a regionally uniform  $R_\mathrm{0}\geq R$, defined as the $R$-value for an optimally oriented fault segment, allows us to constrain the amplitude of deviatoric stresses relative to the frictional strength drop, while naturally incorporating stress variability due to the megathrust geometry.

\subsubsection*{Ambient prestress} \label{subsubsec:bij}
The ambient prestress tensor $b_{ij}$ is constrained using observed regional stress orientations, and assumed fault-fluid pressure and Mohr–Coulomb frictional failure criteria, following \cite{Ulrich2022Stress}. 
We prescribe a uniform regional stress field orientation based on the inferred principal stress orientations along the Japan subduction zone from the World Stress Map \cite{Heidbach2018World}, with the maximum principal stress oriented at an azimuth of $100^\circ$ and a plunge angle of $8^\circ$. 
The magnitudes of the principal stresses $s_i$ are determined through the stress shape ratio $\nu$ as:
\begin{linenomath*}
\begin{equation}
 \nu = \frac{s_2 - s_3}{s_1 - s_2}\,.
\end{equation}
\end{linenomath*}
We use $\nu=0.5$ in all simulations, again based on the World Stress Map \cite{Heidbach2018World}.

Following the notation of \citet{Aochi20031999}, the Mohr-Coulomb failure criteria is defined as:
\begin{linenomath*}
\begin{equation}
P = (s_1 + s_3)/2 \quad\text{and} \quad ds = (s_1 - s_3)/2\,.
\end{equation}
\end{linenomath*}
with $(P,0)$ being the center of the Mohr-Coulomb circle and $ds$ as its radius. 
Principal stresses $s_i$ are related to $P$, $ds$ and $\nu$ as
\begin{linenomath*}
\begin{equation}
\begin{split}
 s_1 &= P+ds\,,\\
 s_2 &= P-ds + 2\nu ds\,,\\
 s_3 &= P-ds\,. \label{eq:si}
\end{split}
\end{equation}
\end{linenomath*}
The effective mean confining stress $\sigma_c' = (s_1 + s_2 + s_3)/3$ is given by:
\begin{linenomath*}
\begin{equation}
\sigma_c' = P + (2\nu - 1)ds/3 \label{eq:sigma_c}\,.
\end{equation}
\end{linenomath*}
The shear and normal stresses  ($\tau$ and $\sigma_n$) acting on a fault plane oriented at an angle $\Phi$ relative to the maximum principal stress are: 
\begin{linenomath*}
\begin{equation}
\begin{split}
 \tau &= ds\sin{2\Phi}\,,\\
 \sigma_n &= P-ds\cos{2\Phi}\,, \label{eq:mohr_tau_sigma}
\end{split}
\end{equation}
\end{linenomath*}

In this framework, an optimally oriented fault plane is defined as the orientation that, under uniform initial stress and loading rate, reaches frictional failure first, maximizing the shear-to-normal stress ratio to equal the static friction coefficient $\mu_s$. 
Its optimal orientation relative to the maximum principal stress direction is thus:
\begin{linenomath*}
\begin{equation}
\Phi = \pi/4 - 0.5 \arctan(f_0\sigma_n')\,.
\end{equation}
\end{linenomath*}
The deviatoric stress magnitude $ds$ is derived by combining Eqs. \ref{eq:R}, \ref{eq:sigma_c}, and \ref{eq:mohr_tau_sigma}: 
\begin{linenomath*}
\begin{equation}
ds = \frac{\sigma_c'}{\sin{2\Phi}/(f_w+(f_0-f_w)R_0)+(2\nu-1)/3+\cos{2\Phi}}\,. \label{eq:ds}
\end{equation}
\end{linenomath*}
Based on a given regional optimal relative prestress ratio $R_0$, we can compute the principal stress amplitude $s_i$ using Eqs.~\ref{eq:si}, \ref{eq:sigma_c}, and \ref{eq:ds}. 
The orientations of the principal stress axes are constrained by the azimuth $SH_{max}=100^\circ$ and the plunge angle $\theta=8^\circ$.
We systematically vary $R_0$ from 0.1 to 0.2 to identify the preferred rupture model.  This range corresponds to an initial shear stress of 3.12–4.05 MPa at hypocentral depth.

\subsubsection*{Data-informed shear stress heterogeneity} \label{subsubsec:cij}
Prestress heterogeneity may arise from past rupture on the same or nearby fault, unmodelled fault geometrical complexities, local variations of fault strength or pore-fluid pressures, or variations in tectonic loading. The coseismic slip distributions reflect such heterogeneities and have therefore been widely used to constrain the initial stress distribution for dynamic models (e.g., \cite{Weng2019Dynamics, Tinti2021Constraining, Jia2023Complex, Hayek2024}) 

Here, we adopt the median slip distribution derived from 32 published finite-fault models of the Tohoku-Oki earthquake  \cite{Wong2024Quantitative} to inform the heterogeneity pattern. 
This median slip model has a simple slip distribution with a smooth, circular patch up-dip of the hypocenter, showing significant slip extending to the trench. The model robustly captures large-scale slip features common across these models and can successfully reproduce key geodetic and seismic observations when combined with appropriate slip-rate functions.
We compute volumetric stress tensor changes $c_{ij}$ resulting from this imposed slip distribution on the megathrust interface using SeisSol in a dynamic relaxation calculation \cite{Tinti2021Constraining, Glehman2024Partial}, utilizing the same computational mesh and slab geometry as in our subsequent dynamic rupture simulations. 
We impose a regularized Yoffe slip-rate function as an internal boundary condition to compute the stress changes across the slab interface. 
This approach leverages the discontinuous finite-element discretization of SeisSol, accurately capturing displacement discontinuities along the fault interface. 
We perform dynamic relaxation for 200~seconds, sufficient for all seismic waves to exit the computational domain and achieve steady-state stress conditions. 
In contrast to previous methods, which used finite-fault slip models primarily to estimate fault-interface stresses \citep[e.g.,][]{Day1998Dynamic, Tinti2005Earthquake, Causse2014Variability, Yang2019Hypocenter}, our calculation simultaneously estimates both fault-interface and surrounding volumetric stress changes.
High slip gradients can lead to unrealistic stress concentrations, particularly in shallow regions. 
To mitigate this, we include inelastic off-fault plastic yielding during the dynamic relaxation step, employing the same parameters as during dynamic rupture simulations (Section~``\nameref{Off-fault plasticity}''). 
The resulting stress changes on the megathrust interface are shown in Supplementary Fig.~\ref{SFig:Traction_cij}.

\subsection*{Regional stress rupture model without stress heterogeneity}
\label{Regional stress rupture model without stress heterogeneity}
When only using the regionally constrained stress tensor $b_{ij}$ (i.e., $\alpha = 0$), we obtain a laterally homogeneous prestress model with a uniform relative prestress ratio $R$ across the entire megathrust (Fig.~\ref{Fig1:Initial_conditions}b, Supplementary Fig.~\ref{FigE1:Initial stress conditions}a). 
This homogeneity results from the principal stress orientations and overall geometry of the Japan subduction zone being largely uniform along strike. 
To systematically explore dynamic rupture scenarios without imposed stress heterogeneity, we vary the regional relative prestress level $R_0$ within the range 0.56–0.64 in increments of 0.02, consistent with the average relative prestress value within the rupture area of the preferred model (Fig.~\ref{Fig1:Initial_conditions}d).

In all five homogeneous stress scenarios, dynamic rupture propagates along the entire megathrust interface. 
The rupture model with $R_0 = 0.58$ yields an unrealistic magnitude of $M_w$~9.61 and an extended rupture duration of 180~s (Extended Fig.~\ref{FigE7:Compare_arrest}). 
In contrast to the preferred model, the laterally homogeneous prestress model exhibits crack-like rupture reactivation (Extended Fig.~\ref{FigE8:Regional_model}, Supplementary Fig~\ref{SFig:Regional_model_5s_snapshot} and Video~S2), occurring at the downdip healing front of the growing pulse. 
This simulation does not reproduce the distinct updip and downdip rupture propagation speeds and complex rupture evolution documented in back-projection studies \cite{Meng2011Window, Koper2011Alongdip, Yagi2012Smooth, Yao2013Compressive} (Extended~Fig.~\ref{FigE8:Regional_model}b).

\subsection*{Breakdown work density}
\label{Breakdown work density}
Breakdown work is defined as the frictional work that provides an estimate of the irreversible part of the total strain energy change, which does not go into radiated energy \cite{Kammer2024Earthquake}. 
The breakdown work combines fracture energy and restrengthening work \cite{Tinti2005Earthquake, Cocco2023Fracture, Gabriel2024Fault}.
Since multiple rupture episodes occur during most of our simulations, we sum the breakdown work of each rupture episode into the total breakdown work $W_{b}$ (red-shaded areas in Extended Fig.~\ref{FigE3:Profile_comp}).
We then define the breakdown work density $W_b'$ per unit area, defined as the excess of work over the minimum shear stress level achieved during total slip:
\begin{linenomath*}
\begin{equation}
W_{b}' = \int^{t_f}_0(\tau(t) - \tau_{min})\dot\delta(t)dt,
\end{equation}
\end{linenomath*}
where $\dot\delta(t)$ is the slip velocity and $t_f$ is the end time of the rupture defined as the absolute slip-rate decrease less than $0.01 m/s$. 
The calculated breakdown work density of the preferred model is shown in Extended Fig.~\ref{FigE6:Breakdownenergy}.
The breakdown work density exhibits significant spatial variability and depends on the rupture process \cite{Lambert2020Rupturedependent}.
Multiple slip pulses downdip and in the hypocentral region generally increase the breakdown work density, compared to updip regions.
The average breakdown work density of the preferred model is 19.6 $MJ/m^2$, consistent with the estimated and expected average breakdown work density for an $M_w$~9 event \cite{Lay2012Depthvarying, Viesca2015Ubiquitous, Cocco2023Fracture}.

\subsection*{Dynamic rupture scenarios with prestress heterogeneity informed by alternative finite-fault slip models}
\label{Rupture scenarios with heterogeneity informed by finite-fault slip models}
We map the heterogeneity from each kinematic slip model to the initial stress distribution using the same approach as in the preferred model, and retain the same friction parameterization. 
We vary the amplitude of initial stress heterogeneity ($\alpha$) and the ambient regional stress, expressed by the relative prestress level ($R_0$). To save computational cost, we run six simulations per prestress heterogeneity distribution. For each ensemble, the preferred scenario is chosen by maximizing the variance reduction relative to the geodetic observations and SCARDEC seismic moment-rate (\cite{VallA2016New}). The corresponding three alternative dynamic rupture model results are presented in Supplementary Figs. ~\ref{SFig:Kubota_sim}-~\ref{SFig:Yamazaki_SR_snapshot}.
\clearpage

%% file: 6Acknowledgments.tex
\section*{Acknowledgments}
We thank David Schneller, Thomas Ulrich, and the SeisSol team for their help in using SeisSol, and Elisa Tinti, Yoshihiro Kaneko, and Dmitry Garagash for insightful discussions. 

\section*{Data availability}
We use the open-source software package SeisSol version 1.2.0 \cite{Gabriel2025SeisSol}, available at https://github.com/SeisSol/SeisSol, commit $8cddb43c$ on the master branch, to perform all dynamic rupture simulations presented in this study. 
Instructions for downloading, installing, and running the code are available in the SeisSol documentation at \url{https://seissol.readthedocs.io/}.  
All data and model parameter files required to reproduce the dynamic rupture scenarios can be downloaded from \url{https://zenodo.org/records/17382787}.
The onshore geodetic data are provided by the Geospatial Information Authority (GSI) \cite{sagiya2004decade}.  

\subsection*{Funding}
This work was supported by NSF grants EAR-2143413 and EAR-2121568. JWCW and AAG also acknowledge support from Horizon Europe (Geo-INQUIRE, project no. 101058518).
AAG acknowledges additional support from NASA (grant No. 80NSSC20K0495), NSF (grants EAR-2225286, OAC-2139536, OAC-2311208, RISE-2531036), the European Union’s Horizon 2020 Research and Innovation Programme (grant No. 852992), Horizon Europe (grants No. 101093038, 101058129), and the Statewide California Earthquake Center (SCEC award 25341).
We gratefully acknowledge the Texas Advanced Computing Center (TACC, NSF grant No. OAC-2139536), the Gauss Centre for Supercomputing (LRZ, project pn49ha), and the CINECA Supercomputing Centre for providing supercomputing time.
Additional computing resources were provided by the Institute of Geophysics of LMU Munich \cite{oeser2006cluster}.

\subsection*{Author contributions}
JWCW: Data Curation, Methodology, Formal analysis, Investigation, Writing - Original Draft, Visualization.
AAG: Conceptualization, Software, Methodology, Formal Analysis, Resources, Writing - Original Draft, Supervision, Validation, Funding acquisition.
WWF: Conceptualization, Methodology, Resources, Writing - Review \& Editing, Supervision, Validation, Funding acquisition.

\subsection*{Competing interests}
The authors declare no competing interests.

%% file: 8Supplementary.tex
\clearpage
\newpage

\section*{Supplementary Information}

List of supplementary figures, videos and corresponding simulations.

\begin{figure}
\noindent\includegraphics[width=0.7\textwidth]{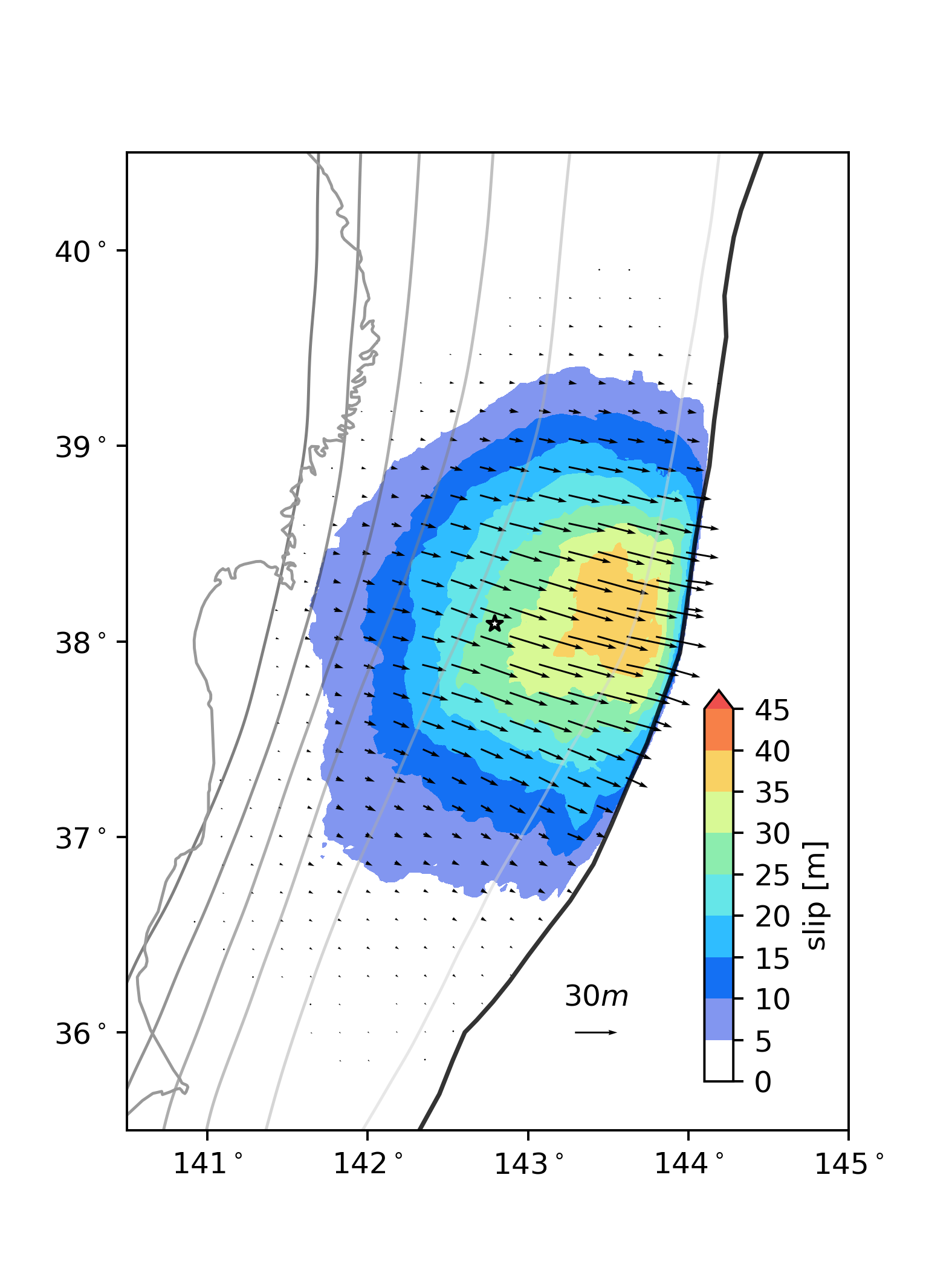}
\caption{\textbf{Median slip distribution computed from 32 finite-fault slip distributions of the Tohoku-Oki earthquake compiled by \cite{Wong2024Quantitative} and projected onto our new slab geometry} (Methods Sec.`Model geometry and mesh'). Colors and vectors represent the amplitude and direction of slip.
Gray contour lines indicate the slab geometry at 10~km depth intervals. 
The USGS hypocenter is indicated as the star \cite{Hayes2011Rapid}. 
The median slip model reveals a smoothly distributed circular slip patch predominantly updip from the hypocenter, confined mostly along strike.
Large slip extends toward the trench, reaching a maximum amplitude of approximately 38.0 m roughly 5 km away from the trench axis. 
This major slip feature has been recognized in \cite{Lay2018Review, Wang2018Learning, Uchida2021Decade}, although previous discussions have been largely qualitative.}
\label{SFig:Kinematic_med_slip_model}
\end{figure}

\begin{figure}
\noindent\includegraphics[width=\textwidth]{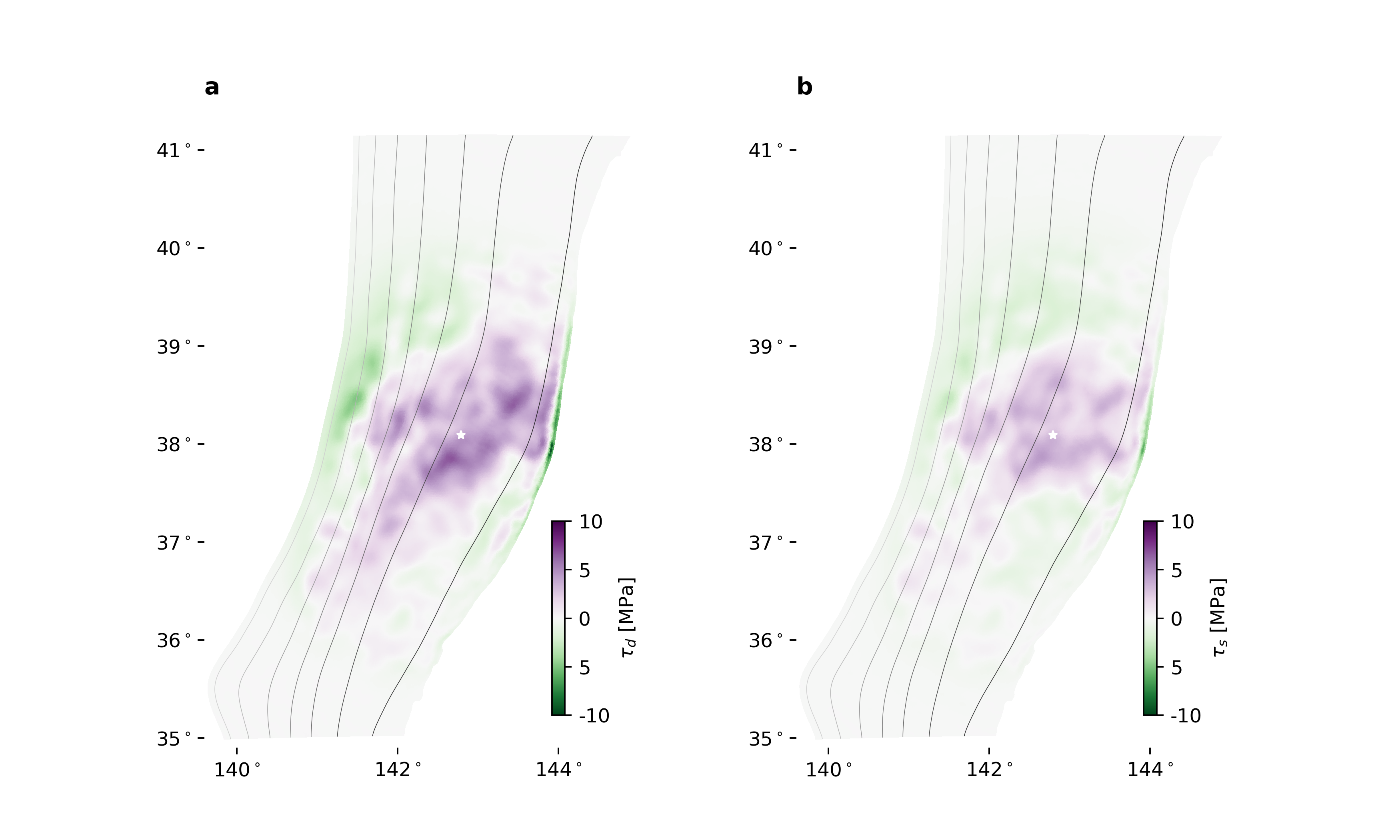}
\caption{\textbf{Stress changes resulting from the median finite-fault slip distribution on the megathrust interface.} (a) Along-dip shear stress change. (b) Along-strike shear stress change. These stress changes serve as the basis for constructing the observationally informed initial stress conditions for dynamic rupture modeling.}
\label{SFig:Traction_cij}
\end{figure}

\begin{figure}[!htb]
\noindent\includegraphics[width=\textwidth]{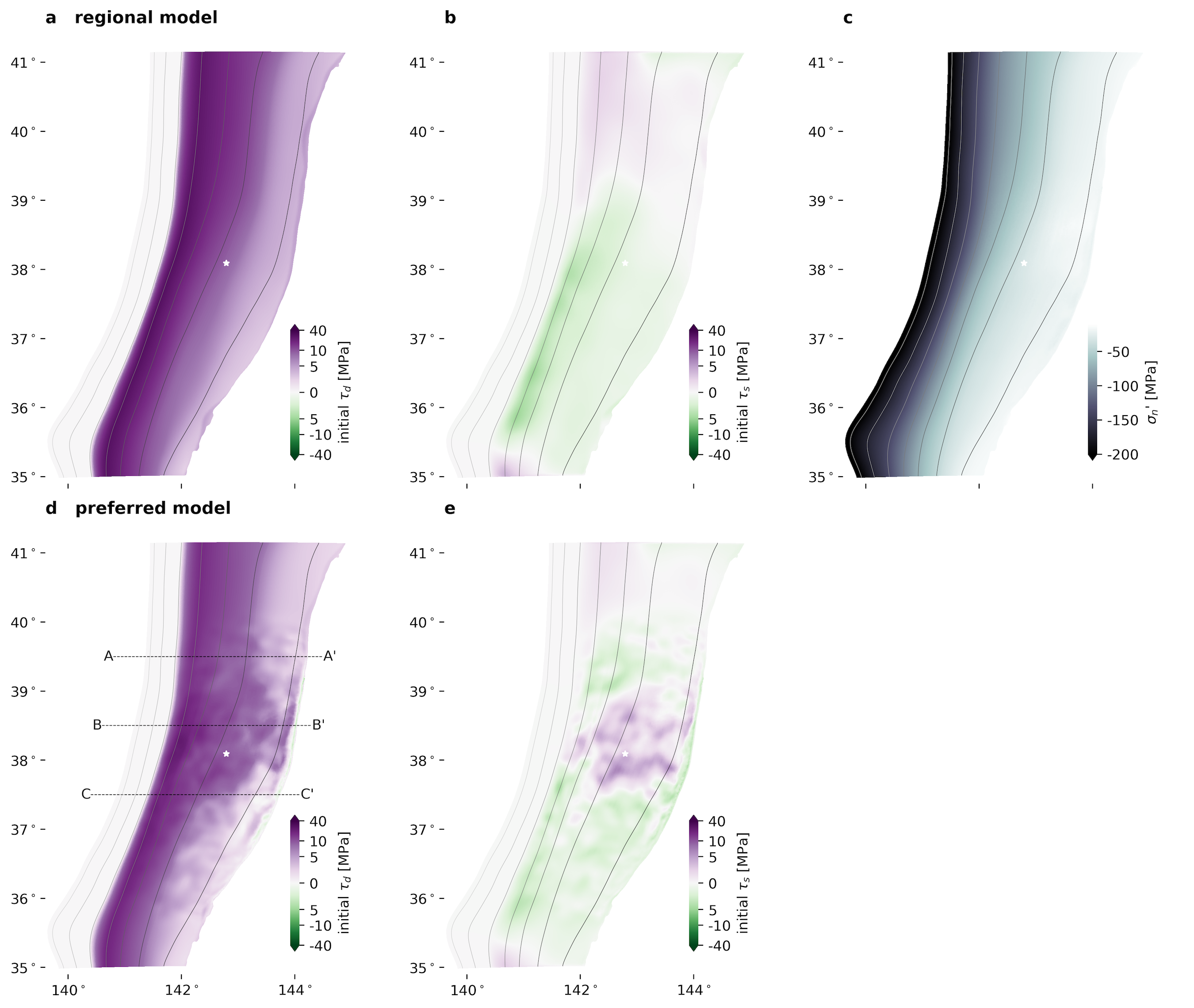}
\caption{\textbf{Initial shear stress and effective normal stress distributions along the megathrust interface. }
(a, b)  Initial shear stress ($\tau_d$, $\tau_s$) distribution for the homogeneous regional stress dynamic rupture model (shown in Fig.~\ref{Fig1:Initial_conditions}b). 
(c) Depth-dependent distribution of effective normal stress ($\sigma_n'$).
(d, e)  Initial shear stress distribution for the preferred model incorporating stress heterogeneity from the median finite-fault model in \citet{Wong2024Quantitative}.  Hypocenter location (star) and depth contours (gray lines, 10 km intervals) are shown in all panels.}
\label{FigE1:Initial stress conditions}
\end{figure}

\begin{figure}
\noindent\includegraphics[width=\textwidth]{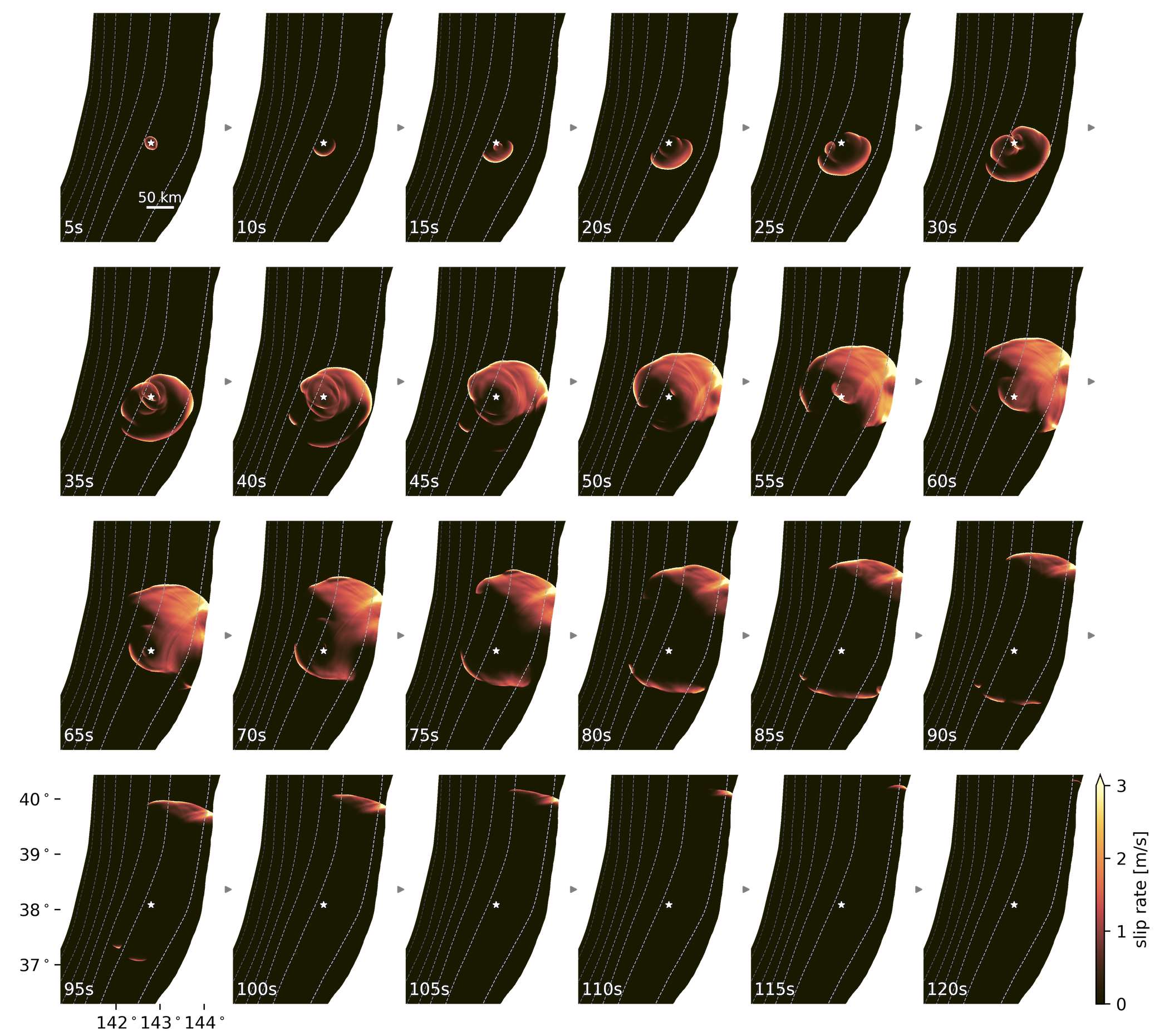}
\caption{\textbf{Dynamic rupture evolution of the preferred model. }Snapshots of slip rate shown in 5~s intervals, see also Supplementary Video~S1. Earthquake rupture initiates as a growing pulse within the first 15~s, followed by a first rupture reactivation initiating at the primary pulse' healing front between 15--25~s. 
Between 25--40~s, reactivated rupture fronts coalesce, ``spiral'' and back-propagate, resulting in complex slip rate patterns and a second major hypocentral slip reactivation at 40~s, taking again the form of a growing pulse. Between 40~s and 50~s rupture time, the primary updip rupture front reaches the seafloor interface, resulting in strong dynamic interactions with the free surface and generating reflected phases.
The third episode of hypocentral rupture reactivation occurs at around 50~s, initiating at the healing front of the secondary propagating pulse-like rupture. A fourth reactivated pulse emerging at 65~s is not sustained. Subsequently, after around 75~s, the rupture simplifies and propagates pulse-like bilaterally along strike, featuring extended shallow rupture in the northern portion of the megathrust between 100--120~s, consistent with slip models inferred from tsunami inversion studies \cite{Satake2013Time, Yamazaki2018Self, Kubota2022New}. 
The white star denotes the hypocenter location.}
\label{FigE2:All_slip_rate_5s}
\end{figure}

\begin{figure}
\noindent\includegraphics[width=\textwidth]{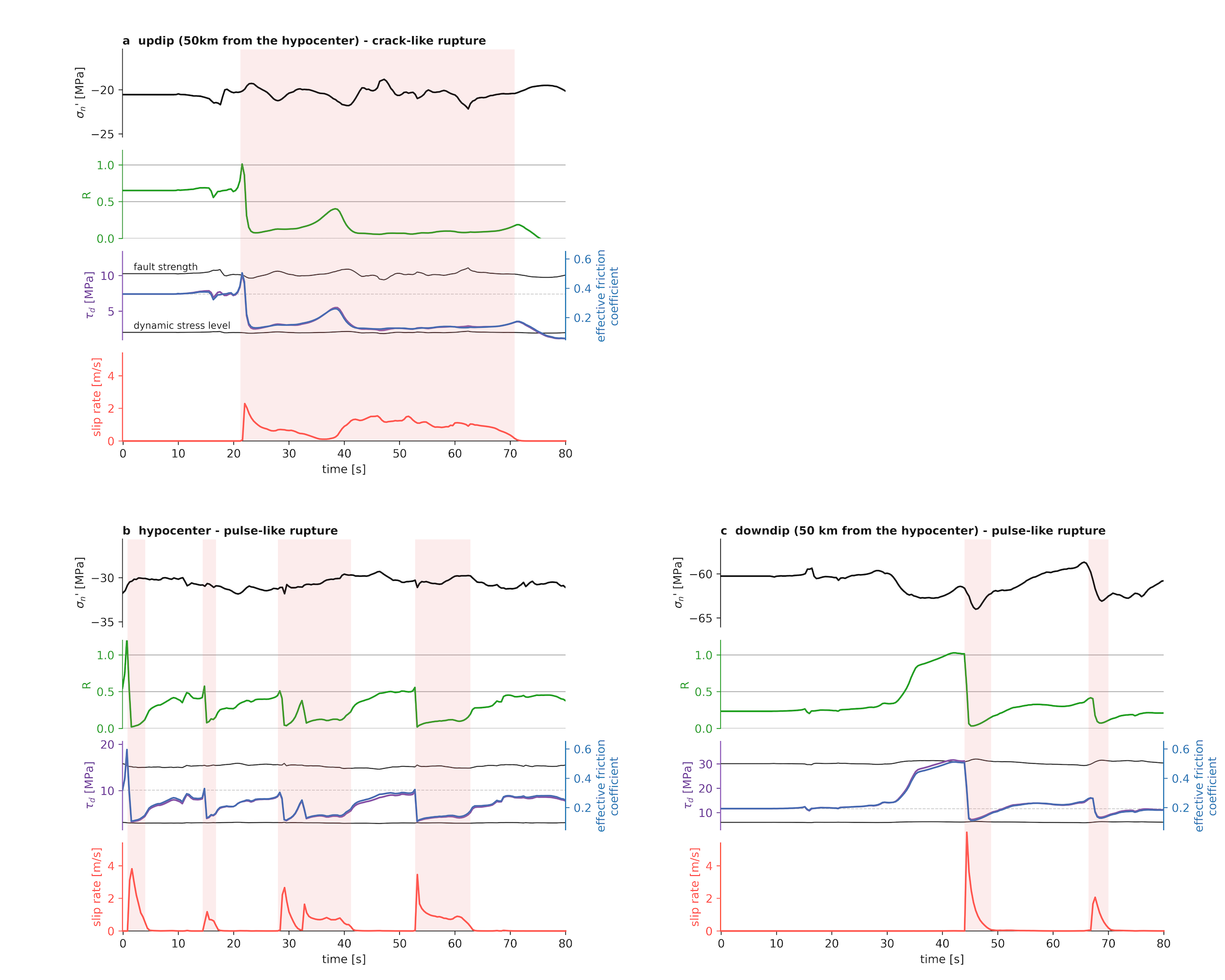}
\caption{\textbf{Temporal evolution of effective normal stress $\sigma_n'$ (black), relative prestress ratio $R$ (green), along-dip shear stress $\tau_d$ (purple), effective friction coefficient (blue), and slip rate (red) of the preferred model.} (a) Evolution in the updip, (b) hypocentral, and (c) downdip regions. 
The shaded red areas denote periods when the slip rate exceeds 0.05~m/s. In the along-dip shear stress panels, the light gray dashed lines represent the initial shear stress, while the solid black lines indicate the fault strength ($f_0\sigma_n'$) and dynamic stress level ($f_w\sigma_n'$). }
\label{FigE3:Profile_comp}
\end{figure}

\begin{figure}
\noindent\includegraphics[width=\textwidth]{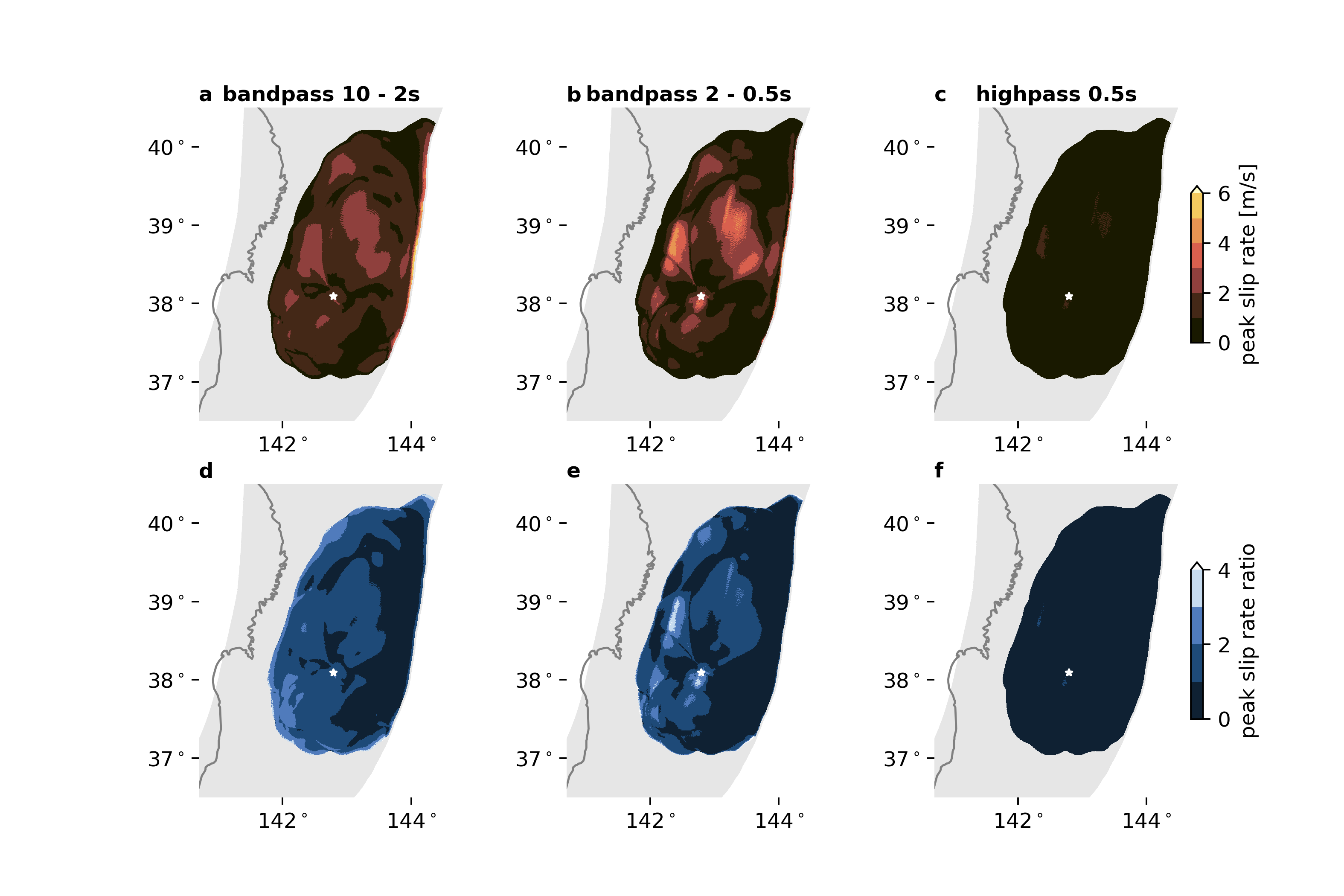}
\caption{\textbf{Comparison of the peak-slip-rate distributions of the preferred model at three frequency ranges.} Top row: filtered peak slip-rate distribution with the same plotting style as Fig.\ref{Fig4:depth_dependent_rupture_style}c. Bottom row: Ratio of peak-slip-rate distribution between the top row and the low-pass filtered at 10~s with the same plotting style as Fig.~\ref{Fig4:depth_dependent_rupture_style}e. (a, d) Comparison with band-pass filtered between 10 and 2~s. (b,e) Band-pass filtered between 2 and 0.5~s. (c,f) High-pass filtered at 0.5~s. The downdip high-frequency radiation is mostly dominated in the back-propagation study seismic frequency range of 2 to 0.5~s.}
\label{FigE4:Depth_dependent_freq_ranges}
\end{figure}

\begin{itemize}
\item Preferred dynamic rupture model with prestress heterogeneity:
\begin{itemize}
    \item Suppl. Fig.~\ref{SFig:Kinematic_med_slip_model}: Median slip distribution computed from 32 finite-fault slip distributions of the Tohoku-Oki earthquake compiled by \cite{Wong2024Quantitative} and projected onto our new slab geometry.
    \item Suppl. Fig~\ref{SFig:Traction_cij}: Stress changes resulting from the median finite-fault slip distribution on the megathrust interface.
    \item Suppl. Fig.~\ref{FigE1:Initial stress conditions}d-e: Initial shear stress and effective normal stress distributions along the megathrust interface.
    \item Suppl. Fig.~\ref{FigE2:All_slip_rate_5s}:  Dynamic rupture evolution of the preferred model with slip rate evolution snapshots.
    \item Suppl. Fig.~\ref{FigE3:Profile_comp}: Temporal evolution of stress, frictional strength, and slip rate at updip, hypocentral, and downdip regions.
    \item Suppl. Fig.~\ref{FigE4:Depth_dependent_freq_ranges}: Comparison of the peak-slip-rate distributions across three frequency ranges.
    \item Suppl. Fig.~\ref{FigE5:disp_field}: Simulated seafloor displacement field.
    \item Suppl. Fig.~\ref{SFig:SMA_comp}: Comparison of observed and synthetic regional strong-ground motion records.
    \item Suppl. Fig.~\ref{SFig:Knet_vel_spec}: Velocity spectra of modeled waveforms at onshore strong-motion K-net stations using the preferred model.
    \item Suppl. Fig.~\ref{SFig:CF_cohesion_profile}: Depth-dependent cohesion (red) and closeness-to-failure profiles.
    \item Suppl. Video~S1: Video of slip rate, along-dip shear stress, and friction.
\end{itemize}

\item Dynamic rupture model with prestress heterogeneity and simple reactivation rupture style:
    \begin{itemize}
    \item    Suppl. Fig.~\ref{SFig:SMA_comp}: Regional strong-ground motion comparison.
    \item    Suppl. Fig.~\ref{SFig:MR_grid_search}: Comparison of moment-rate functions for two distinct rupture styles shown in Figure~\ref{Fig6:Grid_search}.
    \item    Suppl. Fig.~\ref{SFig:grid_search_obs_VR}: Comparison of onshore and offshore geodetic displacement misfits across models with varying prestress heterogeneity amplitude $\alpha$ and regional relative prestress level $R_0$.
    \item     Suppl. Video~S2: Video of slip rate, along-dip shear stress, and friction.
    \end{itemize}

\item Dynamic rupture model with homogeneous regional prestress condition:
    \begin{itemize}
    \item    Suppl. Fig.~\ref{FigE1:Initial stress conditions}a-b: Initial shear stress conditions.
    \item    Suppl. Fig.~\ref{FigE7:Compare_arrest}: Comparison of fault slip distributions.
    \item    Suppl. Fig.~\ref{FigE8:Regional_model}: Slip rate, friction, and shear stress evolution along the hypocentral dip profile.
    \item    Suppl. Fig.~\ref{FigE9:Reg_depth-varying}: Depth-dependent slip rate evolution and corresponding amplitude spectra.
    \item    Suppl. Fig.~\ref{SFig:Regional_model_5s_snapshot}: Slip rate evolution snapshots.
    \item    Suppl. Video~S2: Video of slip rate, along-dip shear stress, and friction.
    \end{itemize}

\item Dynamic rupture model with heterogeneous prestress and uniform state-evolution distance:
    \begin{itemize}
    \item   Suppl. Fig.~\ref{SFig:model_uniformDc}: Alternative dynamic rupture model using a uniform weakening distance $L$ of 0.3~m.
    \end{itemize}

\item Dynamic rupture model with prestress heterogeneity and multiscale heterogeneity in the state-evolution distance:
    \begin{itemize}
    \item     Fig.~\ref{Fig10:Sl0_het_model}: Model setup and overview.
    \item     Suppl. Fig.~\ref{SFig:Sl0_snapshot}: Slip-rate evolution of the heterogeneous-friction dynamic rupture model with multiscale variations in state-evolution distance.
    \item     Suppl. Video~S4: Video of slip rate, along-dip shear stress, and friction.
    \end{itemize}

\item Dynamic rupture model with heterogeneous distribution of fully-weakened dynamic frictional strength and homogeneous depth-dependent initial stress:
    \begin{itemize}
    \item    Suppl. Fig~\ref{SFig:Fw_slip_strength_illustration}: Illustrative diagram of fault-local frictional evolution for heterogeneous friction or prestress setup.
    \item    Suppl. Fig~\ref{SFig:Fw_compare}: Depth-dependent variation of frictional strength and initial stress conditions along the hypocentral dip profile.
    \item    Suppl. Fig.~\ref{SFig:Fw_sim}: Model setup and overview.
    \item    Suppl. Fig.~\ref{SFig:Fw_snapshot}: Slip-rate evolution of the dynamic rupture model with heterogeneous distribution of fully- weakened dynamic frictional strength and homogeneous, depth-dependent initial stress.
    \item    Suppl. Video~S5: Video of slip rate, along-dip shear stress, and friction.
    \end{itemize}

\item Dynamic rupture model using the stress-change pattern derived from the finite-fault model of Kubota et al. (2022):
    \begin{itemize}
    \item    Suppl. Fig.~\ref{SFig:Kubota_sim}: Model setup and overview.
    \item    Suppl. Fig.~\ref{SFig:Kubota_SR_snapshot}:Slip-rate evolution of the dynamic rupture model using stress-change pattern derived from the finite-fault slip model of Kubota et al., 2020.
    \item    Suppl. Video~S6: Video of slip rate, along-dip shear stress, and friction.
    \end{itemize}

\item Dynamic rupture model using the stress-change pattern derived from the finite-fault model of Melgar et al. (2015):
    \begin{itemize}
    \item    Suppl. Fig.~\ref{SFig:Melgar_sim}: Model setup and overview.
    \item    Suppl. Fig.~\ref{SFig:Melgar_SR_snapshot}: Slip-rate evolution of the dynamic rupture model using stress-change pattern derived from the finite-fault slip model of Melgar et al. (2015). 
    \item    Suppl. Video~S7: Video of slip rate, along-dip shear stress, and friction.
    \end{itemize}

\item Dynamic rupture model using the stress-change pattern derived from the finite-fault model of Yamazaki et al. (2018):
    \begin{itemize}
    \item    Suppl. Fig.~\ref{SFig:Yamazaki_sim}: Model setup and overview.
    \item    Suppl. Fig.~\ref{SFig:Yamazaki_SR_snapshot}: Slip-rate evolution of the dynamic rupture model using stress-change pattern derived from the finite-fault slip model of Yamazaki et al. (2018).
    \item    Suppl. Video~S8: Video of slip rate, along-dip shear stress, and friction.
    \end{itemize}
\end{itemize}

\begin{table}[!htb]
\caption{Fault frictional properties assumed in this study. VW: velocity-weakening, VS: velocity-strengthening.}
\centering
\begin{tabular}{l l l l}
\hline
Parameter  & Symbol  & Values & Unit\\ 
\hline
Direct-effect parameter$^*$     & $a$     & VW: 0.01 (0--9~km, $>45$~km) & \\
                                &       & VS: 0.018  (9--45~km) &\\
Evolution-effect parameter      & $b$     & 0.014  &\\
Reference slip rate             & $V_0$ & $10^{-6}$ & m/s \\
Steady-state low-velocity friction coefficient at slip rate $V_0$  & $f_0$  & 0.5 &\\
Weakened slip rate              & $V_W$ & 0.1   & m/s \\
State evolution distance $^\dagger$     & L     &   0.6   & m   \\
Fully weakened friction coefficient & $f_w$ & 0.1&\\
Initial slip velocity           & $V_i$ & $10^{-16}$ & m/s \\
\hline
\multicolumn{4}{l}{$^*$ The $a$ parameter smoothly transitions from a velocity-strengthening (VS) value at depths shallower than 9~km }\\
\multicolumn{4}{l}{and deeper than 45~km to a velocity-weakening (VW) value within the seismogenic zone (9--45~km).}\\ 
\multicolumn{4}{l}{$^\dagger$ State evolution distance $L$ is initially set to $0.2$~m within 6~km radius of the hypocenter location,}\\ 
\multicolumn{4}{l}{increases linearly to 0.6~m within a 12~km radius, and remains constant of 0.6~m elsewhere.}\\ 
\multicolumn{4}{l}{(See Supplementary Section~``\nameref{SM2: Nucleation}'' for details)}\\
\end{tabular}
\label{T:Friction}
\end{table}
\clearpage
\newpage

\subsection*{SM1: Model resolution}
\label{SM: Model domain and resolution}
Numerical convergence of dynamic rupture simulations is governed by the resolution of the process zone \cite{Day2005Comparison}.
We follow \citet{Wollherr2018Fault} to determine the required on-fault resolution of our SeisSol dynamic rupture simulations, which use basis functions of polynomial order $p=5$. 
Our mesh features an element size of 1000~m everywhere along the slab. 
This ensures that we resolve the average process zone width, which we measure to be $\Lambda$=4,500~m in our preferred model.

Off-the slab, we employ a velocity-aware adaptive mesh refinement approach \cite{Breuer2022NextGeneration}, focusing resolution along the slab interface and in onshore regions. 
The target frequency resolved by the mesh is determined by:  
\begin{linenomath*}
\begin{equation}
f\approx V_s/ (\Delta x \times \text{elements per wavelength}),
\end{equation}
\end{linenomath*}
with $\Delta x$ defining the tetrahedral element size, $V_s$ as the S wave speed. We follow the analysis by \cite{Kaser2006Arbitrary} and require at least two elements per wavelength, suitable for polynomial basis functions of order $p =5$ in space and time.
While our mesh is conservatively designed to resolve seismic wave propagation throughout the domain at frequencies up to 1 Hz, it resolves seismic wavefields recorded at seismic stations at frequencies up to 2~Hz (Supplementary Fig.~\ref{SFig:Knet_vel_spec}).

\subsection*{SM2: Nucleation}
\label{SM2: Nucleation} 
The Tohoku-Oki earthquake began with a low initial moment-release rate \cite{Hayes2011Rapid, Lay2011Possible, VallA2016New}, which is challenging to capture in dynamic rupture simulations that cannot account for long-term fault slip evolution. Our models capture the slow initiation behavior using a smooth nucleation procedure and scale-dependent fracture energy in the hypocentral region \cite{Ide2002Estimation, Aochi2011Conceptual, Gabriel2024Fault}. 

Following common practice from community dynamic rupture benchmarks \cite{Harris2018Suite}, we define an overstressed nucleation region with a radius of $r_{nuc}$ of 7~km and an additional shear stress perturbation of 10~MPa to locally reach the yielding stress level. We position the nucleation patch at the hypocenter location provided by the USGS (142.7897$^\circ$E, 38.0919$^\circ$N) \cite{Hayes2011Rapid}.
The stress perturbation is smoothly imposed spatially and temporally, using an exponential spatial function $f(r)$ and a smooth temporal function $g(t)$:
\begin{linenomath*}
\begin{equation}
f(r) = \exp[r^2/(r^2 -r_{crit}^2)]\,,
\end{equation}
\end{linenomath*}
\begin{linenomath*}
\begin{equation}
g(t) = \exp[(t-T)^2/t(t-2T)]\,,
\end{equation}
\end{linenomath*}
with $T = 3s$.

To ensure a realistic, gradual rupture initiation, we impose a spatially variable slip-weakening distance \cite{Ulrich2022Stress}.
The state evolution distance is set to $L=$ 0.2~m within a 6~km radius from the hypocenter, increases linearly to 0.6~m within a 12~km radius, and remains constant at $L=$ 0.6~m and beyond.
To quantify the effects of varying slip-weakening distances, we perform an additional simulation using a uniform slip-weakening distance of 0.3~m. 
This uniform nucleation model reproduces the overall dynamic complexity seen in our preferred model, including multiple rupture reactivation, depth-dependent rupture characteristics, substantial slip to the trench, and spontaneous rupture arrest (Supplementary Fig.~\ref{SFig:model_uniformDc}). However, it results in the peak slip rate being reached early, at 50~s. This dynamic rupture model also does not match the geodetic deformation as closely as our preferred model, with an onshore and offshore geodetic data variance reduction of 76.1\% and 34.1\%, respectively.

\subsection*{SM3: Dynamic stress drop}
\label{SM3: Dynamic stress drop} 
To quantify the modeled spatially varying dynamic stress drop ($\Delta\tau$), defined as the difference between initial and final shear stresses during the rupture, we compute the slip-weighted mean stress drop across the ruptured area: 
\begin{linenomath*}
\begin{equation}
\Delta\sigma_E = \frac{\int_\Sigma\Delta\tau\delta\text{dS}}{\int_\Sigma \delta \text{dS}}\,,
\end{equation}
\end{linenomath*}
where $\Sigma$ is the rupture area and $\delta$ denotes the slip amplitude. Our preferred rupture model yields a slip-weighted average dynamic stress drop of 2.37~MPa, comparable to the estimated stress drop of finite-fault slip models \cite{Brown2015Static}.

\clearpage
\newpage

\begin{table}
\caption{1D velocity model, modified from \cite{Fukuyama1998Automated}}
\centering
\begin{tabular}{c c c c c}
\hline
Depth[km] & Thickness [km] & P-wave velocity [km/s]  & S-wave velocity [km/s] & Density [$kg/m^3$]\\ 
\hline
3   & 3  & 5.5  & 3.14 & 2300 \\
18  & 15  & 6.0  & 3.55 & 2400 \\
33  & 15  & 6.7  & 3.84 & 2800 \\
100 & 67  & 7.8  & 4.46 & 3200 \\
$\infty$ & $\infty$ & 8.0 & 4.57  & 3300\\ 
\hline
\end{tabular}
\label{ST:Velocity_structure}
\end{table}

\clearpage
\newpage

\clearpage
\newpage

\clearpage
\newpage

\clearpage
\newpage

\clearpage
\newpage

\begin{figure}
\noindent\includegraphics[width=0.5\textwidth]{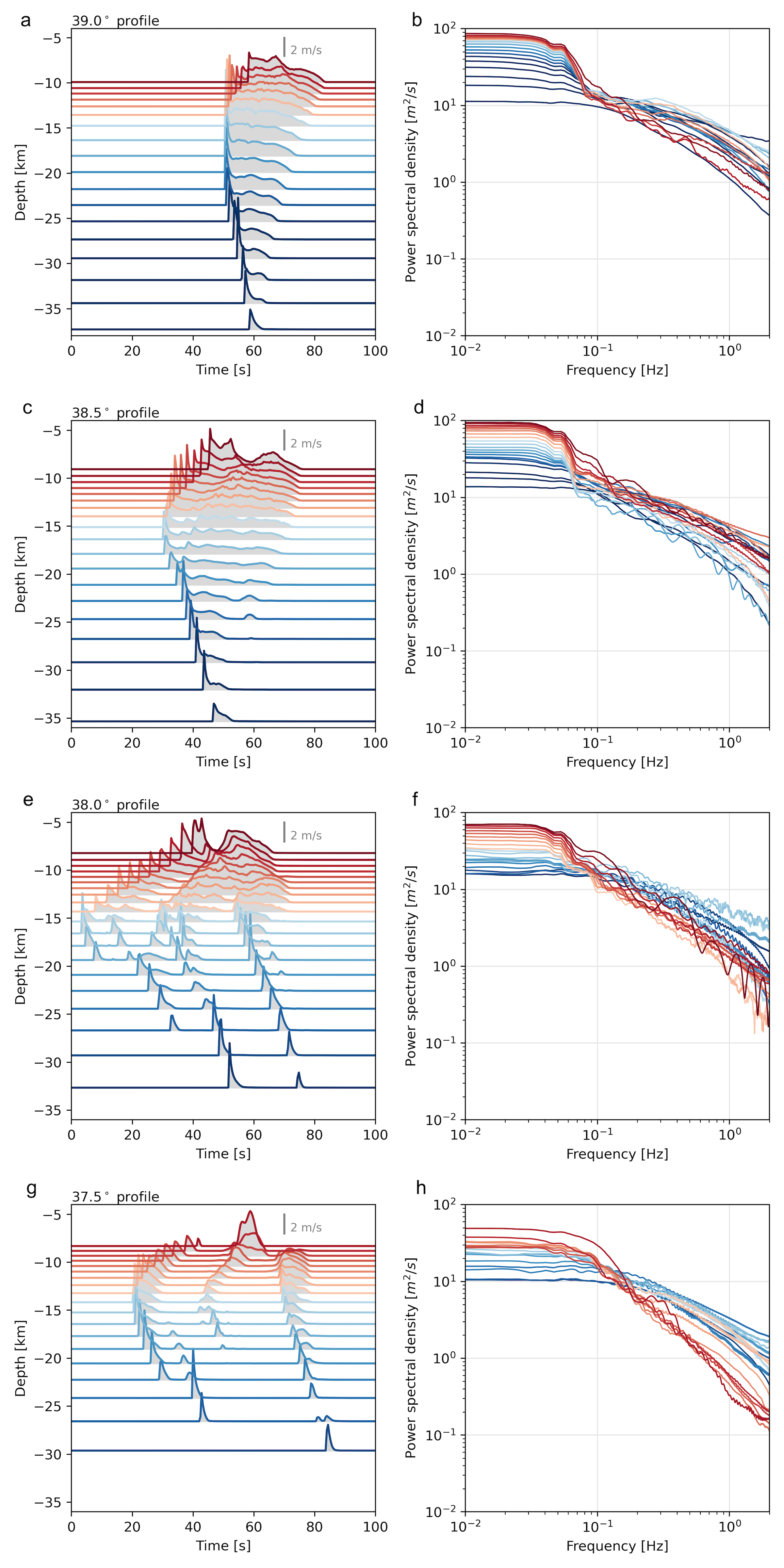}
\caption{\textbf{Along-dip slip-rate time histories and corresponding power spectral densities for the preferred model.} Left column (a,c,e,g): slip-rate evolution with depth along four dip profiles. Right column (b,d,f,h): power spectral density of the slip-rate time series for each profile. Across all profiles, local slip duration (rise time) increases toward shallower depths; the contrast is most pronounced along the 38.0° profile. }
\label{FigS:Depth_dependent_profile_comparison}
\end{figure}

\begin{figure}
\noindent\includegraphics[width=\textwidth]{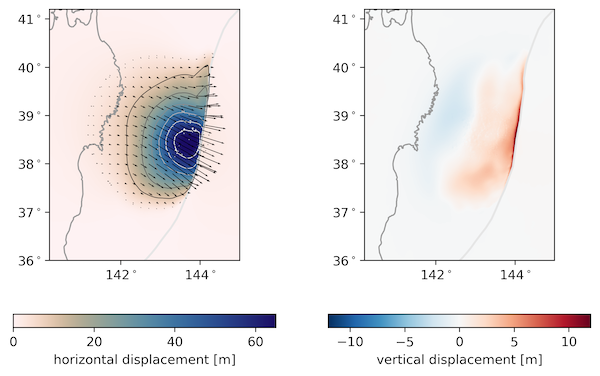}
\caption{\textbf{Simulated seafloor displacement from the preferred model. }(a) Horizontal and (b) vertical displacement fields.
In (a), contours show horizontal displacement amplitudes at 10~m intervals, the gray line denotes the trench location.
The near-trench modeled horizontal and vertical displacements are broadly consistent with differential bathymetry observations \cite{Fujii2011Tsunami, Sun2017Large, Kodaira2020Large, Ueda2023Submarine, Zhang2023Complex}.
The pronounced uplift in the northern near-trench region (in b) agrees with deformation inferred from tsunami waveform inversion \cite{Satake2013Time, Hossen2015Tsunami, Dettmer2016Tsunami, Yamazaki2018Self, Kubota2022New}.}
\label{FigE5:disp_field}
\end{figure}
\clearpage
\newpage

\begin{figure}
\noindent\includegraphics[width=\textwidth]{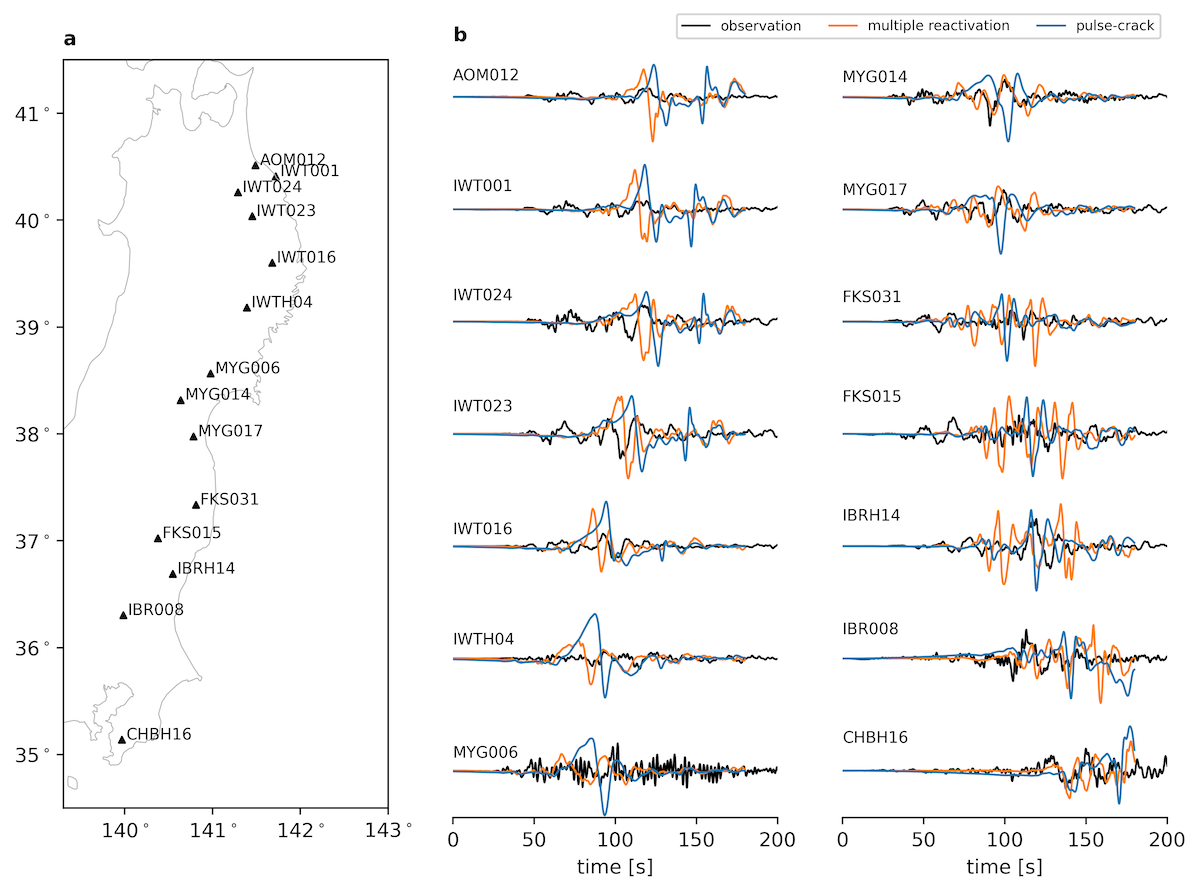}
\caption{\textbf{Comparison of the observed regional strong-ground motion with synthetics from the preferred multiple-reactivation model (orange) and the simple reactivation model (blue)}. (a) Map of the KiK-net and K-NET strong-ground-motion stations used. (b) Waveform comparison between the observations (black) and synthetics from the preferred multiple-reactivation model (orange) and the simple pulse–crack model (blue). The vertical components are shown in velocity, bandpass-filtered between 100 and 1 s period. The synthetics from the preferred model with multiple reactivation exhibit multiple move-out branches, whereas the simple pulse–crack model synthetics display a single dominant phase. The preferred model synthetics match the waveforms at the MYG014 and MYG017 stations, located near the major rupture area.}
\label{SFig:SMA_comp}
\end{figure}
\clearpage
\newpage

\begin{figure}
\noindent\includegraphics[width=\textwidth]{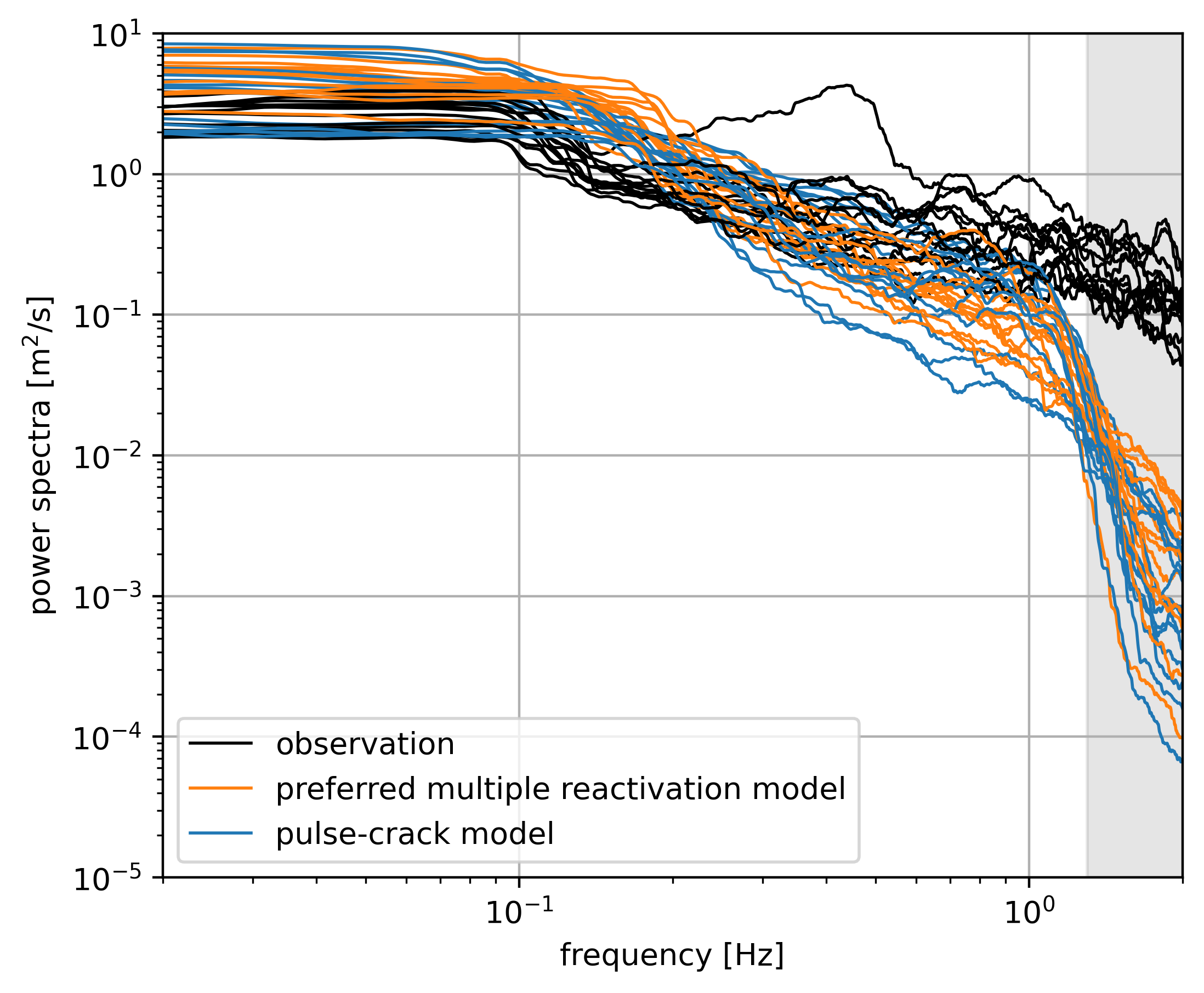}
\caption{\textbf{Comparison of power spectra of modeled waveforms at onshore strong-motion K-net stations with observations (black), preferred model (orange), pulse-crack model (blue).} The shaded area shows the effective maximum frequency resolved in the simulated seismic wavefield at 1.5 Hz. }
\label{SFig:Knet_vel_spec}
\end{figure}
\clearpage
\newpage

\begin{figure}
\noindent\includegraphics[width=0.6\textwidth]{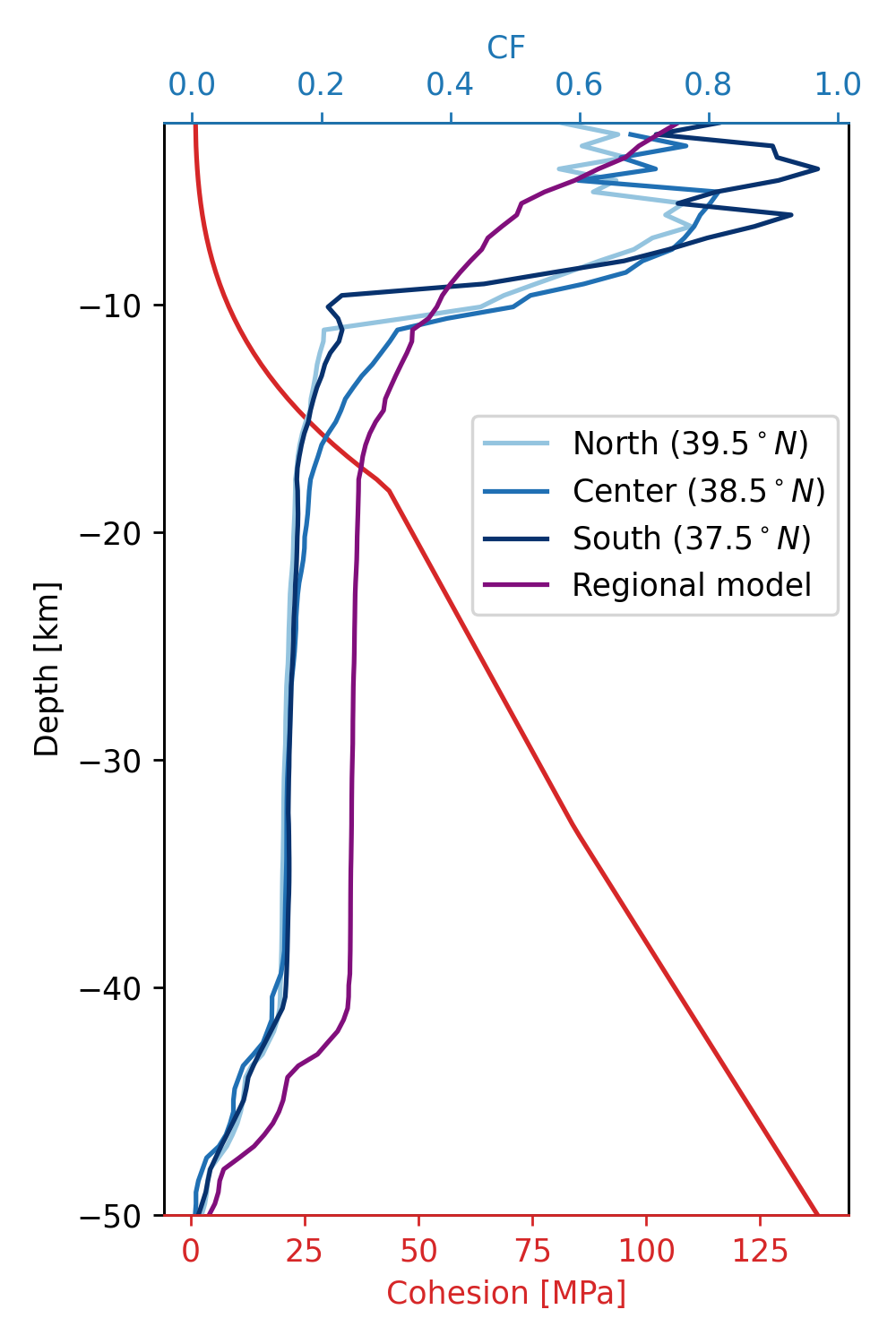}
\caption{\textbf{Depth-dependent cohesion (red) and closeness-to-failure ($CF$) profiles} across the north (light blue), center (blue), and south (dark blue) cross sections as of Fig~\ref{Fig5:trench_slip} of the preferred model, and the laterally homogeneous prestress model (purple). See Methods Sec.~``Off-fault plasticity''.}
\label{SFig:CF_cohesion_profile}
\end{figure}
\clearpage
\newpage

\begin{figure}
\noindent\includegraphics[width=\textwidth]{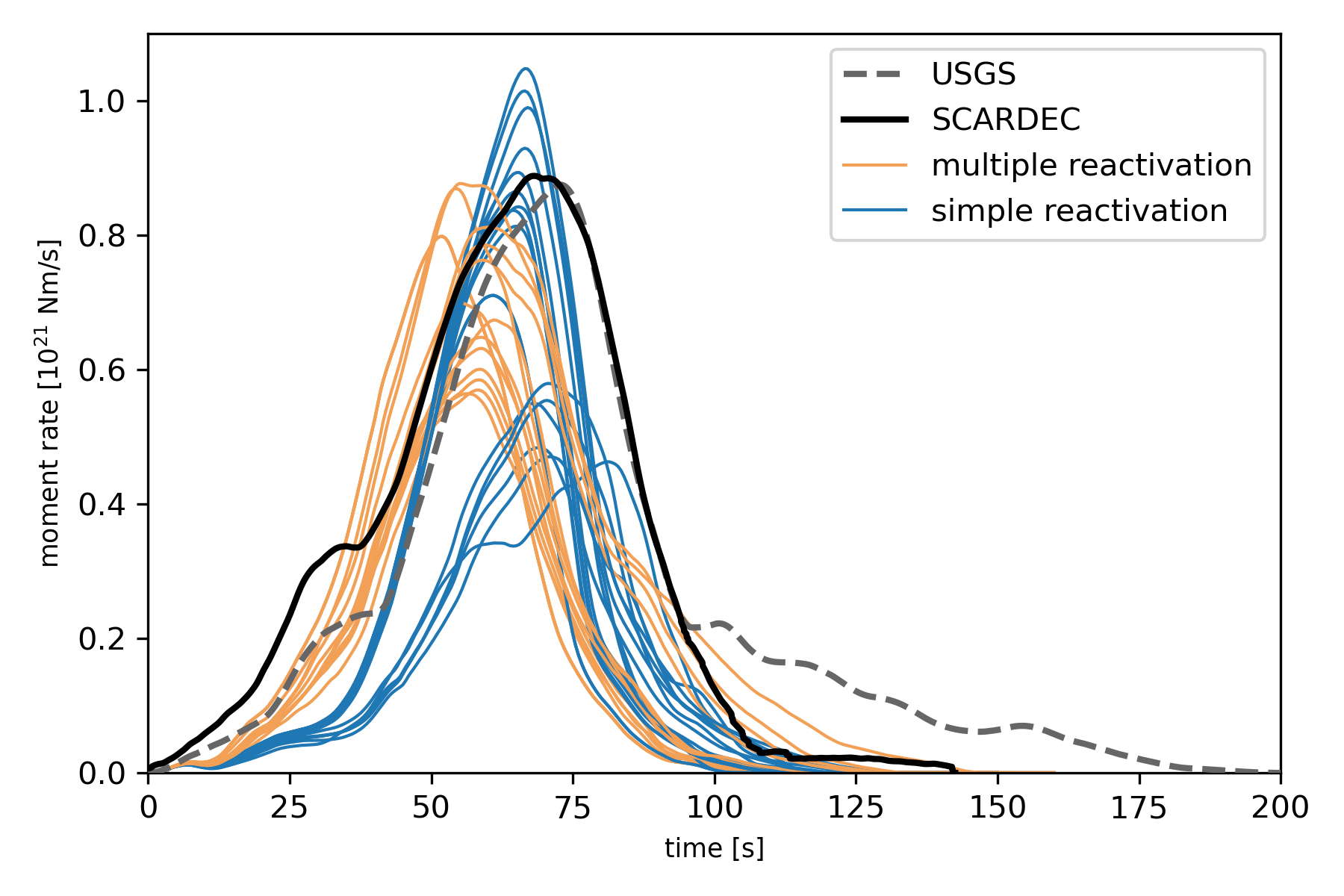}
\caption{\textbf{Comparison of moment-rate functions for two distinct rupture styles shown in Figure~\ref{Fig6:Grid_search}.} Yellow lines represent the moment-rate functions of dynamic rupture models characterized by repeated rupture reactivation near the hypocenter, while blue lines correspond to models dominated by single pulse-like ruptures with free-surface reflection. The reactivation model captures the early moment-rate evolution within the first 40~s, while the simple reactivation rupture model underestimates moment release during the 0--40~s rupture time interval.
}
\label{SFig:MR_grid_search}
\end{figure}
\clearpage
\newpage

\begin{figure}
\noindent\includegraphics[width=\textwidth]{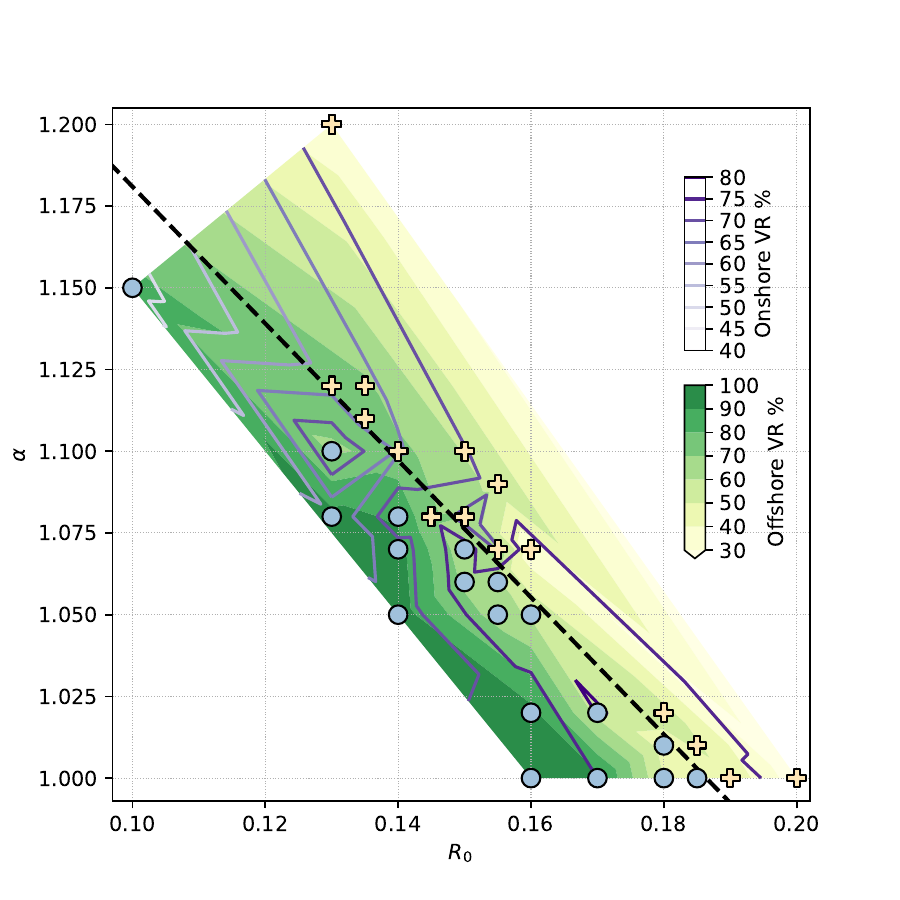}
\caption{\textbf{Comparison of onshore and offshore geodetic displacement misfits across models with varying prestress heterogeneity amplitude ($\alpha$) and regional relative prestress level $R_0$.} Green-filled contours indicate variance reduction for offshore geodetic data, while purple contour lines represent variance reduction for onshore data. Blue circles represent a family of dynamic rupture models dominated by single pulse-like ruptures driven by free-surface reflection, while yellow crosses denote models exhibiting repeated rupture reactivation near the hypocenter. Our results illustrate that stress heterogeneity amplitude ($\alpha$) primarily controls peak slip magnitude, whereas the regional relative stress level $R_0$ predominantly determines rupture extent.}
\label{SFig:grid_search_obs_VR}
\end{figure}
\clearpage
\newpage

\begin{figure}
\noindent\includegraphics[width=\textwidth]{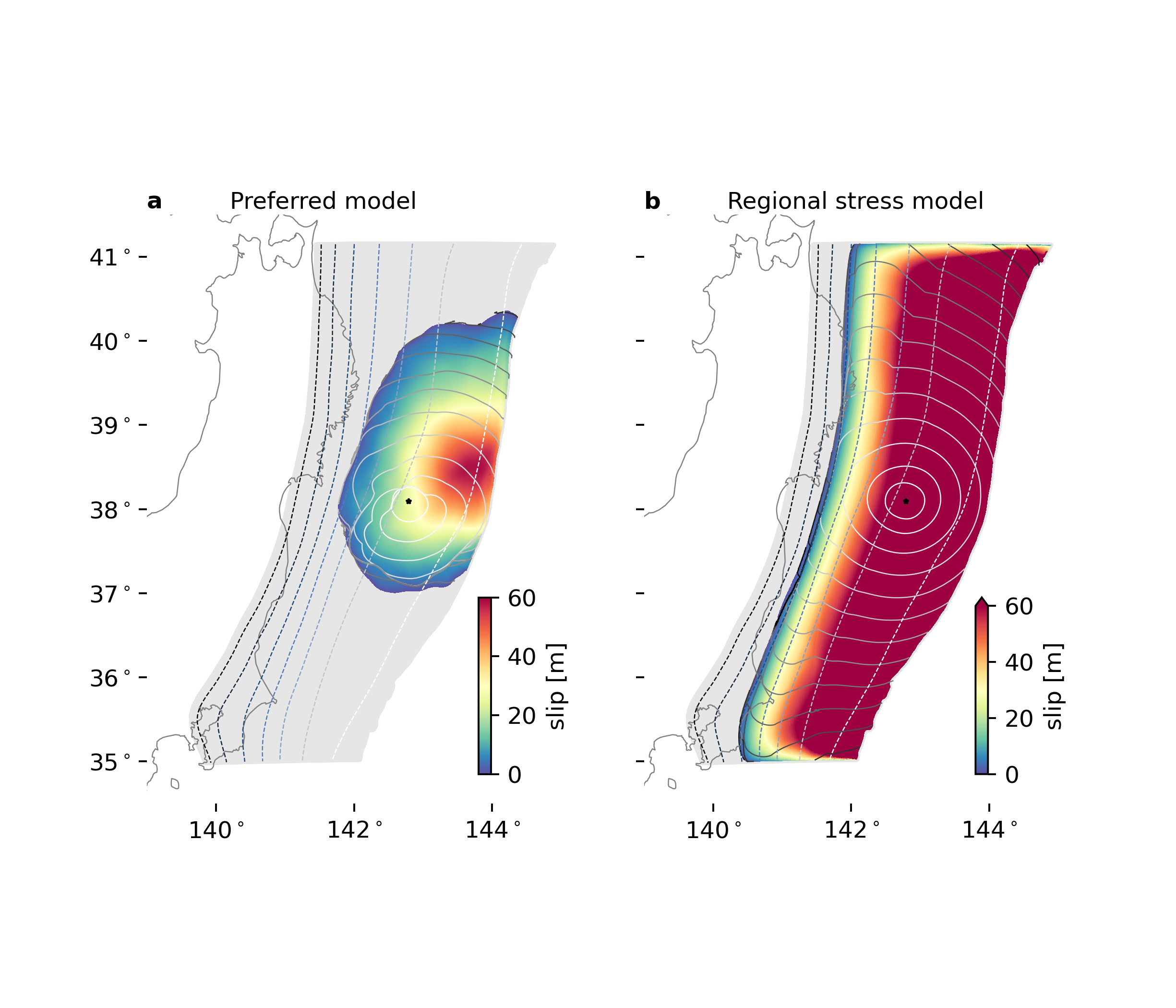}
\caption{\textbf{Comparison of fault slip distribution and rupture evolution between (a) the heterogeneous prestress model (the preferred model) and (b) the laterally homogeneous prestress model.}  
 Solid contour lines indicate rupture front location at 10~s intervals.
The preferred model spontaneously arrests with $M_w$=8.96, whereas the laterally homogeneous prestress model fails to arrest and ruptures the entire fault, reaching $M_w$=9.61. Depth contours (dotted lines, 10~km intervals) and the hypocenter location (star, \cite{Hayes2011Rapid}) are shown in both panels.
}
\label{FigE7:Compare_arrest}
\end{figure}

\begin{figure}
\noindent\includegraphics[width=\textwidth]{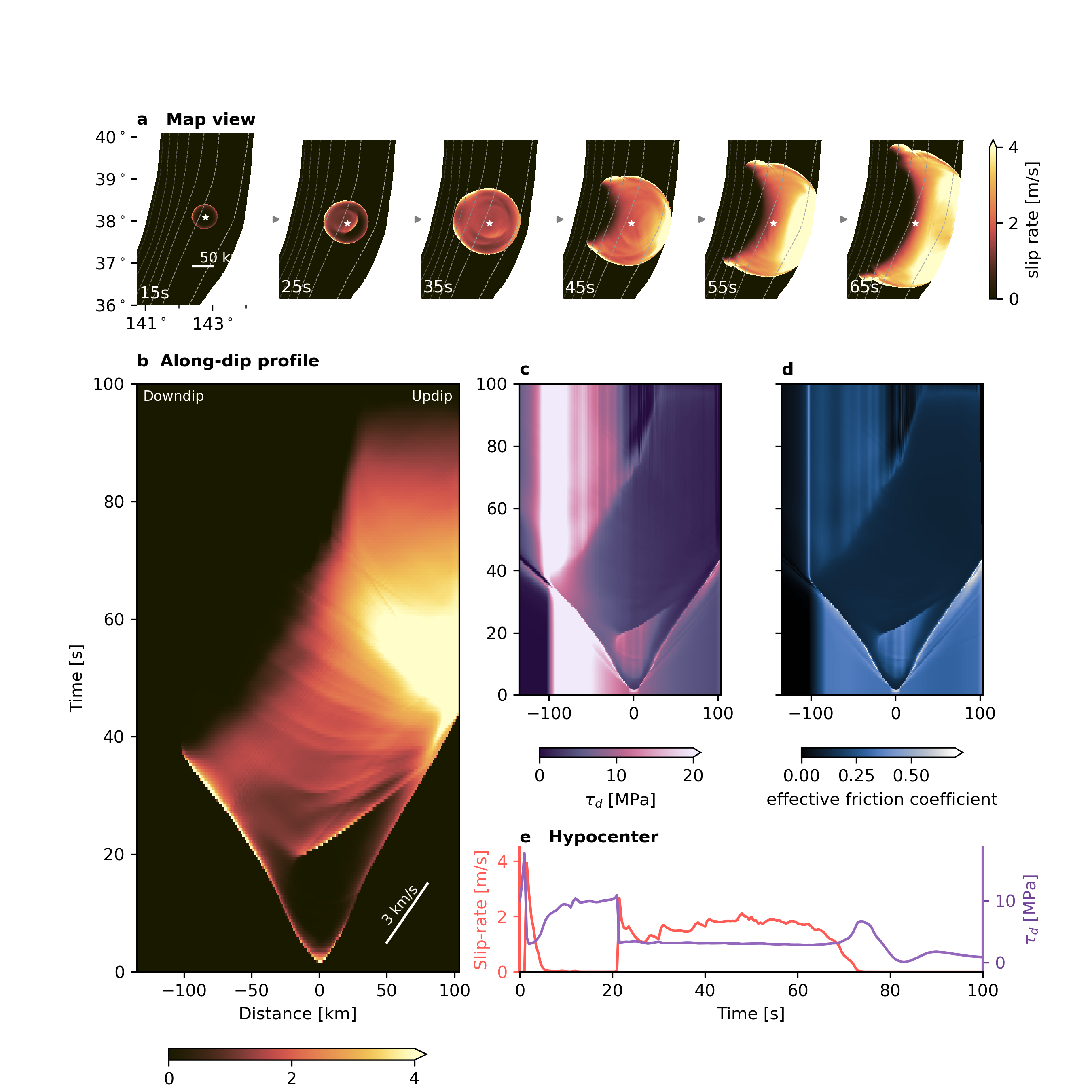}
\caption{\textbf{Rupture dynamics of the laterally homogeneous prestress model}, see also Supplementary Fig.~\ref{SFig:Regional_model_5s_snapshot} 
and Supplementary Video~S2. (a) Map-view snapshots of slip rate evolution at 10~s intervals. A primary growing pulse is followed by crack-like slip reactivation at approximately 20~s, propagating updip. Primary and secondary rupture fronts subsequently merge into sustained crack-like rupture without clear healing fronts distinguishing separate slip episodes. 
The white star indicates the hypocenter location. 
(b)-(d) Temporal evolution of slip rate, along-dip shear stress $\tau_d$ (purple), and effective friction coefficient (blue) along a hypocentral dip profile, highlighting rapid coseismic restrengthening and subsequent rupture reactivation. 
(e) Time series of hypocentral slip rate (red) and along-dip shear stress $\tau_d$ (purple). }
\label{FigE8:Regional_model}
\end{figure}

\begin{figure}
\noindent\includegraphics[width=\textwidth]{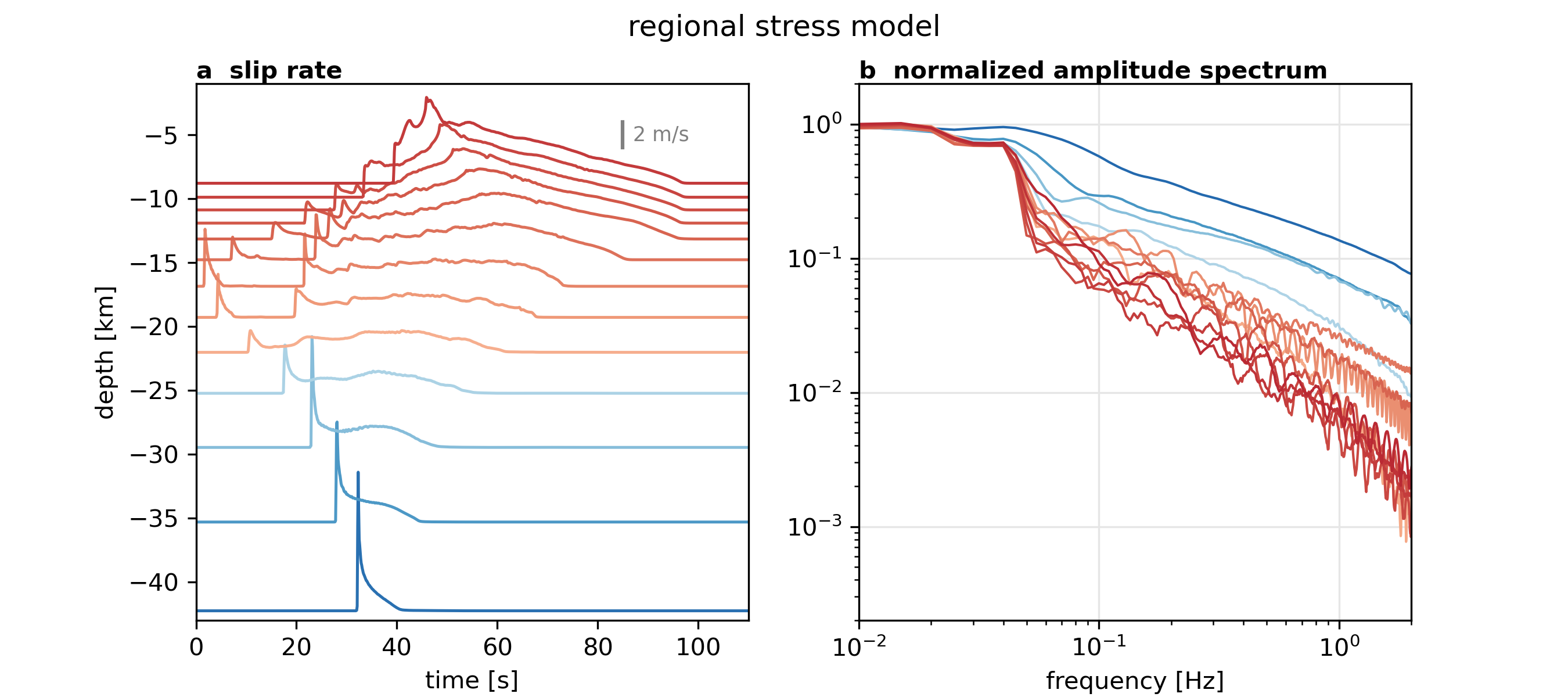}
\caption{\textbf{Depth-dependent slip rate characteristics in the laterally homogeneous prestress model. }(a) Slip rate functions along-dip through the hypocenter at various depths. 
Downdip pulse-like ruptures are highlighted in blue, while updip crack-like ruptures are indicated in red.
(b) Normalized amplitude spectra of the corresponding slip rate functions shown in (a), illustrating distinct frequency content between downdip pulse-like and updip crack-like rupture styles.
}
\label{FigE9:Reg_depth-varying}
\end{figure}

\clearpage

\begin{figure}
\noindent\includegraphics[width=\textwidth]{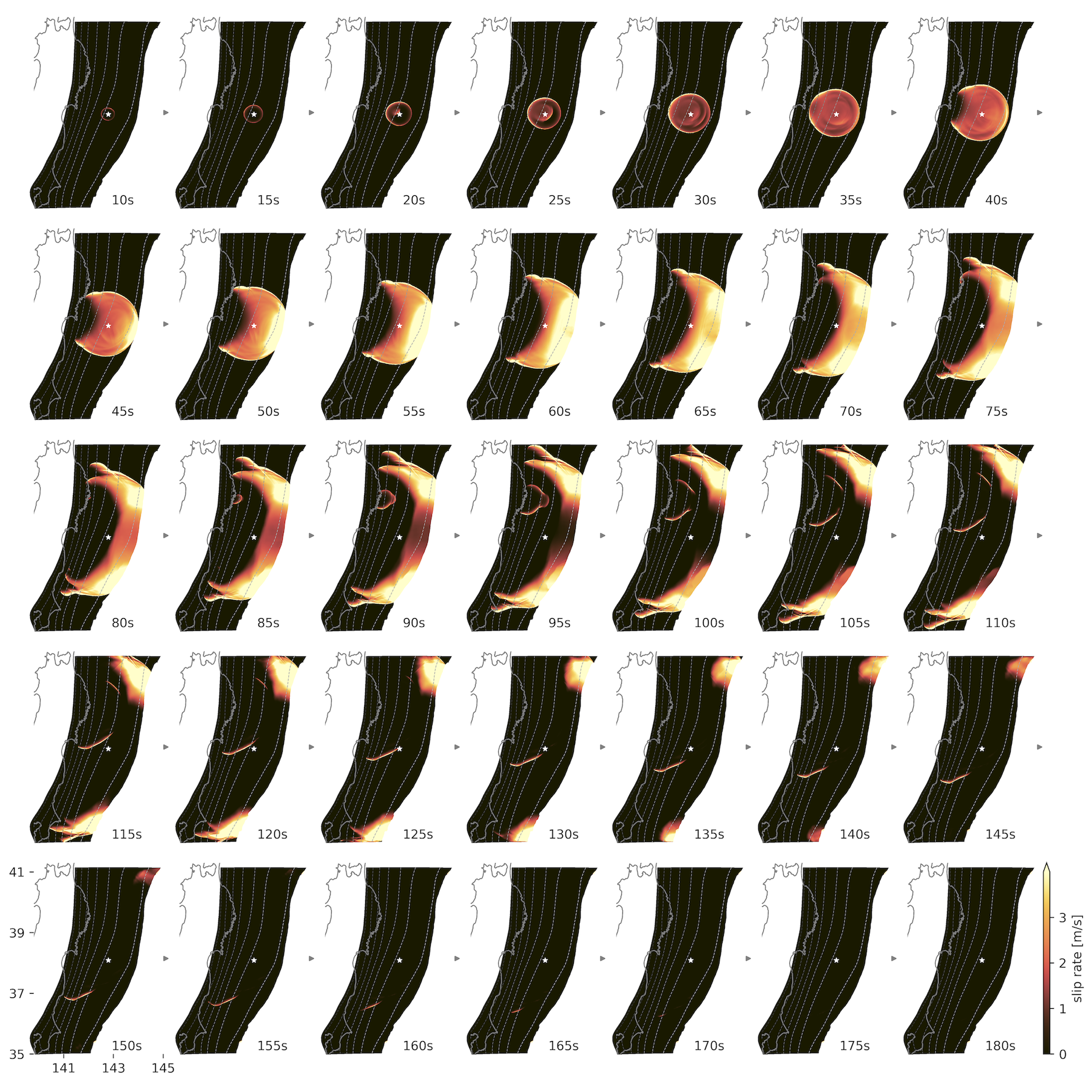}
\caption{\textbf{Dynamic rupture evolution of the laterally homogeneous prestress model} (see also Fig.~\ref{FigE8:Regional_model} and Supplementary Video S2). Snapshots of slip rate are shown in 5~s intervals.
Earthquake rupture initiates as a growing pulse, followed by rupture reactivation initiating at the downdip healing front of the primary growing pulse at 20~s.
Between 20--40~s, primary and secondary rupture fronts subsequently merge into a sustained, crack-like rupture without clear healing fronts separating slip episodes. 
At 40~s, rupture reaches the downdip limit of the seismogenic zone, forming a healing front that propagates updip and progressively shortens central slip rise times toward shallower depths as the rupture expands along strike. 
At the same time, bilateral deep supershear rupture is initiated ahead of the primary rupture front via the ``daughter crack'' mechanism \cite{Andrews1976Rupture} and likely due to higher effective normal stress and a relatively sharp transition to velocity-strengthening friction at depth \cite{Hu2019Sustainability}. 
This local supershear rupture remains confined to depths between 30--50 km.
At 45~s rupture time, about 5~s later compared to the preferred model, the primary updip rupture front reaches the seafloor interface, resulting in reflected phases.
Between 70--80~s, two secondary sub-Rayleigh rupture fronts re-rupture the down-dip part of the slab, including spiraling rupture dynamics and initiating backward-propagating fronts at 76~s (north) and 79~s (south). 
The northern reactivated front expands, triggering a local up-dip supershear rupture at about 90~s and re-rupturing the central slab, while the southern front decays. 
Updip ruptures arrest at the slab boundary around 140~s, whereas the downdip rupture front continues propagating until approximately 180~s simulation time.
}
\label{SFig:Regional_model_5s_snapshot}
\end{figure}
\clearpage
\newpage

\begin{figure}
\noindent\includegraphics[width=\textwidth]{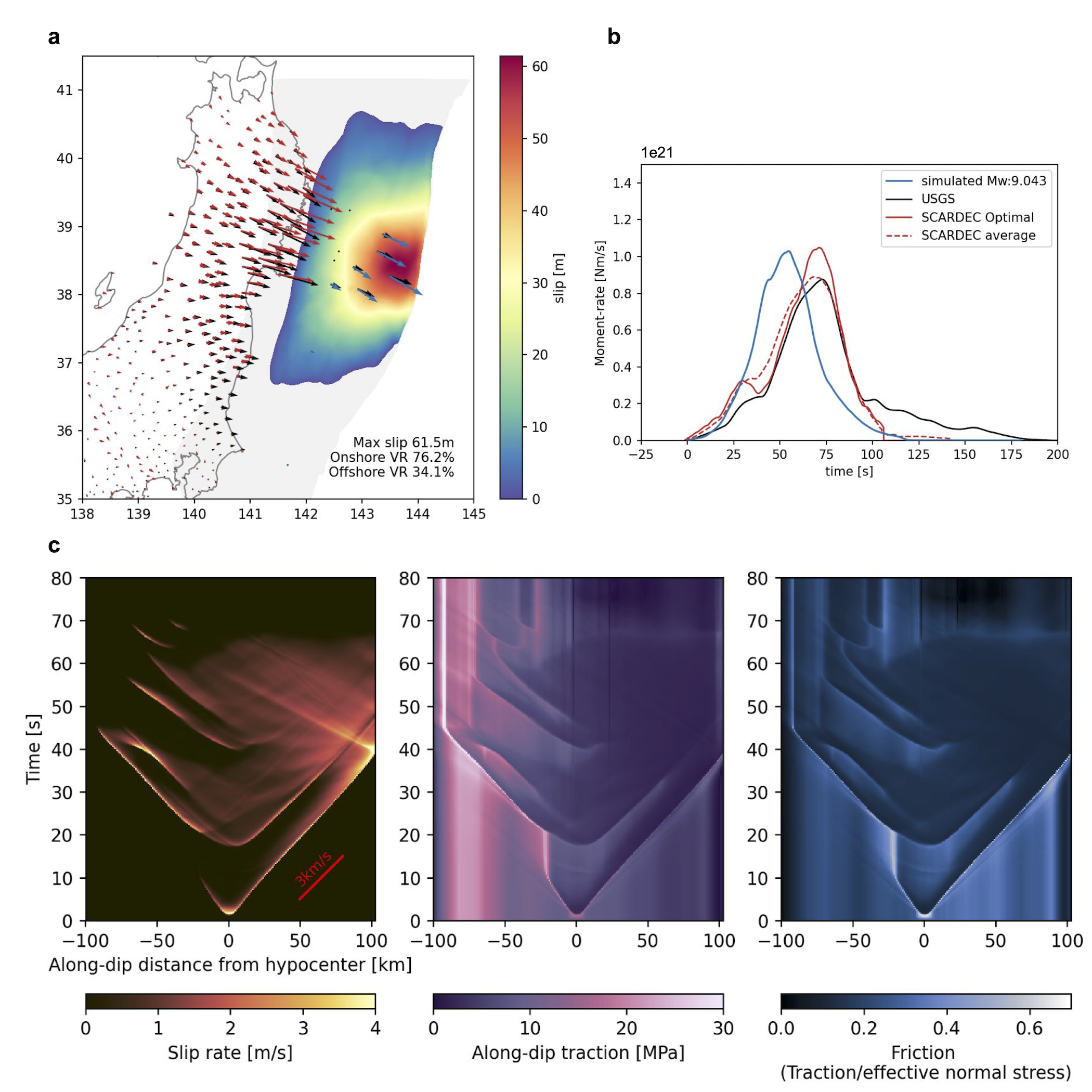}
\caption{\textbf{Alternative dynamic rupture model using a uniform weakening distance $L$ of 0.3~m} (See Supplementary section - SM2: Nucleation for details). (a) Fault slip distribution with geodetic data fit. Observed geodetic displacements are shown as black arrows. Onshore and offshore modeled displacements are shown as red and blue arrows, respectively. The model achieves a variance reduction of 76.2~\% (onshore) and 34.1~\% (offshore). (b) Comparison of modeled moment-rate function and moment-rate estimates from USGS \cite{Hayes2011Rapid} and SCARDEC \cite{VallA2016New}. (c) Temporal evolution of slip rate (red), along-dip shear stress (purple), and effective friction coefficient (blue) along the hypocentral dip profile, highlighting rapid variations coincident with dynamic rupture reactivation. This model features downdip pulse-like rupture and updip crack-like rupture characteristics.}
\label{SFig:model_uniformDc}
\end{figure}
\clearpage
\newpage

\clearpage
\newpage
\begin{figure}
\noindent\includegraphics[width=\textwidth]{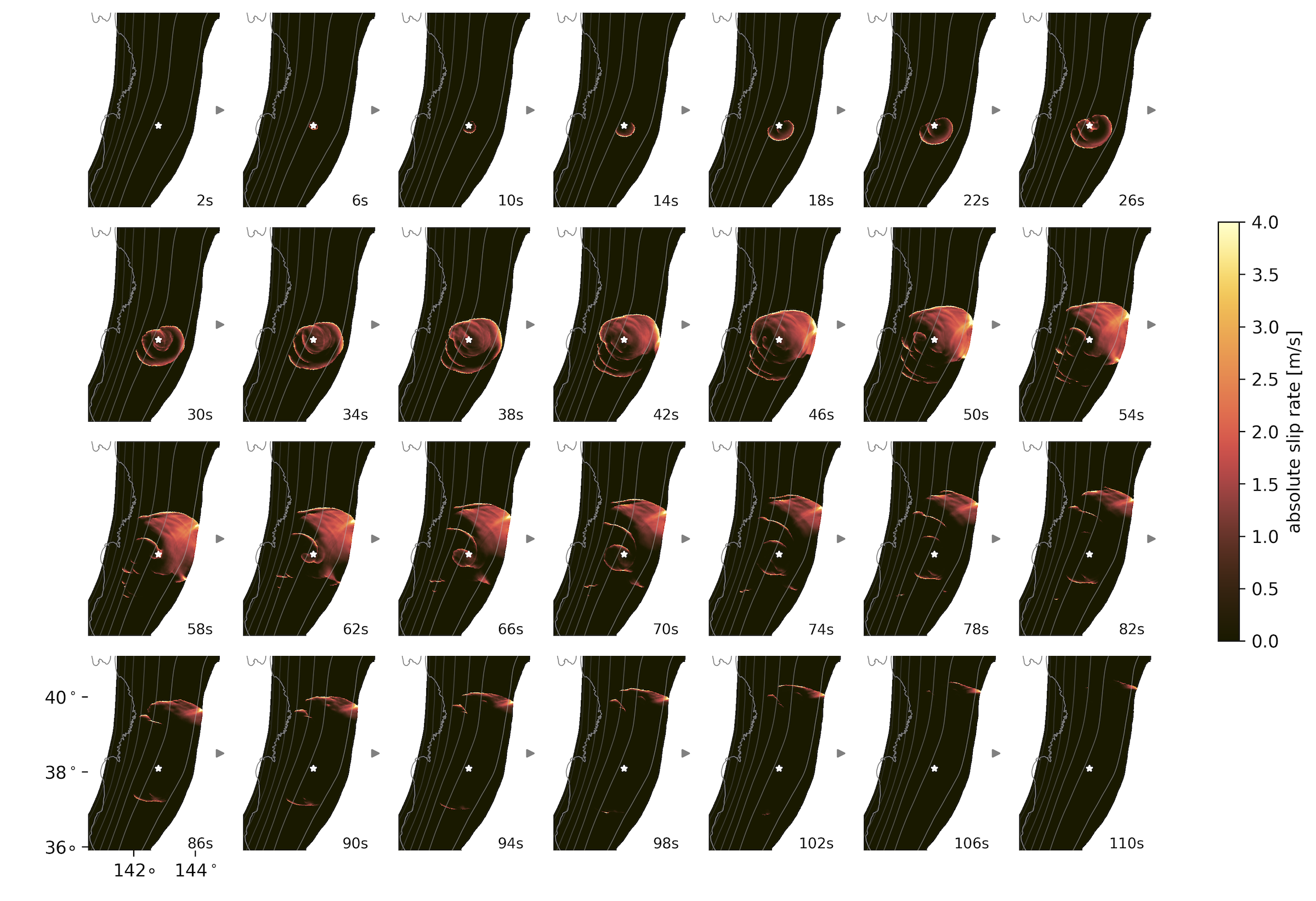}
\caption{\textbf{Slip-rate evolution of the heterogeneous-friction dynamic rupture model with multiscale variations in state-evolution distance.}  Snapshots are shown at 4~s intervals from 2 s to 110~s (left to right, top to bottom). Colors indicate absolute slip rate. Contours outline the slab depth at 10~km intervals. The simulation reproduces repeated slip reactivation and mixed downdip pulse-like and updip crack-like rupture styles. This simulation exhibits more frequent re-nucleation than the six episodes in the preferred model. See also Figure~\ref{Fig10:Sl0_het_model} and Video S4.}
\label{SFig:Sl0_snapshot}
\end{figure}
\clearpage
\newpage

\begin{figure}
\noindent\includegraphics[width=\textwidth]{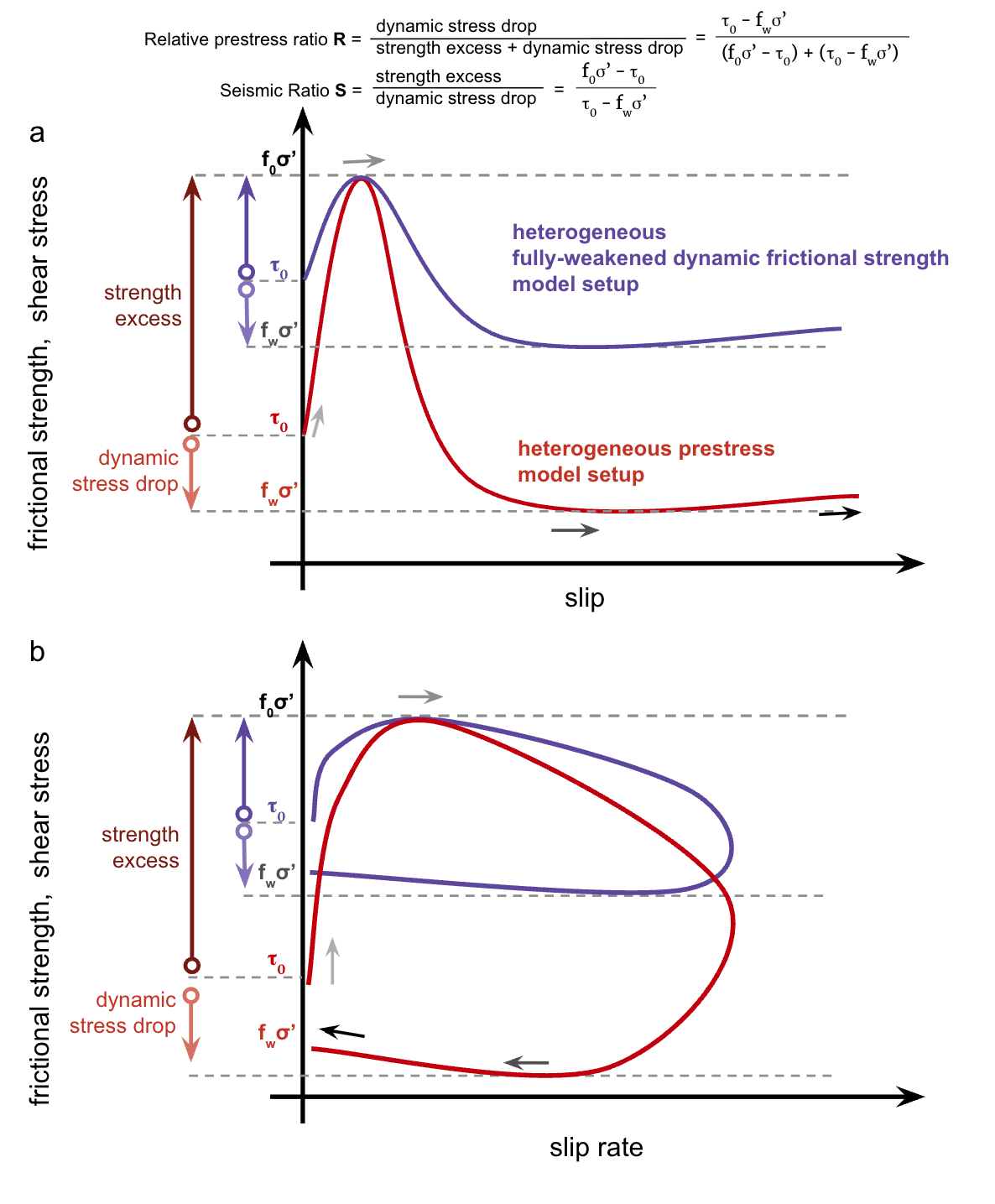}
\caption{\textbf{Modeled fault-local frictional evolution in two alternative model setups with heterogeneity either in friction or in prestress.} a) friction evolution with slip. b) frictional evolution with slip rate. The purple curve shows the frictional strength evolution for the setup with a heterogeneous fully-weakened dynamic frictional strength. The red curve shows the frictional strength evolution for the preferred model with heterogeneous prestress. The grey arrows in both panels show the stages of frictional evolution correspondingly. The darker arrows in the y-axis labels indicate the frictional strength excess, and the lighter arrows indicate the dynamic stress drop. We added the definitions of the seismic ratio S and relative prestress level $R$. Given the same dynamic stress drop value, the heterogeneous fully-weakened dynamic frictional strength model has a lower ratio of strength excess to dynamic stress drop ($S$ ratio, equation \ref{eq:S}) than the heterogeneous prestress model.}
\label{SFig:Fw_slip_strength_illustration}
\end{figure}

\begin{figure}
\noindent\includegraphics[width=\textwidth]{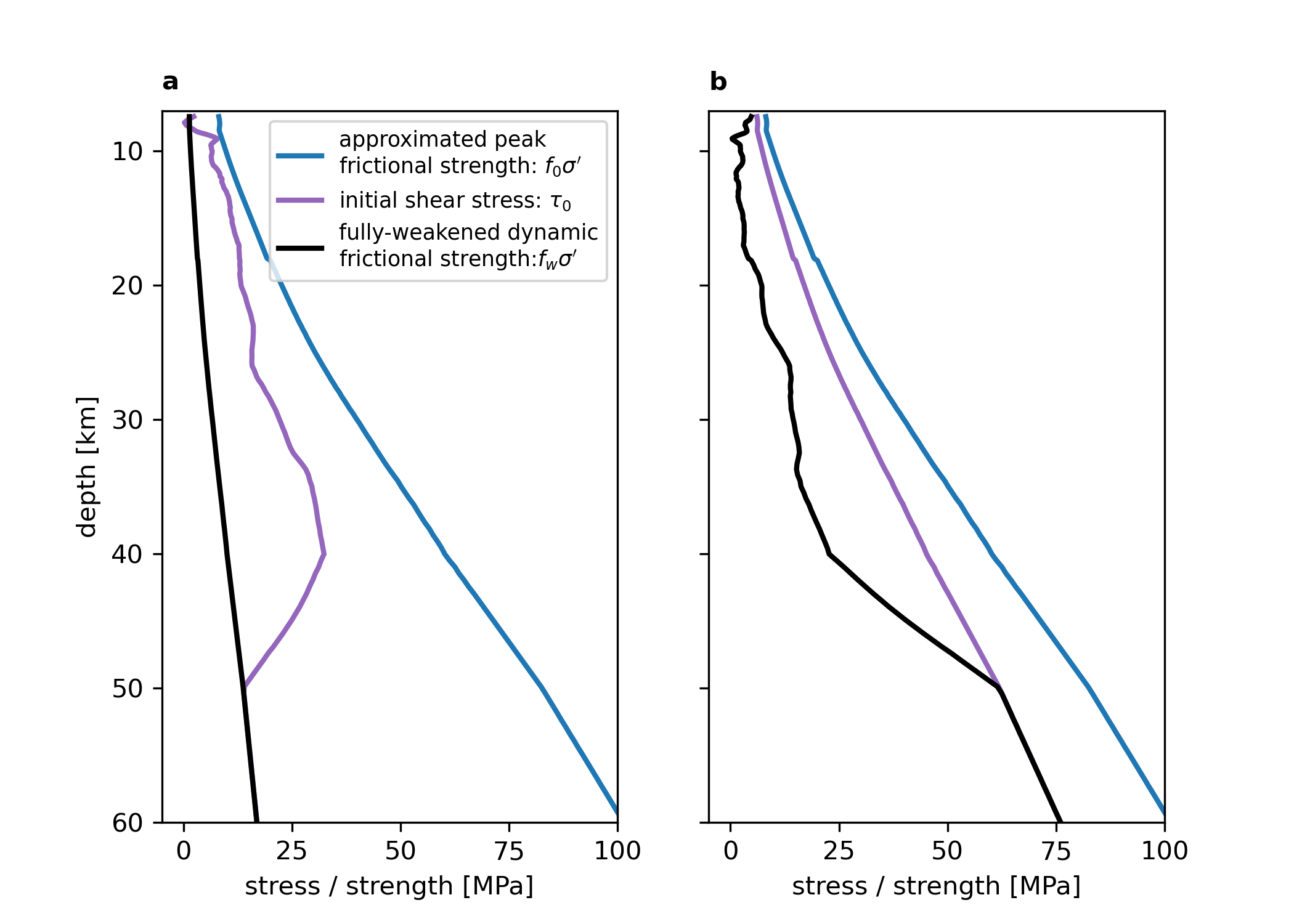}
\caption{\textbf{Depth-dependent variation of frictional strength and initial stress conditions along a hypocentral dip profile using two alternative model setups with heterogeneity in dynamic parameters.} Blue, purple, and black lines represent the approximated peak frictional strength ($f_0\sigma'$), initial shear stress ($\tau_0$), and fully-weakened dynamic frictional strength ($f_w\sigma'$), respectively (Methods section: \nameref{Initial relative prestress condition}). (a) Heterogeneity in pretress. This model setup assumes depth-dependent frictional strength. The initial stress level follows the stress-change pattern inferred from the median slip distribution. (b) Heterogeneity in the fully-weakened dynamic frictional strength. This setup maps the heterogeneity onto the fully-weakened dynamic frictional strength while using a homogeneous depth-dependent initial stress. }
\label{SFig:Fw_compare}
\end{figure}

\begin{figure}
\noindent\includegraphics[width=\textwidth]{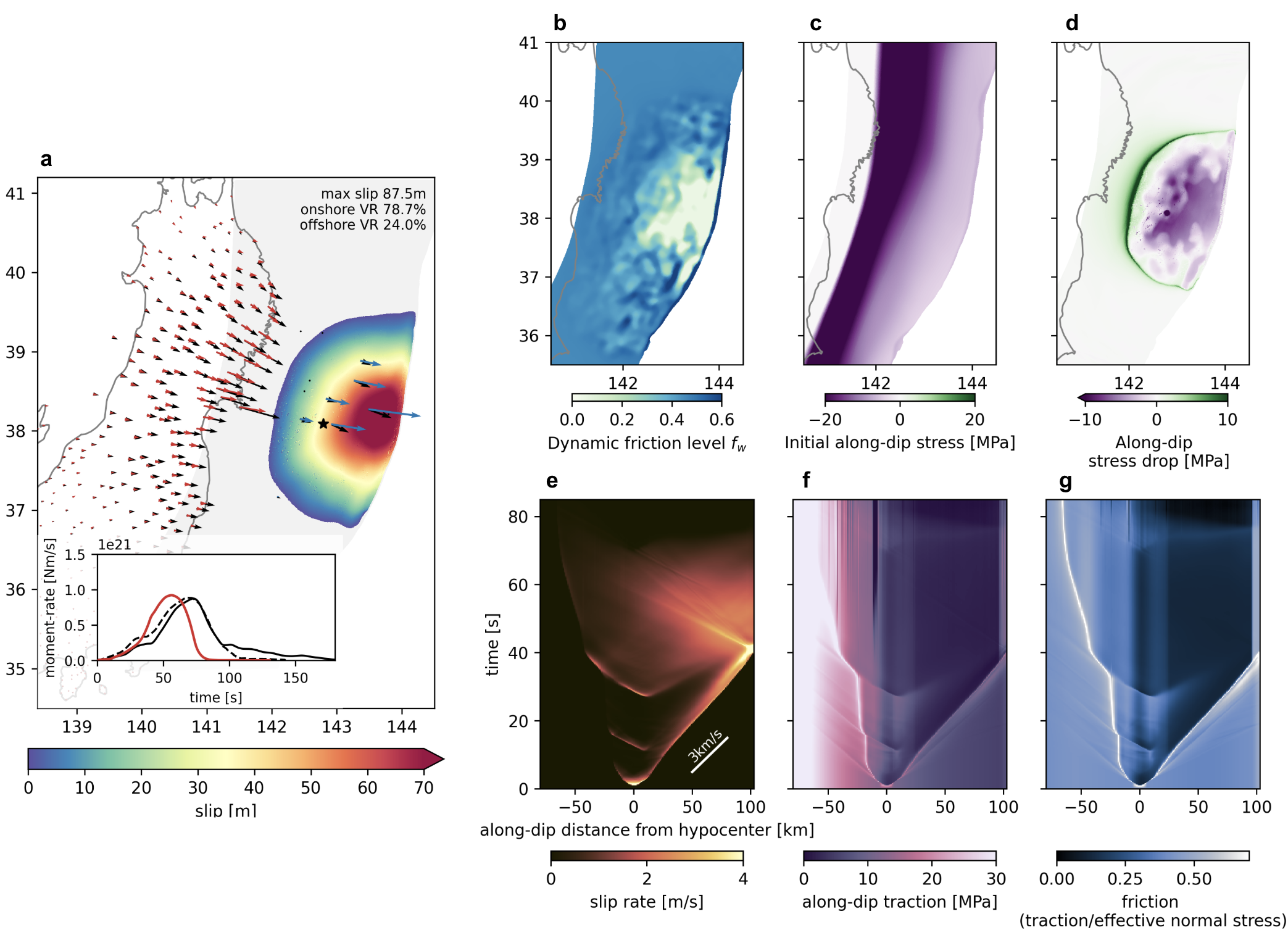}
\caption{\textbf{Dynamic rupture scenario with heterogeneous distribution of fully-weakened dynamic frictional strength and homogeneous, depth-dependent initial stress.} (a) Simulated slip distribution and corresponding geodetic deformation. Red and blue arrows indicate synthetic onshore and offshore displacements, respectively. Black arrows show the observations. The inset compares the simulated moment-rate function (red), with the USGS (solid black), and SCARDEC (dashed black) source model moment-rate functions (\cite{Hayes2011Rapid, VallA2016New}). (b) Spatial distribution of the fully-weakened friction coefficient ($f_w$). (c) Distribution of initial along-dip shear stress. (d) Spatial distribution of the along-dip stress drop. (e-g) Along-dip profiles of slip-rate, along-dip traction, and effective friction coefficient evolution, from left to right, respectively.See also slip-rate evolution in Supplementary Figure~\ref{SFig:Fw_snapshot} and Video~S5.}
\label{SFig:Fw_sim}
\end{figure}

\clearpage
\newpage
\begin{figure}
\noindent\includegraphics[width=\textwidth]{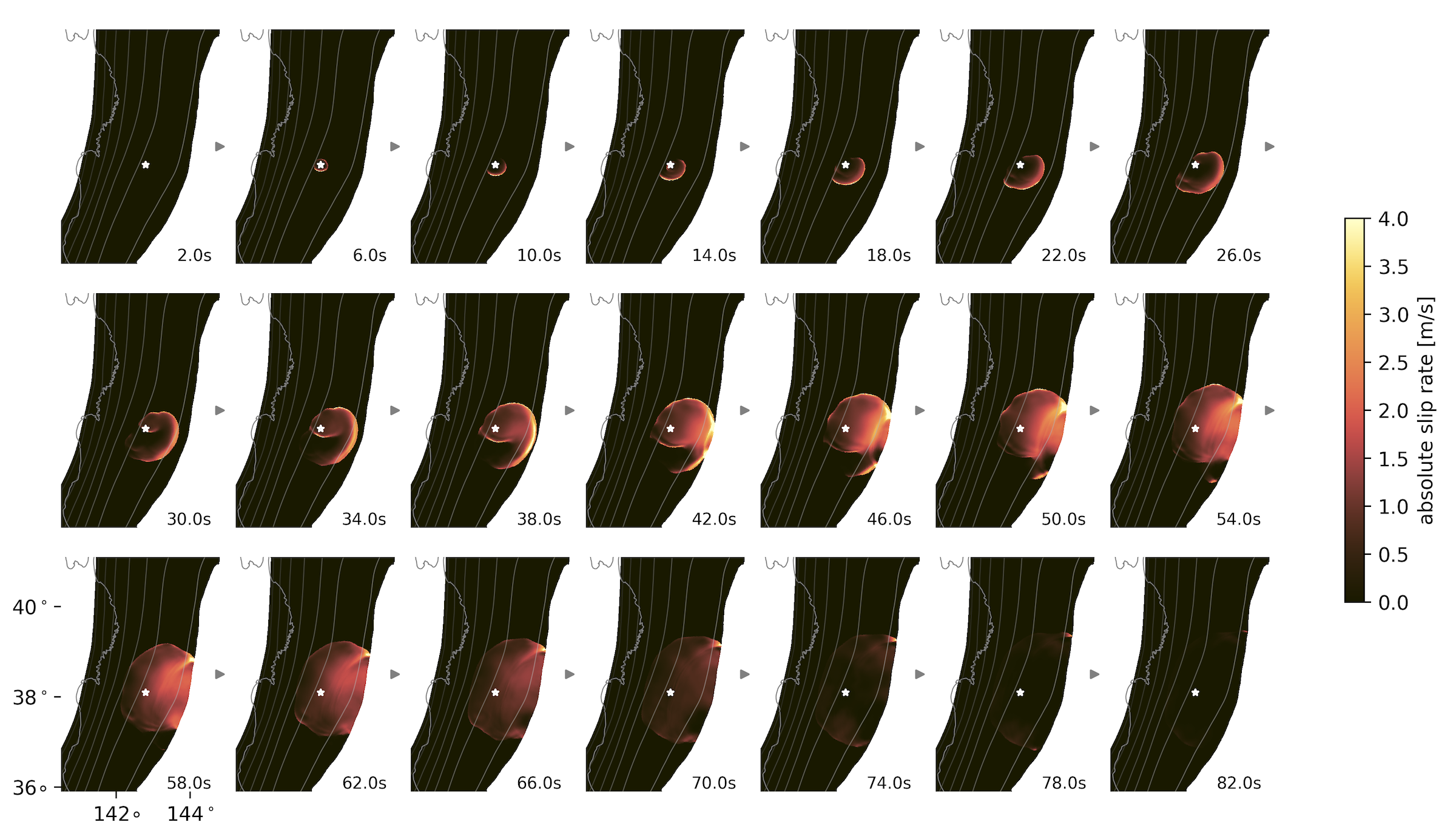}
\caption{\textbf{Slip-rate evolution of the dynamic rupture model with heterogeneous distribution of fully-weakened dynamic frictional strength and homogeneous, depth-dependent initial stress}. Snapshots are shown at 4~s intervals from 2~s to 82~s (left to right, top to bottom). Colors indicate absolute slip rate. Contours outline the slab depth at 10 km intervals. The simulation reproduces slip reactivation, mixed pulse-like and crack-like rupture styles, and large slip to the trench. Compared to the preferred heterogeneous prestress model, the depth-dependent variability of rupture style in this simulation is less clear, and rupture transitions into a pure crack-like style after 50~s simulation time. See also Supplementary Figure~\ref{SFig:Fw_sim} and Video S5.}
\label{SFig:Fw_snapshot}
\end{figure}
\clearpage
\newpage

\begin{figure}
\noindent\includegraphics[width=\textwidth]{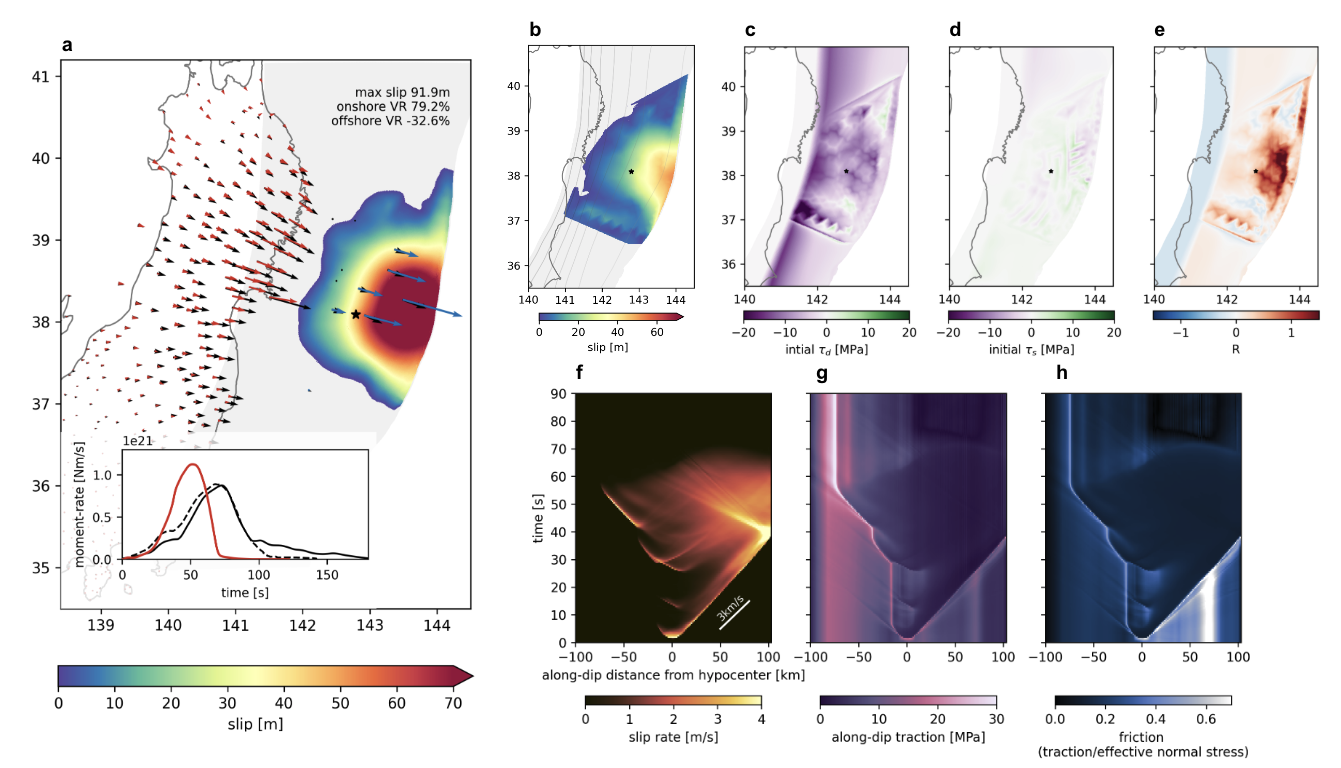}
\caption{\textbf{Dynamic rupture model using the stress-change pattern derived from the finite-fault slip model of Kubota et al. (2022).} (a) Simulated slip distribution and corresponding geodetic deformation. Red and blue arrows indicate synthetic onshore and offshore displacements, respectively. Black arrows show the observations. The inset compares the simulated moment-rate function (red), with the USGS (solid black), and SCARDEC (dashed black) source model moment-rate functions (\cite{Hayes2011Rapid, VallA2016New}). (b) Kubota et al., 2022 finite-fault model slip distribution. (c-e) Distribution of initial along-dip shear stress  ($\tau_d$), along-strike shear stress ($\tau_s$), and relative prestress ratio ($R$), from left to right, respectively. (f-h) Along dip profile of slip rate, along-dip traction, and effective friction coefficient, from left to right, respectively. See also Supplementary Figure~\ref{SFig:Kubota_SR_snapshot} and Video~S6.}
\label{SFig:Kubota_sim}
\end{figure}
\clearpage
\newpage

\begin{figure}
\noindent\includegraphics[width=\textwidth]{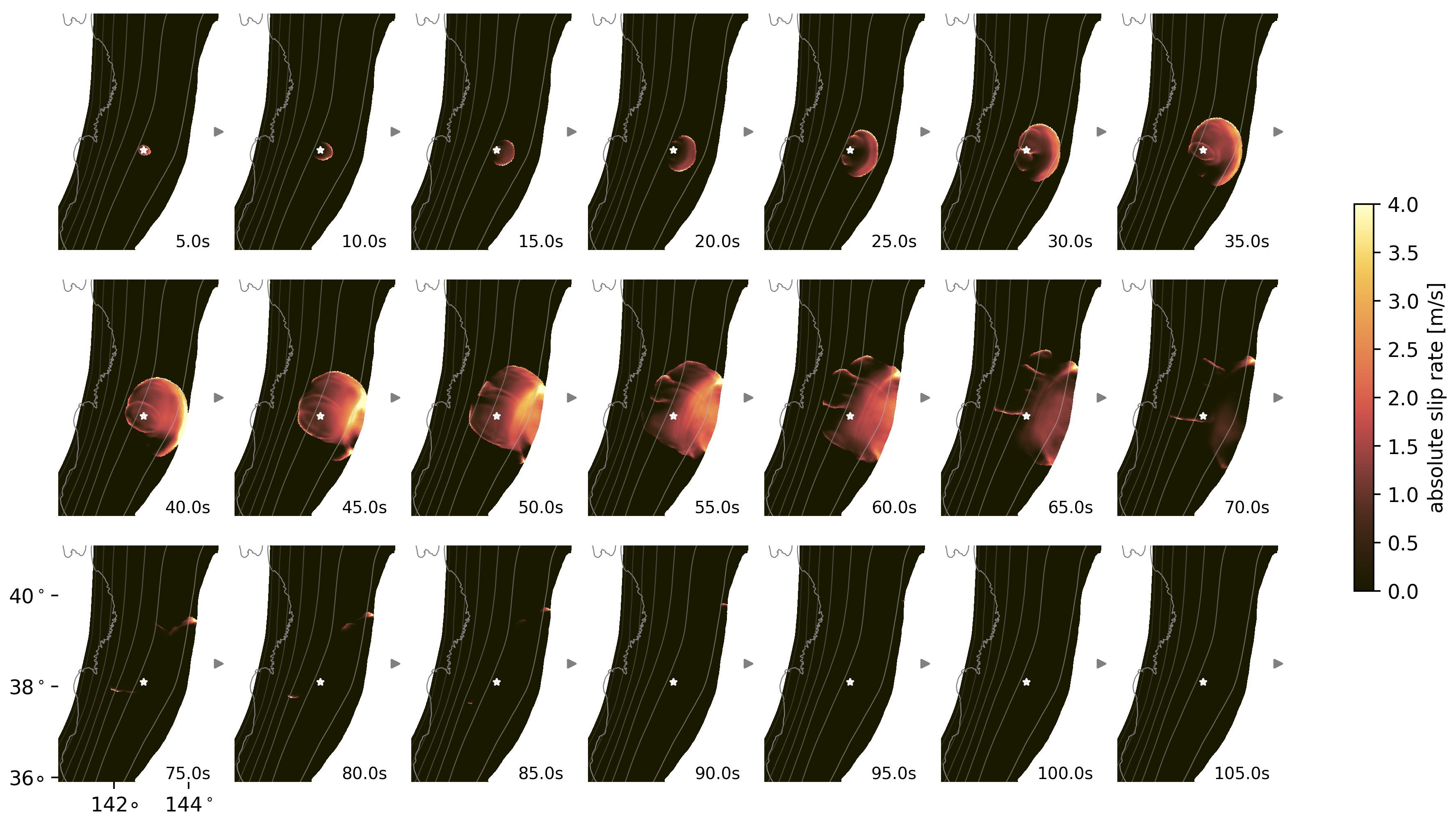}
\caption{\textbf{Slip-rate evolution of dynamic rupture model using stress-change pattern derived from the finite-fault slip model of Kubota et al. (2022).} Snapshots are shown at 5 s intervals from 5 s to 105 s (left to right, top to bottom). Colors indicate absolute slip rate. Contours outline the depth at 10 km intervals. The simulation reproduces repeated slip reactivation and mixed downdip pulse-like and updip crack-like rupture styles. See also Supplementary Figure S11 and Video S4. See also Supplementary Video S5.}
\label{SFig:Kubota_SR_snapshot}
\end{figure}
\clearpage
\newpage

\begin{figure}
\noindent\includegraphics[width=\textwidth]{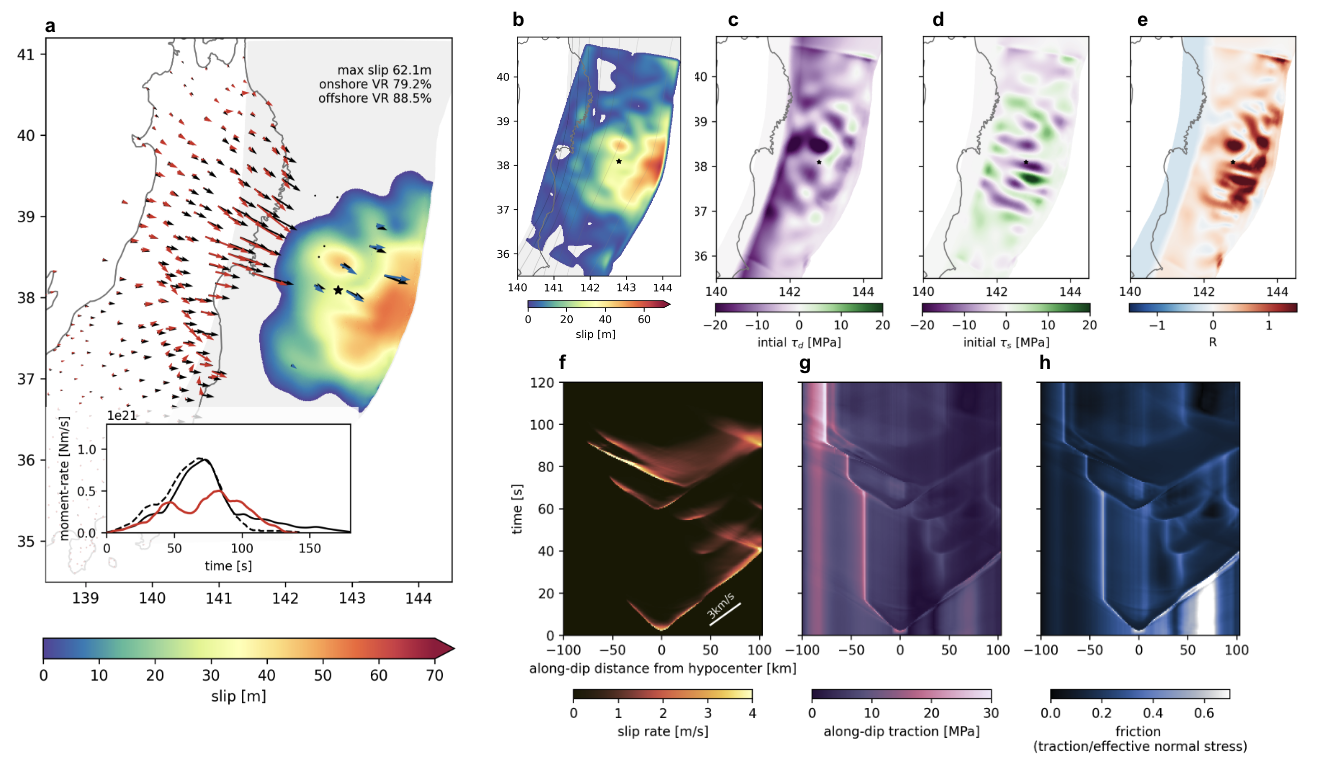}
\caption{\textbf{Dynamic rupture model using stress-change pattern derived from the finite-fault slip model of Melgar \& Bock, 2016.} (a) Simulated slip distribution and corresponding geodetic deformation. Red and blue arrows indicate synthetic onshore and offshore deformation, respectively. Black arrows show the observations. The inset compares the simulated moment-rate function (red), with the USGS (solid black), and SCARDEC (dashed black) source model moment-rate functions \cite{Hayes2011Rapid, Vallee2013Source}. (b) Melgar \& Bock (2016) finite-fault model slip distribution. (c-e) Distribution of initial along-dip shear stress ($\tau_d$), along-strike shear stress ($\tau_s$), and relative prestress ratio ($R$), from left to right, respectively. (f-h) Along dip profile of slip rate, along-dip traction, and effective friction coefficient, from left to right, respectively. See also Supplementary Figure~\ref{SFig:Melgar_SR_snapshot} and Video~S7.}
\label{SFig:Melgar_sim}
\end{figure}
\clearpage
\newpage

\begin{figure}
\noindent\includegraphics[width=\textwidth]{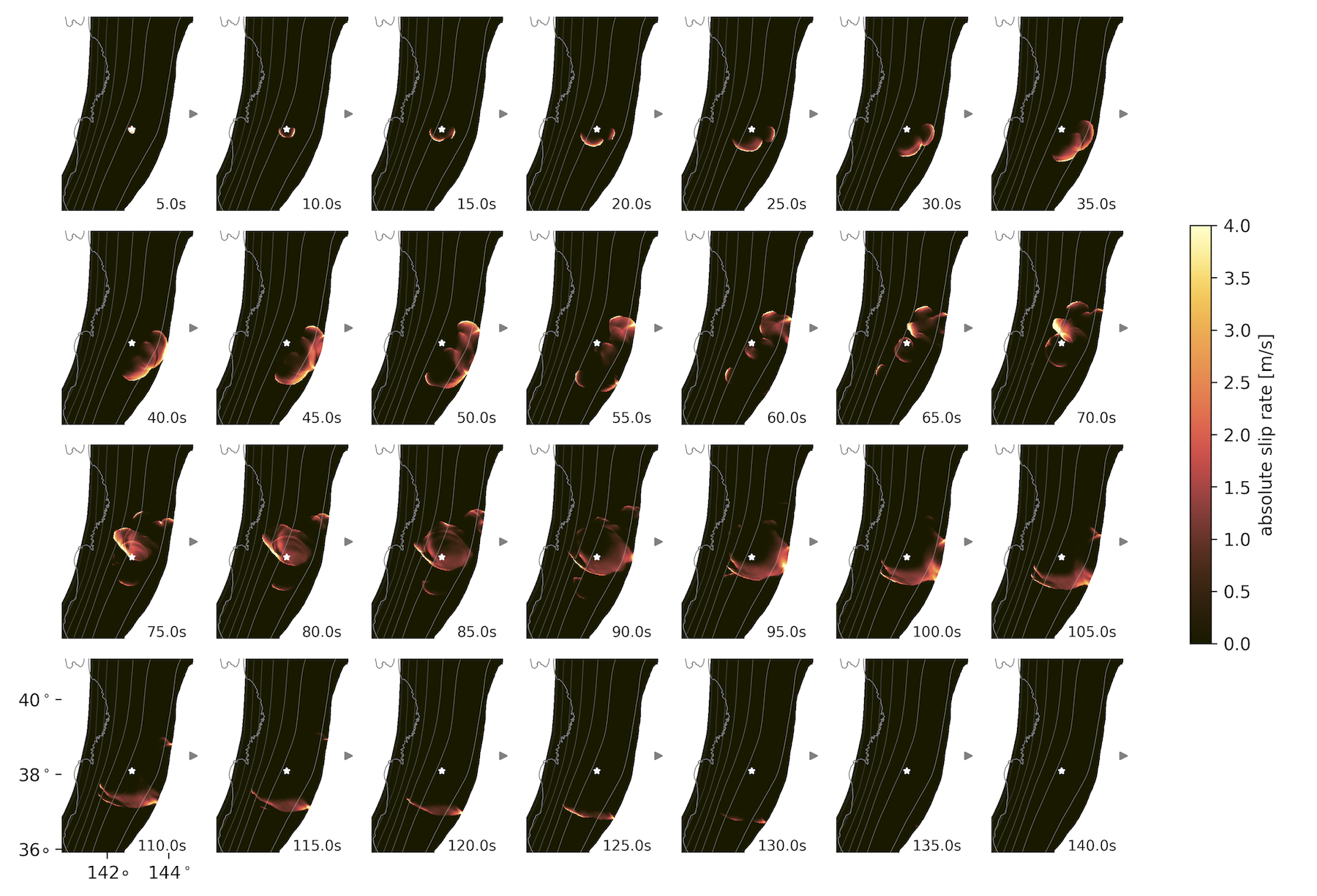}
\caption{\textbf{Slip-rate evolution of dynamic rupture model using stress-change pattern derived from the finite-fault slip model of Melgar \& Bock (2016).} 
Snapshots are shown at 5~s intervals from 5~s to 140~s (left to right, top to bottom). 
Colors indicate absolute slip rate. 
Contours outline the slab depth at 10 km intervals. 
The simulation reproduces repeated slip reactivation and mixed pulse-like and crack-like rupture styles. 
The strong prestress heterogeneity leads to updip reactivated slip pulses and less clear depth-dependent rupture style variability. See also Supplementary Figure~\ref{SFig:Melgar_sim} and Video~S7.}
\label{SFig:Melgar_SR_snapshot}
\end{figure}
\clearpage
\newpage

\begin{figure}
\noindent\includegraphics[width=\textwidth]{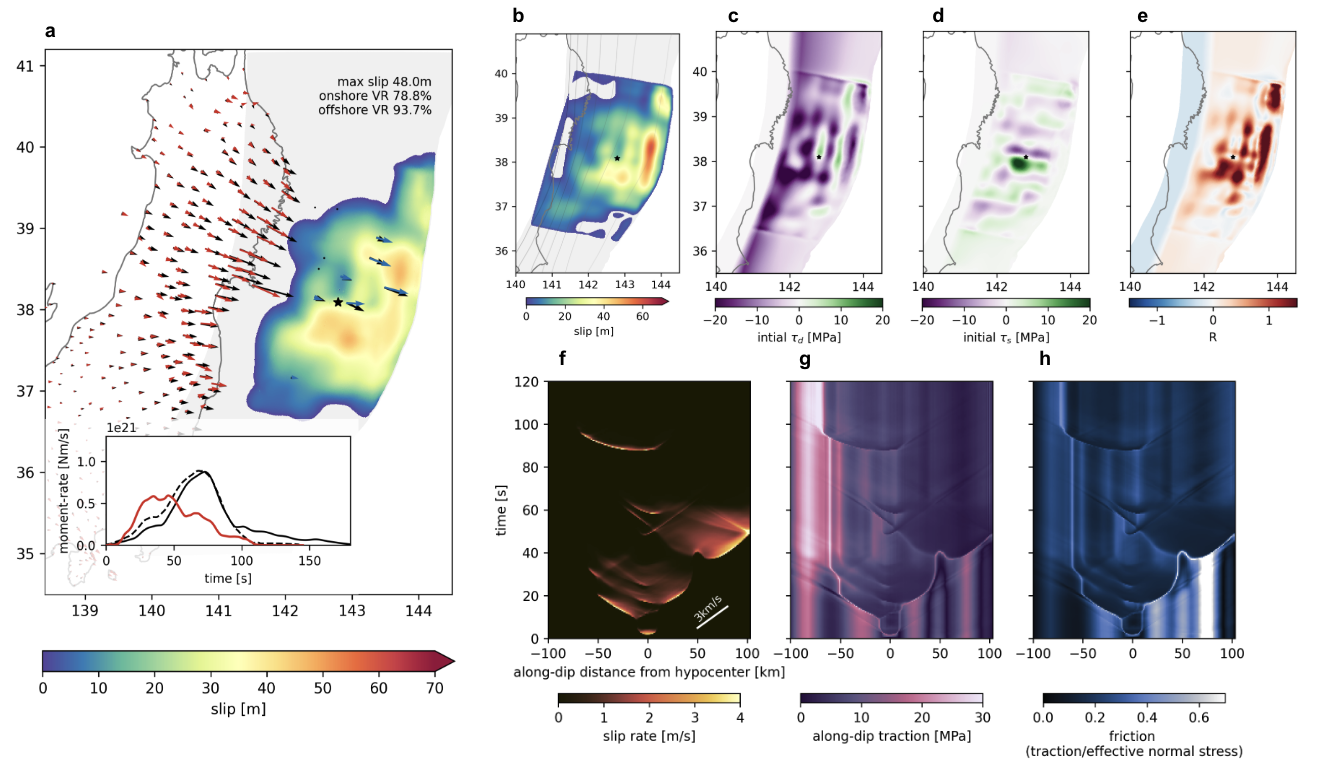}
\caption{\textbf{Dynamic rupture model using stress-change pattern derived from the finite-fault slip model of Yamazaki et al. (2018). }(a) Simulated slip distribution and corresponding geodetic deformation. Red and blue arrows indicate synthetic onshore and offshore deformation, respectively. Black arrows show the observations. 
The inset compares the simulated moment-rate function (red), with the USGS (solid black), and SCARDEC (dashed black) source model moment-rate functions \cite{Hayes2011Rapid, Vallee2013Source}. 
(b) Yamazaki et al. (2018) finite-fault model slip distribution. (c-e) Distribution of initial along-dip shear stress ($\tau_d$), along-strike shear stress ($\tau_s$), and relative prestress ratio ($R$), from left to right, respectively. (f-h) Along dip profile of slip rate, along-dip traction, and effective friction coefficient, from left to right, respectively.See also Supplementary Figure~\ref{SFig:Yamazaki_SR_snapshot} and Video~S8.}
\label{SFig:Yamazaki_sim}
\end{figure}
\clearpage
\newpage

\begin{figure}
\noindent\includegraphics[width=\textwidth]{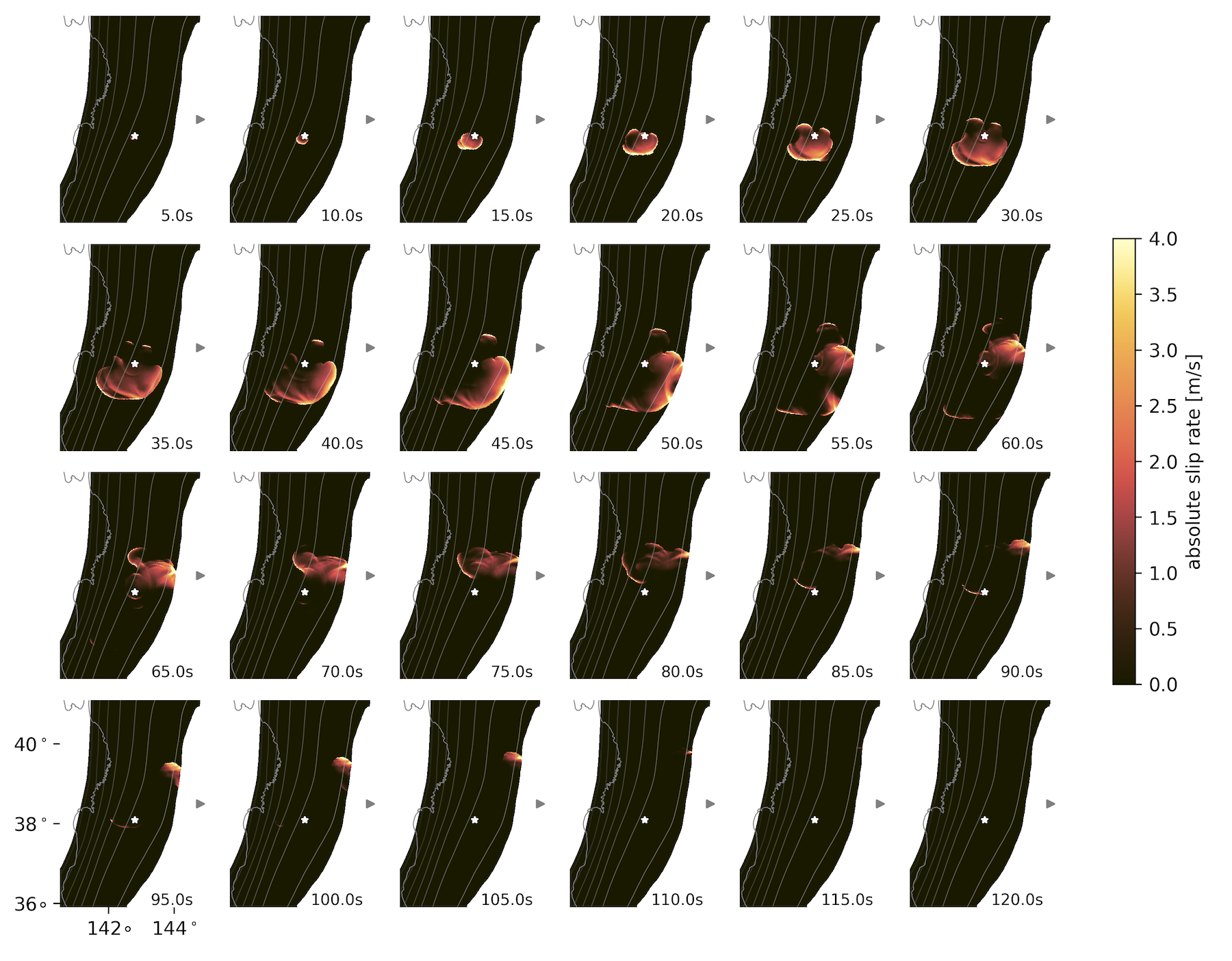}
\caption{\textbf{Slip-rate evolution of dynamic rupture model using stress-change pattern derived from the finite-fault slip model of Yamazaki et al. (2018)}. Snapshots are shown at 5 s intervals from 5 s to 120 s (left to right, top to bottom). Colors indicate absolute slip rate. Contours outline the depth at 10 km intervals. The simulation reproduces repeated slip reactivation and mixed downdip pulse-like and updip crack-like rupture styles. See also Supplementary Figure~\ref{SFig:Yamazaki_sim} and Video S8. }
\label{SFig:Yamazaki_SR_snapshot}
\end{figure}
\clearpage
\newpage

%% file: Tohoku.bib
@article{Ampuero2008Earthquake,
  title = {Earthquake Nucleation on Rate and State Faults -- {{Aging}} and Slip Laws},
  author = {Ampuero, Jean-Paul and Rubin, Allan M.},
  year = {2008},
  month = jan,
  journal = {Journal of Geophysical Research: Solid Earth},
  volume = {113},
  number = {B1},
  pages = {2007JB005082},
  issn = {0148-0227},
  doi = {10.1029/2007JB005082}
}

@article{Andrews2005Rupture,
  title = {Rupture Dynamics with Energy Loss Outside the Slip Zone},
  author = {Andrews, D. J.},
  year = {2005},
  journal = {Journal of Geophysical Research: Solid Earth},
  volume = {110},
  number = {B1},
  issn = {2156-2202},
  doi = {10.1029/2004JB003191}
}

@article{Aochi20031999,
  title = {The 1999 {{Izmit}}, {{Turkey}}, {{Earthquake}}: {{Nonplanar Fault Structure}}, {{Dynamic Rupture Process}}, and {{Strong Ground Motion}}},
  author = {Aochi, H.},
  year = {2003},
  month = jun,
  journal = {Bulletin of the Seismological Society of America},
  volume = {93},
  number = {3},
  pages = {1249--1266},
  issn = {0037-1106},
  doi = {10.1785/0120020167}
}

@article{Aochi2011Conceptual,
  title = {Conceptual Multi-Scale Dynamic Rupture Model for the 2011 off the {{Pacific}} Coast of {{Tohoku Earthquake}}},
  author = {Aochi, Hideo and Ide, Satoshi},
  year = {2011},
  month = jul,
  journal = {Earth, Planets and Space},
  volume = {63},
  number = {7},
  pages = {761--765},
  publisher = {SpringerOpen},
  issn = {1880-5981},
  doi = {10.5047/eps.2011.05.008}
}

@article{Barbot2019Slowslip,
  title = {Slow-Slip, Slow Earthquakes, Period-Two Cycles, Full and Partial Ruptures, and Deterministic Chaos in a Single Asperity Fault},
  author = {Barbot, Sylvain},
  year = {2019},
  month = oct,
  journal = {Tectonophysics},
  volume = {768},
  pages = {228171},
  issn = {00401951},
  doi = {10.1016/j.tecto.2019.228171}
}

@article{Barras2023How,
  title = {How {{Do Earthquakes Stop}}? {{Insights From}} a {{Minimal Model}} of {{Frictional Rupture}}},
  author = {Barras, Fabian and Th{\o}gersen, Kjetil and Aharonov, Einat and Renard, Fran{\c c}ois},
  year = {2023},
  month = aug,
  journal = {Journal of Geophysical Research: Solid Earth},
  volume = {128},
  number = {8},
  pages = {e2022JB026070},
  issn = {2169-9313, 2169-9356},
  doi = {10.1029/2022JB026070}
}

@article{Bassett2015Gravity,
  title = {Gravity Anomalies, Crustal Structure, and Seismicity at Subduction Zones: 2. {{Interrelationships}} between Fore-Arc Structure and Seismogenic Behavior},
  author = {Bassett, Dan and Watts, Anthony B.},
  year = {2015},
  journal = {Geochemistry, Geophysics, Geosystems},
  volume = {16},
  number = {5},
  pages = {1541--1576},
  issn = {1525-2027},
  doi = {10.1002/2014GC005685}
}

@article{Bassett2025Variation,
  title = {Variation in Slip Behaviour along Megathrusts Controlled by Multiple Physical Properties},
  author = {Bassett, Dan and Shillington, Donna J. and Wallace, Laura M. and Elliott, Julie L.},
  year = {2025},
  month = jan,
  journal = {Nature Geoscience},
  issn = {1752-0894, 1752-0908},
  doi = {10.1038/s41561-024-01617-9}
}

@article{Beeler2008Constitutive,
  title = {Constitutive Relationships and Physical Basis of Fault Strength Due to Flash Heating},
  author = {Beeler, N. M. and Tullis, T. E. and Goldsby, D. L.},
  year = {2008},
  journal = {Journal of Geophysical Research: Solid Earth},
  volume = {113},
  number = {B1},
  issn = {2156-2202},
  doi = {10.1029/2007JB004988}
}

@article{Bizzarri2006Thermal,
  title = {A Thermal Pressurization Model for the Spontaneous Dynamic Rupture Propagation on a Three-dimensional Fault: 1. {{Methodological}} Approach},
  author = {Bizzarri, A. and Cocco, M.},
  year = {2006},
  month = may,
  journal = {Journal of Geophysical Research: Solid Earth},
  volume = {111},
  number = {B5},
  pages = {2005JB003862},
  issn = {0148-0227},
  doi = {10.1029/2005JB003862}
}

@inproceedings{Breuer2022NextGeneration,
  title = {Next-{{Generation Local Time Stepping}} for the {{ADER-DG Finite Element Method}}},
  booktitle = {2022 {{IEEE International Parallel}} and {{Distributed Processing Symposium}} ({{IPDPS}})},
  author = {Breuer, Alexander and Heinecke, Alexander},
  year = {2022},
  month = may,
  pages = {402--413},
  publisher = {IEEE},
  address = {Lyon, France},
  doi = {10.1109/IPDPS53621.2022.00046},
  isbn = {978-1-66548-106-9}
}

@article{Brodsky2020State,
  title = {The {{State}} of {{Stress}} on the {{Fault Before}}, {{During}}, and {{After}} a {{Major Earthquake}}},
  author = {Brodsky, Emily E. and Mori, James J. and Anderson, Louise and Chester, Frederick M. and Conin, Marianne and Dunham, Eric M. and Eguchi, Nobu and Fulton, Patrick M. and Hino, Ryota and Hirose, Takehiro and Ikari, Matt J. and Ishikawa, Tsuyoshi and Jeppson, Tamara and Kano, Yasuyuki and Kirkpatrick, James and Kodaira, Shuichi and Lin, Weiren and Nakamura, Yasuyuki and Rabinowitz, Hannah S. and Regalla, Christine and Remitti, Francesca and Rowe, Christie and Saffer, Demian M. and Saito, Saneatsu and Sample, James and Sanada, Yoshinori and Savage, Heather M. and Sun, Tianhaozhe and Toczko, Sean and Ujiie, Kohtaro and {Wolfson-Schwehr}, Monica and Yang, Tao},
  year = {2020},
  month = may,
  journal = {Annual Review of Earth and Planetary Sciences},
  volume = {48},
  number = {1},
  pages = {49--74},
  issn = {0084-6597, 1545-4495},
  doi = {10.1146/annurev-earth-053018-060507}
}

@article{Brown2015Static,
  title = {Static Stress Drop in the {\emph{ }}{{{\emph{M}}}}{\emph{ }}{\emph{{\textsubscript{w}}}}{\emph{ }} 9 {{Tohoku}}-oki Earthquake: {{Heterogeneous}} Distribution and Low Average Value},
  author = {Brown, Lonn and Wang, Kelin and Sun, Tianhaozhe},
  year = {2015},
  month = dec,
  journal = {Geophysical Research Letters},
  volume = {42},
  number = {24},
  issn = {0094-8276, 1944-8007},
  doi = {10.1002/2015GL066361}
}

@article{Cattania2019Complex,
  title = {Complex {{Earthquake Sequences On Simple Faults}}},
  author = {Cattania, C.},
  year = {2019},
  month = sep,
  journal = {Geophysical Research Letters},
  volume = {46},
  number = {17-18},
  pages = {10384--10393},
  issn = {0094-8276, 1944-8007},
  doi = {10.1029/2019GL083628}
}

@article{Chan2023Impact,
  title = {Impact of {{Hypocenter Location}} on {{Rupture Extent}} and {{Ground Motion}}: {{A Case Study}} of {{Southern Cascadia}}},
  author = {Chan, Yuk Po Bowie and Yao, Suli and Yang, Hongfeng},
  year = {2023},
  month = aug,
  journal = {Journal of Geophysical Research: Solid Earth},
  volume = {128},
  number = {8},
  pages = {e2023JB026371},
  issn = {2169-9313, 2169-9356},
  doi = {10.1029/2023JB026371}
}

@article{Cocco2023Fracture,
  title = {Fracture {{Energy}} and {{Breakdown Work During Earthquakes}}},
  author = {Cocco, Massimo and Aretusini, Stefano and Cornelio, Chiara and Nielsen, Stefan B. and Spagnuolo, Elena and Tinti, Elisa and Di Toro, Giulio},
  year = {2023},
  month = may,
  journal = {Annual Review of Earth and Planetary Sciences},
  volume = {51},
  number = {1},
  pages = {217--252},
  issn = {0084-6597, 1545-4495},
  doi = {10.1146/annurev-earth-071822-100304}
}

@article{Cochard2024Propagation,
  title = {Propagation of Extended Fractures by Local Nucleation and Rapid Transverse Expansion of Crack-Front Distortion},
  author = {Cochard, T. and Svetlizky, I. and Albertini, G. and Viesca, R. C. and Rubinstein, S. M. and Spaepen, F. and Yuan, C. and Denolle, M. and Song, Y-Q. and Xiao, L. and Weitz, D. A.},
  year = {2024},
  month = apr,
  journal = {Nature Physics},
  volume = {20},
  number = {4},
  pages = {660--665},
  issn = {1745-2473, 1745-2481},
  doi = {10.1038/s41567-023-02365-0}
}

@article{Das1977Numerical,
  title = {A Numerical Study of Two-Dimensional Spontaneous Rupture Propagation},
  author = {Das, Shamita and Aki, Keiiti},
  year = {1977},
  month = sep,
  journal = {Geophysical Journal International},
  volume = {50},
  number = {3},
  pages = {643--668},
  issn = {0956-540X},
  doi = {10.1111/j.1365-246X.1977.tb01339.x}
}

@article{Day2005Comparison,
  title = {Comparison of Finite Difference and Boundary Integral Solutions to Three-Dimensional Spontaneous Rupture},
  author = {Day, Steven M. and Dalguer, Luis A. and Lapusta, Nadia and Liu, Yi},
  year = {2005},
  journal = {Journal of Geophysical Research: Solid Earth},
  volume = {110},
  number = {B12},
  issn = {2156-2202},
  doi = {10.1029/2005JB003813}
}

@article{Dettmer2016Tsunami,
  title = {Tsunami Source Uncertainty Estimation: {{The}} 2011 {{Japan}} Tsunami},
  author = {Dettmer, Jan and Hawkins, Rhys and Cummins, Phil R. and Hossen, Jakir and Sambridge, Malcolm and Hino, Ryota and Inazu, Daisuke},
  year = {2016},
  month = jun,
  journal = {Journal of Geophysical Research: Solid Earth},
  volume = {121},
  number = {6},
  pages = {4483--4505},
  issn = {2169-9313, 2169-9356},
  doi = {10.1002/2015JB012764}
}

@article{Dieterich1979Modeling,
  title = {Modeling of Rock Friction: 1. {{Experimental}} Results and Constitutive Equations},
  author = {Dieterich, James H.},
  year = {1979},
  month = may,
  journal = {Journal of Geophysical Research: Solid Earth},
  volume = {84},
  number = {B5},
  pages = {2161--2168},
  issn = {0148-0227},
  doi = {10.1029/JB084iB05p02161}
}

@article{Dieterich1992Earthquake,
  title = {Earthquake Nucleation on Faults with Rate-and State-Dependent Strength},
  author = {Dieterich, James H.},
  year = {1992},
  month = sep,
  journal = {Tectonophysics},
  volume = {211},
  number = {1},
  pages = {115--134},
  issn = {0040-1951},
  doi = {10.1016/0040-1951(92)90055-B}
}

@article{Dieterich1994Direct,
  title = {Direct Observation of Frictional Contacts: {{New}} Insights for State-Dependent Properties},
  author = {Dieterich, James H. and Kilgore, Brian D.},
  year = {1994},
  journal = {Pure and Applied Geophysics PAGEOPH},
  volume = {143},
  number = {1-3},
  pages = {283--302},
  issn = {0033-4553, 1420-9136},
  doi = {10.1007/BF00874332}
}

@article{DiToro2011Fault,
  title = {Fault Lubrication during Earthquakes},
  author = {Di Toro, G. and Han, R. and Hirose, T. and De Paola, N. and Nielsen, S. and Mizoguchi, K. and Ferri, F. and Cocco, M. and Shimamoto, T.},
  year = {2011},
  month = mar,
  journal = {Nature},
  volume = {471},
  number = {7339},
  pages = {494--498},
  publisher = {Nature Publishing Group},
  issn = {1476-4687},
  doi = {10.1038/nature09838}
}

@article{Duan2012Dynamic,
  title = {Dynamic Rupture of the 2011 {{Mw}} 9.0 {{Tohoku-Oki}} Earthquake: {{Roles}} of a Possible Subducting Seamount},
  author = {Duan, Benchun},
  year = {2012},
  journal = {Journal of Geophysical Research: Solid Earth},
  volume = {117},
  number = {B5},
  issn = {2156-2202},
  doi = {10.1029/2011JB009124}
}

@article{Dumbser2006Arbitrary,
  title = {An Arbitrary High-Order Discontinuous {{Galerkin}} Method for Elastic Waves on Unstructured Meshes --- {{II}}. {{The}} Three-Dimensional Isotropic Case},
  author = {Dumbser, Michael and K{\"a}ser, Martin},
  year = {2006},
  month = oct,
  journal = {Geophysical Journal International},
  volume = {167},
  number = {1},
  pages = {319--336},
  issn = {0956-540X},
  doi = {10.1111/j.1365-246X.2006.03120.x}
}

@article{Dunham2011Earthquake,
  title = {Earthquake {{Ruptures}} with {{Strongly Rate-Weakening Friction}} and {{Off-Fault Plasticity}}, {{Part}} 1: {{Planar Faults}}},
  author = {Dunham, E. M. and Belanger, D. and Cong, L. and Kozdon, J. E.},
  year = {2011},
  month = oct,
  journal = {Bulletin of the Seismological Society of America},
  volume = {101},
  number = {5},
  pages = {2296--2307},
  issn = {0037-1106},
  doi = {10.1785/0120100075}
}

@article{Fujii2011Tsunami,
  title = {Tsunami Source of the 2011 off the {{Pacific}} Coast of {{Tohoku Earthquake}}},
  author = {Fujii, Yushiro and Satake, Kenji and Sakai, Shin'ichi and Shinohara, Masanao and Kanazawa, Toshihiko},
  year = {2011},
  month = jul,
  journal = {Earth, Planets and Space},
  volume = {63},
  number = {7},
  pages = {815--820},
  issn = {1343-8832, 1880-5981},
  doi = {10.5047/eps.2011.06.010}
}

@article{Fujiwara20112011,
  title = {The 2011 {{Tohoku-Oki Earthquake}}: {{Displacement Reaching}} the {{Trench Axis}}},
  author = {Fujiwara, Toshiya and Kodaira, Shuichi and No, Tetsuo and Kaiho, Yuka and Takahashi, Narumi and Kaneda, Yoshiyuki},
  year = {2011},
  month = dec,
  journal = {Science},
  volume = {334},
  number = {6060},
  pages = {1240--1240},
  issn = {0036-8075, 1095-9203},
  doi = {10.1126/science.1211554}
}

@article{Fulton2013Low,
  title = {Low {{Coseismic Friction}} on the {{Tohoku-Oki Fault Determined}} from {{Temperature Measurements}}},
  author = {Fulton, P. M. and Brodsky, E. E. and Kano, Y. and Mori, J. and Chester, F. and Ishikawa, T. and Harris, R. N. and Lin, W. and Eguchi, N. and Toczko, S. and {Expedition 343, 343T, and KR13-08 Scientists}},
  year = {2013},
  month = dec,
  journal = {Science},
  volume = {342},
  number = {6163},
  pages = {1214--1217},
  issn = {0036-8075, 1095-9203},
  doi = {10.1126/science.1243641}
}

@article{Gabriel2012Transition,
  title = {The Transition of Dynamic Rupture Styles in Elastic Media under Velocity-Weakening Friction},
  author = {Gabriel, A.-A. and Ampuero, J.-P. and Dalguer, L. A. and Mai, P. M.},
  year = {2012},
  journal = {Journal of Geophysical Research: Solid Earth},
  volume = {117},
  number = {B9},
  issn = {2156-2202},
  doi = {10.1029/2012JB009468}
}

@article{Gabriel2013Source,
  title = {Source Properties of Dynamic Rupture Pulses with Off-fault Plasticity},
  author = {Gabriel, A.-A. and Ampuero, J.-P. and Dalguer, L. A. and Mai, P. M.},
  year = {2013},
  month = aug,
  journal = {Journal of Geophysical Research: Solid Earth},
  volume = {118},
  number = {8},
  pages = {4117--4126},
  issn = {2169-9313, 2169-9356},
  doi = {10.1002/jgrb.50213}
}

@article{Gabriel2024Fault,
  title = {Fault Size--Dependent Fracture Energy Explains Multiscale Seismicity and Cascading Earthquakes},
  author = {Gabriel, Alice-Agnes and Garagash, Dmitry I. and Palgunadi, Kadek H. and Mai, P. Martin},
  year = {2024},
  month = jul,
  journal = {Science},
  volume = {385},
  number = {6707},
  pages = {eadj9587},
  publisher = {American Association for the Advancement of Science},
  doi = {10.1126/science.adj9587}
}

@article{Gallovic2019Bayesian,
  title = {Bayesian {{Dynamic Finite-Fault Inversion}}: 1. {{Method}} and {{Synthetic Test}}},
  author = {Gallovi{\v c}, F. and Valentov{\'a}, {\v L}. and Ampuero, J.-P. and Gabriel, A.-A.},
  year = {2019},
  journal = {Journal of Geophysical Research: Solid Earth},
  volume = {124},
  number = {7},
  pages = {6949--6969},
  issn = {2169-9356},
  doi = {10.1029/2019JB017510}
}

@article{Galvez2014Dynamic,
  title = {Dynamic Earthquake Rupture Modelled with an Unstructured 3-{{D}} Spectral Element Method Applied to the 2011 {{M9 Tohoku}} Earthquake},
  author = {Galvez, P. and Ampuero, J.-P. and Dalguer, L. A. and Somala, S. N. and {Nissen-Meyer}, T.},
  year = {2014},
  month = aug,
  journal = {Geophysical Journal International},
  volume = {198},
  number = {2},
  pages = {1222--1240},
  issn = {1365-246X, 0956-540X},
  doi = {10.1093/gji/ggu203}
}

@article{Galvez2016Rupture,
  title = {Rupture {{Reactivation}} during the 2011 {{{\emph{M}}}} {\textsubscript{w}} 9.0 {{Tohoku Earthquake}}: {{Dynamic Rupture}} and {{Ground}}-{{Motion Simulations}}},
  author = {Galvez, Percy and Dalguer, Luis A. and Ampuero, Jean-Paul and Giardini, Domenico},
  year = {2016},
  month = jun,
  journal = {Bulletin of the Seismological Society of America},
  volume = {106},
  number = {3},
  pages = {819--831},
  issn = {0037-1106, 1943-3573},
  doi = {10.1785/0120150153}
}

@article{Galvez2020Dynamic,
  title = {Dynamic {{Source Model}} for the 2011 {{Tohoku Earthquake}} in a {{Wide Period Range Combining Slip Reactivation}} with the {{Short-Period Ground Motion Generation Process}}},
  author = {Galvez, Percy and Petukhin, Anatoly and Irikura, Kojiro and Somerville, Paul},
  year = {2020},
  month = may,
  journal = {Pure and Applied Geophysics},
  volume = {177},
  number = {5},
  pages = {2143--2161},
  issn = {0033-4553, 1420-9136},
  doi = {10.1007/s00024-019-02210-7}
}

@article{Glehman2024Partial,
  title = {Partial Ruptures Governed by the Complex Interplay between Geodetic Slip Deficit, Rigidity, and Pore Fluid Pressure in {{3D Cascadia}} Dynamic Rupture Simulations},
  author = {Glehman, Jonatan and Gabriel, Alice-Agnes and Ulrich, Thomas and Ramos, Marlon Dale and Huang, Yihe and Lindsey, Eric O.},
  year = {2024},
  month = aug,
  publisher = {EarthArXiv}
}

@article{Goldsby2011Flash,
  title = {Flash {{Heating Leads}} to {{Low Frictional Strength}} of {{Crustal Rocks}} at {{Earthquake Slip Rates}}},
  author = {Goldsby, David L. and Tullis, Terry E.},
  year = {2011},
  month = oct,
  journal = {Science},
  volume = {334},
  number = {6053},
  pages = {216--218},
  issn = {0036-8075, 1095-9203},
  doi = {10.1126/science.1207902}
}

@article{Hardebeck2018Creeping,
  title = {Creeping Subduction Zones Are Weaker than Locked Subduction Zones},
  author = {Hardebeck, Jeanne L. and Loveless, John P.},
  year = {2018},
  month = jan,
  journal = {Nature Geoscience},
  volume = {11},
  number = {1},
  pages = {60--64},
  issn = {1752-0894, 1752-0908},
  doi = {10.1038/s41561-017-0032-1}
}

@article{Hardebeck2018Temporal,
  title = {Temporal {{Stress Changes Caused}} by {{Earthquakes}}: {{A Review}}},
  author = {Hardebeck, Jeanne L. and Okada, Tomomi},
  year = {2018},
  month = feb,
  journal = {Journal of Geophysical Research: Solid Earth},
  volume = {123},
  number = {2},
  pages = {1350--1365},
  issn = {2169-9313, 2169-9356},
  doi = {10.1002/2017JB014617}
}

@article{Harris2018Suite,
  title = {A {{Suite}} of {{Exercises}} for {{Verifying Dynamic Earthquake Rupture Codes}}},
  author = {Harris, Ruth A. and Barall, Michael and Aagaard, Brad and Ma, Shuo and Roten, Daniel and Olsen, Kim and Duan, Benchun and Liu, Dunyu and Luo, Bin and Bai, Kangchen and Ampuero, Jean-Paul and Kaneko, Yoshihiro and Gabriel, Alice-Agnes and Duru, Kenneth and Ulrich, Thomas and Wollherr, Stephanie and Shi, Zheqiang and Dunham, Eric and Bydlon, Sam and Zhang, Zhenguo and Chen, Xiaofei and Somala, Surendra Nadh and Pelties, Christian and Tago, Josu{\'e} and Cruz-Atienza, Victor Manuel and Kozdon, Jeremy and Daub, Eric and Aslam, Khurram and Kase, Yuko and Withers, Kyle and Dalguer, Luis},
  year = {2018},
  month = may,
  journal = {Seismological Research Letters},
  volume = {89},
  number = {3},
  pages = {1146--1162},
  issn = {0895-0695, 1938-2057},
  doi = {10.1785/0220170222}
}

@article{Hayek2024NonTypical,
  title = {Non-{{Typical Supershear Rupture}}: {{Fault Heterogeneity}} and {{Segmentation Govern Unilateral Supershear}} and {{Cascading Multi}}-{{Fault Rupture}} in the 2021 {{Mw}} \$\{\vphantom\}{{M}}\vphantom\{\}\_\{w\}\$7.4 {{Maduo Earthquake}}},
  author = {Hayek, J. N. and Marchandon, M. and Li, D. and Pousse-Beltran, L. and Hollingsworth, J. and Li, T. and Gabriel, A.-A.},
  year = {2024},
  month = oct,
  journal = {Geophysical Research Letters},
  volume = {51},
  number = {20},
  pages = {e2024GL110128},
  issn = {0094-8276, 1944-8007},
  doi = {10.1029/2024GL110128}
}

@article{Hayes2011Rapid,
  title = {Rapid Source Characterization of the 2011 {{M}} w 9.0 off the {{Pacific}} Coast of {{Tohoku Earthquake}}},
  author = {Hayes, Gavin P.},
  year = {2011},
  month = jul,
  journal = {Earth, Planets and Space},
  volume = {63},
  number = {7},
  pages = {529--534},
  issn = {1343-8832, 1880-5981},
  doi = {10.5047/eps.2011.05.012}
}

@article{Heaton1990Evidence,
  title = {Evidence for and Implications of Self-Healing Pulses of Slip in Earthquake Rupture},
  author = {Heaton, Thomas H.},
  year = {1990},
  month = nov,
  journal = {Physics of the Earth and Planetary Interiors},
  volume = {64},
  number = {1},
  pages = {1--20},
  issn = {00319201},
  doi = {10.1016/0031-9201(90)90002-F}
}

@article{Hok2011Dynamic,
  title = {Dynamic Rupture Scenarios of Anticipated {{Nankai-Tonankai}} Earthquakes, Southwest {{Japan}}},
  author = {Hok, S{\'e}bastien and Fukuyama, Eiichi and Hashimoto, Chihiro},
  year = {2011},
  month = dec,
  journal = {Journal of Geophysical Research},
  volume = {116},
  number = {B12},
  pages = {B12319},
  issn = {0148-0227},
  doi = {10.1029/2011JB008492}
}

@article{Hossen2015Tsunami,
  title = {Tsunami Waveform Inversion for Sea Surface Displacement Following the 2011 {{Tohoku}} Earthquake: {{Importance}} of Dispersion and Source Kinematics},
  author = {Hossen, M. Jakir and Cummins, Phil R. and Dettmer, Jan and Baba, Toshitaka},
  year = {2015},
  month = sep,
  journal = {Journal of Geophysical Research: Solid Earth},
  volume = {120},
  number = {9},
  pages = {6452--6473},
  issn = {2169-9313, 2169-9356},
  doi = {10.1002/2015JB011942}
}

@article{Huang2012Dynamic,
  title = {A Dynamic Model of the Frequency-Dependent Rupture Process of the 2011 {{Tohoku-Oki}} Earthquake},
  author = {Huang, Yihe and Meng, Lingsen and Ampuero, Jean-Paul},
  year = {2012},
  month = dec,
  journal = {Earth, Planets and Space},
  volume = {64},
  number = {12},
  pages = {1061--1066},
  issn = {1343-8832, 1880-5981},
  doi = {10.5047/eps.2012.05.011}
}

@article{Huang2014SlipWeakening,
  title = {Slip-{{Weakening Models}} of the 2011 {{Tohoku-Oki Earthquake}} and {{Constraints}} on {{Stress Drop}} and {{Fracture Energy}}},
  author = {Huang, Yihe and Ampuero, Jean-Paul and Kanamori, Hiroo},
  year = {2014},
  month = oct,
  journal = {Pure and Applied Geophysics},
  volume = {171},
  number = {10},
  pages = {2555--2568},
  issn = {0033-4553, 1420-9136},
  doi = {10.1007/s00024-013-0718-2}
}

@article{Ide2002Estimation,
  title = {Estimation of {{Radiated Energy}} of {{Finite-Source Earthquake Models}}},
  author = {Ide, S.},
  year = {2002},
  month = dec,
  journal = {Bulletin of the Seismological Society of America},
  volume = {92},
  number = {8},
  pages = {2994--3005},
  issn = {0037-1106},
  doi = {10.1785/0120020028}
}

@article{Ide2005Earthquakes,
  title = {Earthquakes as Multiscale Dynamic Ruptures with Heterogeneous Fracture Surface Energy},
  author = {Ide, Satoshi and Aochi, Hideo},
  year = {2005},
  journal = {Journal of Geophysical Research: Solid Earth},
  volume = {110},
  number = {B11},
  issn = {2156-2202},
  doi = {10.1029/2004JB003591}
}

@article{Ide2011Shallow,
  title = {Shallow {{Dynamic Overshoot}} and {{Energetic Deep Rupture}} in the 2011 {{{\emph{M}}}} {\textsubscript{w}} 9.0 {{Tohoku-Oki Earthquake}}},
  author = {Ide, Satoshi and Baltay, Annemarie and Beroza, Gregory C.},
  year = {2011},
  month = jun,
  journal = {Science},
  volume = {332},
  number = {6036},
  pages = {1426--1429},
  issn = {0036-8075, 1095-9203},
  doi = {10.1126/science.1207020}
}

@article{Ide2013Historical,
  title = {Historical Seismicity and Dynamic Rupture Process of the 2011 {{Tohoku-Oki}} Earthquake},
  author = {Ide, Satoshi and Aochi, Hideo},
  year = {2013},
  month = jul,
  journal = {Tectonophysics},
  series = {Great {{Earthquakes}} along {{Subduction Zones}}},
  volume = {600},
  pages = {1--13},
  issn = {0040-1951},
  doi = {10.1016/j.tecto.2012.10.018}
}

@article{Ikari2011Cohesive,
  title = {Cohesive Strength of Clay-Rich Sediment},
  author = {Ikari, Matt J. and Kopf, Achim J.},
  year = {2011},
  journal = {Geophysical Research Letters},
  volume = {38},
  number = {16},
  issn = {1944-8007},
  doi = {10.1029/2011GL047918}
}

@article{Ikari2015Strength,
  title = {Strength Characteristics of {{Japan Trench}} Borehole Samples in the High-Slip Region of the 2011 {{Tohoku-Oki}} Earthquake},
  author = {Ikari, Matt J. and Kameda, Jun and Saffer, Demian M. and Kopf, Achim J.},
  year = {2015},
  month = feb,
  journal = {Earth and Planetary Science Letters},
  volume = {412},
  pages = {35--41},
  issn = {0012821X},
  doi = {10.1016/j.epsl.2014.12.014}
}

@article{Ito2011Frontal,
  title = {Frontal Wedge Deformation near the Source Region of the 2011 {{Tohoku-Oki}} Earthquake: {{FRONTAL WEDGE DEFORMATION OF JPN TRENCH}}},
  author = {Ito, Yoshihiro and Tsuji, Takeshi and Osada, Yukihito and Kido, Motoyuki and Inazu, Daisuke and Hayashi, Yutaka and Tsushima, Hiroaki and Hino, Ryota and Fujimoto, Hiromi},
  year = {2011},
  month = apr,
  journal = {Geophysical Research Letters},
  volume = {38},
  number = {7},
  pages = {n/a-n/a},
  issn = {00948276},
  doi = {10.1029/2011GL048355}
}

@article{JamaliHondori2022Connection,
  title = {Connection between High Pore-Fluid Pressure and Frictional Instability at Tsunamigenic Plate Boundary Fault of 2011 {{Tohoku-Oki}} Earthquake},
  author = {Jamali Hondori, Ehsan and Park, Jin-Oh},
  year = {2022},
  month = aug,
  journal = {Scientific Reports},
  volume = {12},
  number = {1},
  pages = {12556},
  publisher = {Nature Publishing Group},
  issn = {2045-2322},
  doi = {10.1038/s41598-022-16578-5}
}

@article{Jia2023Complex,
  title = {The Complex Dynamics of the 2023 {{Kahramanmara{\c s}}}, {{Turkey}}, {{{\emph{M}}}}{\textsubscript{w}} 7.8-7.7 Earthquake Doublet},
  author = {Jia, Zhe and Jin, Zeyu and Marchandon, Mathilde and Ulrich, Thomas and Gabriel, Alice-Agnes and Fan, Wenyuan and Shearer, Peter and Zou, Xiaoyu and Rekoske, John and Bulut, Fatih and Garagon, Asl{\i} and Fialko, Yuri},
  year = {2023},
  month = sep,
  journal = {Science},
  volume = {381},
  number = {6661},
  pages = {985--990},
  issn = {0036-8075, 1095-9203},
  doi = {10.1126/science.adi0685}
}

@article{Kammer2024Earthquake,
  title = {Earthquake Energy Dissipation in a Fracture Mechanics Framework},
  author = {Kammer, David S. and McLaskey, Gregory C. and Abercrombie, Rachel E. and Ampuero, Jean-Paul and Cattania, Camilla and Cocco, Massimo and Dal Zilio, Luca and Dresen, Georg and Gabriel, Alice-Agnes and Ke, Chun-Yu and Marone, Chris and Selvadurai, Paul Antony and Tinti, Elisa},
  year = {2024},
  month = jun,
  journal = {Nature Communications},
  volume = {15},
  number = {1},
  pages = {4736},
  issn = {2041-1723},
  doi = {10.1038/s41467-024-47970-6}
}

@article{Kaser2006Arbitrary,
  title = {An Arbitrary High-Order Discontinuous {{Galerkin}} Method for Elastic Waves on Unstructured Meshes --- {{I}}. {{The}} Two-Dimensional Isotropic Case with External Source Terms},
  author = {K{\"a}ser, Martin and Dumbser, Michael},
  year = {2006},
  month = aug,
  journal = {Geophysical Journal International},
  volume = {166},
  number = {2},
  pages = {855--877},
  issn = {0956-540X},
  doi = {10.1111/j.1365-246X.2006.03051.x}
}

@article{Ke2018Rupture,
  title = {Rupture {{Termination}} in {{Laboratory}}-{{Generated Earthquakes}}},
  author = {Ke, Chun-Yu and McLaskey, Gregory C. and Kammer, David S.},
  year = {2018},
  month = dec,
  journal = {Geophysical Research Letters},
  volume = {45},
  number = {23},
  issn = {0094-8276, 1944-8007},
  doi = {10.1029/2018GL080492}
}

@article{Kodaira2020Large,
  title = {Large {{Coseismic Slip}} to the {{Trench During}} the 2011 {{Tohoku-Oki Earthquake}}},
  author = {Kodaira, Shuichi and Fujiwara, Toshiya and Fujie, Gou and Nakamura, Yasuyuki and Kanamatsu, Toshiya},
  year = {2020},
  month = may,
  journal = {Annual Review of Earth and Planetary Sciences},
  volume = {48},
  number = {1},
  pages = {321--343},
  issn = {0084-6597, 1545-4495},
  doi = {10.1146/annurev-earth-071719-055216}
}

@article{Koper2011Alongdip,
  title = {Along-Dip Variation of Teleseismic Short-Period Radiation from the 11 {{March}} 2011 {{Tohoku}} Earthquake ({{Mw}} 9.0)},
  author = {Koper, K. D. and Hutko, A. R. and Lay, T.},
  year = {2011},
  journal = {Geophysical Research Letters},
  volume = {38},
  number = {21},
  issn = {1944-8007},
  doi = {10.1029/2011GL049689}
}

@article{Kozdon2013Rupture,
  title = {Rupture to the {{Trench}}: {{Dynamic Rupture Simulations}} of the 11 {{March}} 2011 {{Tohoku Earthquake}}},
  author = {Kozdon, J. E. and Dunham, E. M.},
  year = {2013},
  month = may,
  journal = {Bulletin of the Seismological Society of America},
  volume = {103},
  number = {2B},
  pages = {1275--1289},
  issn = {0037-1106},
  doi = {10.1785/0120120136}
}

@inproceedings{Krenz20213D,
  title = {{{3D}} Acoustic-Elastic Coupling with Gravity: The Dynamics of the 2018 {{Palu}}, {{Sulawesi}} Earthquake and Tsunami},
  booktitle = {Proceedings of the {{International Conference}} for {{High Performance Computing}}, {{Networking}}, {{Storage}} and {{Analysis}}},
  author = {Krenz, Lukas and Uphoff, Carsten and Ulrich, Thomas and Gabriel, Alice-Agnes and Abrahams, Lauren S. and Dunham, Eric M. and Bader, Michael},
  year = {2021},
  month = nov,
  pages = {1--14},
  publisher = {ACM},
  address = {St. Louis Missouri},
  doi = {10.1145/3458817.3476173},
  isbn = {978-1-4503-8442-1}
}

@article{Kubota2022New,
  title = {A New Mechanical Perspective on a Shallow Megathrust Near-Trench Slip from the High-Resolution Fault Model of the 2011 {{Tohoku-Oki}} Earthquake},
  author = {Kubota, Tatsuya and Saito, Tatsuhiko and Hino, Ryota},
  year = {2022},
  month = dec,
  journal = {Progress in Earth and Planetary Science},
  volume = {9},
  number = {1},
  pages = {68},
  issn = {2197-4284},
  doi = {10.1186/s40645-022-00524-0}
}

@article{Kurahashi2013ShortPeriod,
  title = {Short-{{Period Source Model}} of the 2011 {{Mw}}~9.0 {{Off}} the {{Pacific Coast}} of {{Tohoku Earthquake}}},
  author = {Kurahashi, Susumu and Irikura, Kojiro},
  year = {2013},
  month = may,
  journal = {Bulletin of the Seismological Society of America},
  volume = {103},
  number = {2B},
  pages = {1373--1393},
  issn = {0037-1106},
  doi = {10.1785/0120120157}
}

@article{Lambert2021Propagation,
  title = {Propagation of Large Earthquakes as Self-Healing Pulses or Mild Cracks},
  author = {Lambert, Val{\`e}re and Lapusta, Nadia and Perry, Stephen},
  year = {2021},
  month = mar,
  journal = {Nature},
  volume = {591},
  number = {7849},
  pages = {252--258},
  issn = {0028-0836, 1476-4687},
  doi = {10.1038/s41586-021-03248-1}
}

@article{Lambert2023Absolute,
  title = {Absolute Stress Levels in Models of Low-Heat Faults: {{Links}} to Geophysical Observables and Differences for Crack-like Ruptures and Self-Healing Pulses},
  author = {Lambert, Val{\`e}re and Lapusta, Nadia},
  year = {2023},
  month = sep,
  journal = {Earth and Planetary Science Letters},
  volume = {618},
  pages = {118277},
  issn = {0012821X},
  doi = {10.1016/j.epsl.2023.118277}
}

@article{Lapusta2003Nucleation,
  title = {Nucleation and Early Seismic Propagation of Small and Large Events in a Crustal Earthquake Model},
  author = {Lapusta, Nadia and Rice, James R.},
  year = {2003},
  month = apr,
  journal = {Journal of Geophysical Research: Solid Earth},
  volume = {108},
  number = {B4},
  pages = {2001JB000793},
  issn = {0148-0227},
  doi = {10.1029/2001JB000793}
}

@article{Lay2011Possible,
  title = {Possible Large Near-Trench Slip during the 2011 {{M}} w 9.0 off the {{Pacific}} Coast of {{Tohoku Earthquake}}},
  author = {Lay, Thorne and Ammon, Charles J. and Kanamori, Hiroo and Xue, Lian and Kim, Marina J.},
  year = {2011},
  month = jul,
  journal = {Earth, Planets and Space},
  volume = {63},
  number = {7},
  pages = {687--692},
  issn = {1343-8832, 1880-5981},
  doi = {10.5047/eps.2011.05.033}
}

@article{Lay2012Depthvarying,
  title = {Depth-Varying Rupture Properties of Subduction Zone Megathrust Faults},
  author = {Lay, Thorne and Kanamori, Hiroo and Ammon, Charles J. and Koper, Keith D. and Hutko, Alexander R. and Ye, Lingling and Yue, Han and Rushing, Teresa M.},
  year = {2012},
  journal = {Journal of Geophysical Research: Solid Earth},
  volume = {117},
  number = {B4},
  issn = {2156-2202},
  doi = {10.1029/2011JB009133}
}

@article{Lay2018Review,
  title = {A Review of the Rupture Characteristics of the 2011 {{Tohoku-oki Mw}} 9.1 Earthquake},
  author = {Lay, Thorne},
  year = {2018},
  month = may,
  journal = {Tectonophysics},
  volume = {733},
  pages = {4--36},
  issn = {00401951},
  doi = {10.1016/j.tecto.2017.09.022}
}

@article{Lee2011Evidence,
  title = {Evidence of Large Scale Repeating Slip during the 2011 {{Tohoku-Oki}} Earthquake: {{REPEATING SLIP DURING TOHOKU EARTHQUAKE}}},
  author = {Lee, Shiann-Jong and Huang, Bor-Shouh and Ando, Masataka and Chiu, Hung-Chie and Wang, Jeen-Hwa},
  year = {2011},
  month = oct,
  journal = {Geophysical Research Letters},
  volume = {38},
  number = {19},
  pages = {n/a-n/a},
  issn = {00948276},
  doi = {10.1029/2011GL049580}
}

@article{Li2024Linking,
  title = {Linking {{3D Long-Term Slow-Slip Cycle Models With Rupture Dynamics}}: {{The Nucleation}} of the 2014 {{Mw}} 7.3 {{Guerrero}}, {{Mexico Earthquake}}},
  author = {Li, Duo and Gabriel, Alice-Agnes},
  year = {2024},
  journal = {AGU Advances},
  volume = {5},
  number = {2},
  pages = {e2023AV000979},
  issn = {2576-604X},
  doi = {10.1029/2023AV000979}
}

@article{Loveless2016Two,
  title = {Two Decades of Spatiotemporal Variations in Subduction Zone Coupling Offshore {{Japan}}},
  author = {Loveless, John P. and Meade, Brendan J.},
  year = {2016},
  month = feb,
  journal = {Earth and Planetary Science Letters},
  volume = {436},
  pages = {19--30},
  issn = {0012821X},
  doi = {10.1016/j.epsl.2015.12.033}
}

@article{Ma2008Physical,
  title = {A Physical Model for Widespread Near-Surface and Fault Zone Damage Induced by Earthquakes},
  author = {Ma, Shuo},
  year = {2008},
  journal = {Geochemistry, Geophysics, Geosystems},
  volume = {9},
  number = {11},
  issn = {1525-2027},
  doi = {10.1029/2008GC002231}
}

@article{Ma2019Dynamic,
  title = {Dynamic {{Wedge Failure}} and {{Along}}-{{Arc Variations}} of {{Tsunamigenesis}} in the {{Japan Trench Margin}}},
  author = {Ma, Shuo and Nie, Shiying},
  year = {2019},
  month = aug,
  journal = {Geophysical Research Letters},
  volume = {46},
  number = {15},
  pages = {8782--8790},
  issn = {0094-8276, 1944-8007},
  doi = {10.1029/2019GL083148}
}

@article{Ma2023Wedge,
  title = {Wedge Plasticity and a Minimalist Dynamic Rupture Model for the 2011 {{MW}} 9.1 {{Tohoku-Oki}} Earthquake and Tsunami},
  author = {Ma, Shuo},
  year = {2023},
  month = dec,
  journal = {Tectonophysics},
  volume = {869},
  pages = {230146},
  issn = {00401951},
  doi = {10.1016/j.tecto.2023.230146}
}

@article{Madden2022State,
  title = {The {{State}} of {{Pore Fluid Pressure}} and 3-{{D Megathrust Earthquake Dynamics}}},
  author = {Madden, Elizabeth H. and Ulrich, Thomas and Gabriel, Alice-Agnes},
  year = {2022},
  month = apr,
  journal = {Journal of Geophysical Research: Solid Earth},
  volume = {127},
  number = {4},
  pages = {e2021JB023382},
  issn = {2169-9313, 2169-9356},
  doi = {10.1029/2021JB023382}
}

@article{Melgar2015Kinematic,
  title = {Kinematic Earthquake Source Inversion and Tsunami Runup Prediction with Regional Geophysical Data},
  author = {Melgar, D. and Bock, Y.},
  year = {2015},
  month = may,
  journal = {Journal of Geophysical Research: Solid Earth},
  volume = {120},
  number = {5},
  pages = {3324--3349},
  issn = {2169-9313, 2169-9356},
  doi = {10.1002/2014JB011832}
}

@article{Meng2011Window,
  title = {A Window into the Complexity of the Dynamic Rupture of the 2011 {{Mw}} 9 {{Tohoku-Oki}} Earthquake: {{THE}} 2011 {{TOHOKU-OKI EARTHQUAKE}}},
  author = {Meng, Lingsen and Inbal, Asaf and Ampuero, Jean-Paul},
  year = {2011},
  month = apr,
  journal = {Geophysical Research Letters},
  volume = {38},
  number = {7},
  pages = {n/a-n/a},
  issn = {00948276},
  doi = {10.1029/2011GL048118}
}

@article{Moore2015Sediment,
  title = {Sediment Provenance and Controls on Slip Propagation: {{Lessons}} Learned from the 2011 {{Tohoku}} and Other Great Earthquakes of the Subducting Northwest {{Pacific}} Plate},
  author = {Moore, J. C. and Plank, T. A. and Chester, F. M. and Polissar, P. J. and Savage, H. M.},
  year = {2015},
  month = jun,
  journal = {Geosphere},
  volume = {11},
  number = {3},
  pages = {533--541},
  issn = {1553-040X},
  doi = {10.1130/GES01099.1}
}

@article{Nielsen2000Rupture,
  title = {Rupture {{Pulse Characterization}}: {{Self-Healing}}, {{Self-Similar}}, {{Expanding Solutions}} in a {{Continuum Model}} of {{Fault Dynamics}}},
  author = {Nielsen, S. B. and Carlson, J. M.},
  year = {2000},
  month = dec,
  journal = {Bulletin of the Seismological Society of America},
  volume = {90},
  number = {6},
  pages = {1480--1497},
  issn = {0037-1106},
  doi = {10.1785/0120000021}
}

@article{Nielsen2003SelfHealing,
  title = {On the {{Self-Healing Fracture Mode}}},
  author = {Nielsen, S. and Madariaga, R.},
  year = {2003},
  month = dec,
  journal = {Bulletin of the Seismological Society of America},
  volume = {93},
  number = {6},
  pages = {2375--2388},
  issn = {0037-1106},
  doi = {10.1785/0120020090}
}

@article{Nishikawa2019Slow,
  title = {The Slow Earthquake Spectrum in the {{Japan Trench}} Illuminated by the {{S-net}} Seafloor Observatories},
  author = {Nishikawa, T. and Matsuzawa, T. and Ohta, K. and Uchida, N. and Nishimura, T. and Ide, S.},
  year = {2019},
  month = aug,
  journal = {Science},
  volume = {365},
  number = {6455},
  pages = {808--813},
  issn = {0036-8075, 1095-9203},
  doi = {10.1126/science.aax5618}
}

@article{Nishikawa2023Review,
  title = {A Review on Slow Earthquakes in the {{Japan Trench}}},
  author = {Nishikawa, Tomoaki and Ide, Satoshi and Nishimura, Takuya},
  year = {2023},
  month = jan,
  journal = {Progress in Earth and Planetary Science},
  volume = {10},
  number = {1},
  pages = {1},
  issn = {2197-4284},
  doi = {10.1186/s40645-022-00528-w}
}

@article{Noda2009Earthquakea,
  title = {Earthquake Ruptures with Thermal Weakening and the Operation of Major Faults at Low Overall Stress Levels},
  author = {Noda, Hiroyuki and Dunham, Eric M. and Rice, James R.},
  year = {2009},
  month = jul,
  journal = {Journal of Geophysical Research: Solid Earth},
  volume = {114},
  number = {B7},
  pages = {2008JB006143},
  issn = {0148-0227},
  doi = {10.1029/2008JB006143}
}

@article{Noda2013Stable,
  title = {Stable Creeping Fault Segments Can Become Destructive as a Result of Dynamic Weakening},
  author = {Noda, Hiroyuki and Lapusta, Nadia},
  year = {2013},
  month = jan,
  journal = {Nature},
  volume = {493},
  number = {7433},
  pages = {518--521},
  issn = {0028-0836, 1476-4687},
  doi = {10.1038/nature11703}
}

@article{Okuwaki2014Relationship,
  title = {Relationship between {{High-frequency Radiation}} and {{Asperity Ruptures}}, {{Revealed}} by {{Hybrid Back-projection}} with a {{Non-planar Fault Model}}},
  author = {Okuwaki, Ryo and Yagi, Yuji and Hirano, Shiro},
  year = {2014},
  month = nov,
  journal = {Scientific Reports},
  volume = {4},
  number = {1},
  pages = {7120},
  issn = {2045-2322},
  doi = {10.1038/srep07120}
}

@article{Palgunadi2024Rupture,
  title = {Rupture {{Dynamics}} of {{Cascading Earthquakes}} in a {{Multiscale Fracture Network}}},
  author = {Palgunadi, Kadek Hendrawan and Gabriel, Alice-Agnes and Garagash, Dmitry Igor and Ulrich, Thomas and Mai, Paul Martin},
  year = {2024},
  journal = {Journal of Geophysical Research: Solid Earth},
  volume = {129},
  number = {3},
  pages = {e2023JB027578},
  issn = {2169-9356},
  doi = {10.1029/2023JB027578}
}

@article{Pelties2012Threedimensional,
  title = {Three-Dimensional Dynamic Rupture Simulation with a High-Order Discontinuous {{Galerkin}} Method on Unstructured Tetrahedral Meshes},
  author = {Pelties, Christian and {de la Puente}, Josep and Ampuero, Jean-Paul and Brietzke, Gilbert B. and K{\"a}ser, Martin},
  year = {2012},
  journal = {Journal of Geophysical Research: Solid Earth},
  volume = {117},
  number = {B2},
  issn = {2156-2202},
  doi = {10.1029/2011JB008857}
}

@article{Pelties2014Verification,
  title = {Verification of an {{ADER-DG}} Method for Complex Dynamic Rupture Problems},
  author = {Pelties, C. and Gabriel, A.-A. and Ampuero, J.-P.},
  year = {2014},
  month = may,
  journal = {Geoscientific Model Development},
  volume = {7},
  number = {3},
  pages = {847--866},
  publisher = {Copernicus GmbH},
  issn = {1991-959X},
  doi = {10.5194/gmd-7-847-2014}
}

@article{Perrin1995Selfhealing,
  title = {Self-Healing Slip Pulse on a Frictional Surface},
  author = {Perrin, Gilles and Rice, James R. and Zheng, Gutuan},
  year = {1995},
  month = sep,
  journal = {Journal of the Mechanics and Physics of Solids},
  volume = {43},
  number = {9},
  pages = {1461--1495},
  issn = {0022-5096},
  doi = {10.1016/0022-5096(95)00036-I}
}

@article{Perry2020Nearly,
  title = {Nearly {{Magnitude}}-{{Invariant Stress Drops}} in {{Simulated Crack}}-{{Like Earthquake Sequences}} on {{Rate}}-and-{{State Faults}} with {{Thermal Pressurization}} of {{Pore Fluids}}},
  author = {Perry, Stephen M. and Lambert, Val{\`e}re and Lapusta, Nadia},
  year = {2020},
  month = mar,
  journal = {Journal of Geophysical Research: Solid Earth},
  volume = {125},
  number = {3},
  pages = {e2019JB018597},
  issn = {2169-9313, 2169-9356},
  doi = {10.1029/2019JB018597}
}

@article{Premus2022Bridging,
  title = {Bridging Time Scales of Faulting: {{From}} Coseismic to Postseismic Slip of the {{Mw}} 6.0 2014 {{South Napa}}, {{California}} Earthquake},
  author = {Premus, Jan and Gallovi{\v c}, Franti{\v s}ek and Ampuero, Jean-Paul},
  year = {2022},
  month = sep,
  journal = {Science Advances},
  volume = {8},
  number = {38},
  pages = {eabq2536},
  publisher = {American Association for the Advancement of Science},
  doi = {10.1126/sciadv.abq2536}
}

@misc{Propagation,
  title = {Propagation of Extended Fractures by Local Nucleation and Rapid Transverse Expansion of Crack-Front Distortion {\textbar} {{Nature Physics}}},
  howpublished = {https://www.nature.com/articles/s41567-023-02365-0}
}

@article{Ramos2021Assessing,
  title = {Assessing {{Margin}}-{{Wide Rupture Behaviors Along}} the {{Cascadia Megathrust With}} 3-{{D Dynamic Rupture Simulations}}},
  author = {Ramos, Marlon D. and Huang, Yihe and Ulrich, Thomas and Li, Duo and Gabriel, Alice-Agnes and Thomas, Amanda M.},
  year = {2021},
  month = jul,
  journal = {Journal of Geophysical Research: Solid Earth},
  volume = {126},
  number = {7},
  pages = {e2021JB022005},
  issn = {2169-9313, 2169-9356},
  doi = {10.1029/2021JB022005}
}

@article{Ramos2022Working,
  title = {Working with {{Dynamic Earthquake Rupture Models}}: {{A Practical Guide}}},
  author = {Ramos, Marlon D. and Thakur, Prithvi and Huang, Yihe and Harris, Ruth A. and Ryan, Kenny J.},
  year = {2022},
  month = jul,
  journal = {Seismological Research Letters},
  volume = {93},
  number = {4},
  pages = {2096--2110},
  issn = {0895-0695, 1938-2057},
  doi = {10.1785/0220220022}
}

@article{Rice2006Heating,
  title = {Heating and Weakening of Faults during Earthquake Slip},
  author = {Rice, James R.},
  year = {2006},
  journal = {Journal of Geophysical Research: Solid Earth},
  volume = {111},
  number = {B5},
  issn = {2156-2202},
  doi = {10.1029/2005JB004006}
}

@article{Rubino2022Intermittent,
  title = {Intermittent Lab Earthquakes in Dynamically Weakening Fault Gouge},
  author = {Rubino, V. and Lapusta, N. and Rosakis, A. J.},
  year = {2022},
  month = jun,
  journal = {Nature},
  volume = {606},
  number = {7916},
  pages = {922--929},
  issn = {0028-0836, 1476-4687},
  doi = {10.1038/s41586-022-04749-3}
}

@article{Ruina1983Slip,
  title = {Slip Instability and State Variable Friction Laws},
  author = {Ruina, Andy},
  year = {1983},
  month = dec,
  journal = {Journal of Geophysical Research: Solid Earth},
  volume = {88},
  number = {B12},
  pages = {10359--10370},
  issn = {0148-0227},
  doi = {10.1029/JB088iB12p10359}
}

@article{Saffer2003Comparison,
  title = {Comparison of Smectite- and Illite-Rich Gouge Frictional Properties: Application to the Updip Limit of the Seismogenic Zone along Subduction Megathrusts},
  author = {Saffer, Demian M and Marone, Chris},
  year = {2003},
  month = oct,
  journal = {Earth and Planetary Science Letters},
  volume = {215},
  number = {1-2},
  pages = {219--235},
  issn = {0012821X},
  doi = {10.1016/S0012-821X(03)00424-2}
}

@article{Saffer2011Hydrogeology,
  title = {Hydrogeology and {{Mechanics}} of {{Subduction Zone Forearcs}}: {{Fluid Flow}} and {{Pore Pressure}}},
  author = {Saffer, Demian M. and Tobin, Harold J.},
  year = {2011},
  month = may,
  journal = {Annual Review of Earth and Planetary Sciences},
  volume = {39},
  number = {Volume 39, 2011},
  pages = {157--186},
  publisher = {Annual Reviews},
  issn = {0084-6597, 1545-4495},
  doi = {10.1146/annurev-earth-040610-133408}
}

@article{Sallares2019Upperplate,
  title = {Upper-Plate Rigidity Determines Depth-Varying Rupture Behaviour of Megathrust Earthquakes},
  author = {Sallar{\`e}s, Valent{\'i} and Ranero, C{\'e}sar R.},
  year = {2019},
  month = dec,
  journal = {Nature},
  volume = {576},
  number = {7785},
  pages = {96--101},
  issn = {0028-0836, 1476-4687},
  doi = {10.1038/s41586-019-1784-0}
}

@article{Satake2013Time,
  title = {Time and {{Space Distribution}} of {{Coseismic Slip}} of the 2011 {{Tohoku Earthquake}} as {{Inferred}} from {{Tsunami Waveform Data}}},
  author = {Satake, K. and Fujii, Y. and Harada, T. and Namegaya, Y.},
  year = {2013},
  month = may,
  journal = {Bulletin of the Seismological Society of America},
  volume = {103},
  number = {2B},
  pages = {1473--1492},
  issn = {0037-1106},
  doi = {10.1785/0120120122}
}

@article{Schliwa2024Linked,
  title = {The {{Linked Complexity}} of {{Coseismic}} and {{Postseismic Faulting Revealed}} by {{Seismo}}-{{Geodetic Dynamic Inversion}} of the 2004 {{Parkfield Earthquake}}},
  author = {Schliwa, Nico and Gabriel, Alice-Agnes and Premus, Jan and Gallovi{\v c}, Franti{\v s}ek},
  year = {2024},
  month = dec,
  journal = {Journal of Geophysical Research: Solid Earth},
  volume = {129},
  number = {12},
  pages = {e2024JB029410},
  issn = {2169-9313, 2169-9356},
  doi = {10.1029/2024JB029410}
}

@article{Schmitt2015Nucleation,
  title = {Nucleation and Dynamic Rupture on Weakly Stressed Faults Sustained by Thermal Pressurization},
  author = {Schmitt, Stuart V. and Segall, Paul and Dunham, Eric M.},
  year = {2015},
  month = nov,
  journal = {Journal of Geophysical Research: Solid Earth},
  volume = {120},
  number = {11},
  pages = {7606--7640},
  issn = {2169-9313, 2169-9356},
  doi = {10.1002/2015JB012322}
}

@article{Shi2006Dynamic,
  title = {Dynamic Rupture on a Bimaterial Interface Governed by Slip-Weakening Friction},
  author = {Shi, Zheqiang and {Ben-Zion}, Yehuda},
  year = {2006},
  month = may,
  journal = {Geophysical Journal International},
  volume = {165},
  number = {2},
  pages = {469--484},
  issn = {0956540X, 1365246X},
  doi = {10.1111/j.1365-246X.2006.02853.x}
}

@article{Sun2017Large,
  title = {Large Fault Slip Peaking at Trench in the 2011 {{Tohoku-oki}} Earthquake},
  author = {Sun, Tianhaozhe and Wang, Kelin and Fujiwara, Toshiya and Kodaira, Shuichi and He, Jiangheng},
  year = {2017},
  month = jan,
  journal = {Nature Communications},
  volume = {8},
  number = {1},
  pages = {14044},
  issn = {2041-1723},
  doi = {10.1038/ncomms14044}
}

@article{Taufiqurrahman2022Broadband,
  title = {Broadband {{Dynamic Rupture Modeling With Fractal Fault Roughness}}, {{Frictional Heterogeneity}}, {{Viscoelasticity}} and {{Topography}}: {{The}} 2016 {{{\emph{M}}}} {\textsubscript{w}} 6.2 {{Amatrice}}, {{Italy Earthquake}}},
  author = {Taufiqurrahman, T. and Gabriel, A.-A. and Ulrich, T. and Valentov{\'a}, L. and Gallovi{\v c}, F.},
  year = {2022},
  month = nov,
  journal = {Geophysical Research Letters},
  volume = {49},
  number = {22},
  pages = {e2022GL098872},
  issn = {0094-8276, 1944-8007},
  doi = {10.1029/2022GL098872}
}

@article{Taufiqurrahman2023Dynamics,
  title = {Dynamics, Interactions and Delays of the 2019 {{Ridgecrest}} Rupture Sequence},
  author = {Taufiqurrahman, Taufiq and Gabriel, Alice-Agnes and Li, Duo and Ulrich, Thomas and Li, Bo and Carena, Sara and Verdecchia, Alessandro and Gallovi{\v c}, Franti{\v s}ek},
  year = {2023},
  month = jun,
  journal = {Nature},
  volume = {618},
  number = {7964},
  pages = {308--315},
  issn = {0028-0836, 1476-4687},
  doi = {10.1038/s41586-023-05985-x}
}

@article{Templeton2008Fault,
  title = {Off-Fault Plasticity and Earthquake Rupture Dynamics: 1. {{Dry}} Materials or Neglect of Fluid Pressure Changes},
  author = {Templeton, Elizabeth L. and Rice, James R.},
  year = {2008},
  journal = {Journal of Geophysical Research: Solid Earth},
  volume = {113},
  number = {B9},
  issn = {2156-2202},
  doi = {10.1029/2007JB005529}
}

@article{Tinti2005Earthquake,
  title = {Earthquake Fracture Energy Inferred from Kinematic Rupture Models on Extended Faults},
  author = {Tinti, E. and Spudich, P. and Cocco, M.},
  year = {2005},
  month = dec,
  journal = {Journal of Geophysical Research: Solid Earth},
  volume = {110},
  number = {B12},
  pages = {2005JB003644},
  issn = {0148-0227},
  doi = {10.1029/2005JB003644}
}

@article{Tinti2021Constraining,
  title = {Constraining Families of Dynamic Models Using Geological, Geodetic and Strong Ground Motion Data: {{The Mw}} 6.5, {{October}} 30th, 2016, {{Norcia}} Earthquake, {{Italy}}},
  author = {Tinti, Elisa and Casarotti, Emanuele and Ulrich, Thomas and Taufiqurrahman, Taufiq and {li}, Dou and Gabriel, Alice-Agnes},
  year = {2021},
  journal = {Earth and Planetary Science Letters},
  volume = {576}
}

@article{Tsuru2002Alongarc,
  title = {Along-Arc Structural Variation of the Plate Boundary at the {{Japan Trench}} Margin: {{Implication}} of Interplate Coupling},
  author = {Tsuru, Tetsuro and Park, Jin-Oh and Miura, Seiichi and Kodaira, Shuichi and Kido, Yukari and Hayashi, Tsutomu},
  year = {2002},
  journal = {Journal of Geophysical Research: Solid Earth},
  volume = {107},
  number = {B12},
  pages = {ESE 11-1-ESE 11-15},
  issn = {2156-2202},
  doi = {10.1029/2001JB001664}
}

@article{Uchida2021Decade,
  title = {A {{Decade}} of {{Lessons Learned}} from the 2011 {{Tohoku}}-{{Oki Earthquake}}},
  author = {Uchida, N. and B{\"u}rgmann, R.},
  year = {2021},
  month = jun,
  journal = {Reviews of Geophysics},
  volume = {59},
  number = {2},
  pages = {e2020RG000713},
  issn = {8755-1209, 1944-9208},
  doi = {10.1029/2020RG000713}
}

@article{Ueda2023Submarine,
  title = {The Submarine Fault Scarp of the 2011 {{Tohoku-oki Earthquake}} in the {{Japan Trench}}},
  author = {Ueda, Hayato and Kitazato, Hiroshi and Jamieson, Alan and {Pressure Drop Ring of Fire Expedition 2022 Japan Cruise Leg2 science team} and Bond, Todd and Cardigos, Sara and Funaki, Masayoshi and Maroni, Paige J. and Nanbu, Hiroyasu and O'Callaghan, Joanne M. and Onishi, Takuma and Pedersen, Silje W. and Roperez, Jaya and Tsuruzono, Hiroumi and Watanabe, Hiromi K. and Yasuda, Tetsuro},
  year = {2023},
  month = dec,
  journal = {Communications Earth \& Environment},
  volume = {4},
  number = {1},
  pages = {476},
  issn = {2662-4435},
  doi = {10.1038/s43247-023-01118-4}
}

@article{Ujiie2010Highvelocity,
  title = {High-Velocity Frictional Properties of Clay-Rich Fault Gouge in a Megasplay Fault Zone, {{Nankai}} Subduction Zone},
  author = {Ujiie, Kohtaro and Tsutsumi, Akito},
  year = {2010},
  journal = {Geophysical Research Letters},
  volume = {37},
  number = {24},
  issn = {1944-8007},
  doi = {10.1029/2010GL046002}
}

@article{Ujiie2013Low,
  title = {Low {{Coseismic Shear Stress}} on the {{Tohoku-Oki Megathrust Determined}} from {{Laboratory Experiments}}},
  author = {Ujiie, Kohtaro and Tanaka, Hanae and Saito, Tsubasa and Tsutsumi, Akito and Mori, James J. and Kameda, Jun and Brodsky, Emily E. and Chester, Frederick M. and Eguchi, Nobuhisa and Toczko, Sean and {Expedition 343 and 343T Scientists}},
  year = {2013},
  month = dec,
  journal = {Science},
  volume = {342},
  number = {6163},
  pages = {1211--1214},
  issn = {0036-8075, 1095-9203},
  doi = {10.1126/science.1243485}
}

@article{Ulrich2019Dynamic,
  title = {Dynamic Viability of the 2016 {{Mw}} 7.8 {{Kaik{\=o}ura}} Earthquake Cascade on Weak Crustal Faults},
  author = {Ulrich, Thomas and Gabriel, Alice-Agnes and Ampuero, Jean-Paul and Xu, Wenbin},
  year = {2019},
  month = mar,
  journal = {Nature Communications},
  volume = {10},
  number = {1},
  pages = {1213},
  issn = {2041-1723},
  doi = {10.1038/s41467-019-09125-w}
}

@article{Ulrich2022Stress,
  title = {Stress, Rigidity and Sediment Strength Control Megathrust Earthquake and Tsunami Dynamics},
  author = {Ulrich, Thomas and Gabriel, Alice-Agnes and Madden, Elizabeth H.},
  year = {2022},
  month = jan,
  journal = {Nature Geoscience},
  volume = {15},
  number = {1},
  pages = {67--73},
  issn = {1752-0894, 1752-0908},
  doi = {10.1038/s41561-021-00863-5}
}

@inproceedings{Uphoff2017Extreme,
  title = {Extreme Scale Multi-Physics Simulations of the Tsunamigenic 2004 Sumatra Megathrust Earthquake},
  booktitle = {Proceedings of the {{International Conference}} for {{High Performance Computing}}, {{Networking}}, {{Storage}} and {{Analysis}}},
  author = {Uphoff, Carsten and Rettenberger, Sebastian and Bader, Michael and Madden, Elizabeth H. and Ulrich, Thomas and Wollherr, Stephanie and Gabriel, Alice-Agnes},
  year = {2017},
  month = nov,
  pages = {1--16},
  publisher = {ACM},
  address = {Denver Colorado},
  doi = {10.1145/3126908.3126948},
  isbn = {978-1-4503-5114-0}
}

@article{VallA2016New,
  title = {A New Database of Source Time Functions ({{STFs}}) Extracted from the {{SCARDEC}} Method},
  author = {Vall{\~A}, Martin},
  year = {2016},
  journal = {Physics of the Earth and Planetary Interiors}
}

@article{Vallee2013Source,
  title = {Source Time Function Properties Indicate a Strain Drop Independent of Earthquake Depth and Magnitude},
  author = {Vall{\'e}e, Martin},
  year = {2013},
  month = oct,
  journal = {Nature Communications},
  volume = {4},
  number = {1},
  pages = {2606},
  publisher = {Nature Publishing Group},
  issn = {2041-1723},
  doi = {10.1038/ncomms3606}
}

@article{Vallee2023Selfreactivated,
  title = {Self-Reactivated Rupture during the 2019 {{M}} = 8 Northern {{Peru}} Intraslab Earthquake},
  author = {Vall{\'e}e, Martin and Xie, Yuqing and Grandin, Rapha{\"e}l and {Villegas-Lanza}, Juan Carlos and Nocquet, Jean-Mathieu and Vaca, Sandro and Meng, Lingsen and Ampuero, Jean Paul and Mothes, Patricia and Jarrin, Paul and Sierra Farf{\'a}n, Ciro and Rolandone, Fr{\'e}d{\'e}rique},
  year = {2023},
  month = jan,
  journal = {Earth and Planetary Science Letters},
  volume = {601},
  pages = {117886},
  issn = {0012821X},
  doi = {10.1016/j.epsl.2022.117886}
}

@article{Viesca2015Ubiquitous,
  title = {Ubiquitous Weakening of Faults Due to Thermal Pressurization},
  author = {Viesca, Robert C. and Garagash, Dmitry I.},
  year = {2015},
  month = nov,
  journal = {Nature Geoscience},
  volume = {8},
  number = {11},
  pages = {875--879},
  publisher = {Nature Publishing Group},
  issn = {1752-0908},
  doi = {10.1038/ngeo2554}
}

@article{Wang2017Seismic,
  title = {Seismic Source Spectral Properties of Crack-like and Pulse-like Modes of Dynamic Rupture},
  author = {Wang, Yongfei and Day, Steven M.},
  year = {2017},
  month = aug,
  journal = {Journal of Geophysical Research: Solid Earth},
  volume = {122},
  number = {8},
  pages = {6657--6684},
  issn = {2169-9313, 2169-9356},
  doi = {10.1002/2017JB014454}
}

@article{Wang2018Learning,
  title = {Learning from Crustal Deformation Associated with the {{M9}} 2011 {{Tohoku-oki}} Earthquake},
  author = {Wang, Kelin and Sun, Tianhaozhe and Brown, Lonn and Hino, Ryota and Tomita, Fumiaki and Kido, Motoyuki and Iinuma, Takeshi and Kodaira, Shuichi and Fujiwara, Toshiya},
  year = {2018},
  month = apr,
  journal = {Geosphere},
  volume = {14},
  number = {2},
  pages = {552--571},
  issn = {1553-040X},
  doi = {10.1130/GES01531.1}
}

@article{Weng2018Constraining,
  title = {Constraining {{Frictional Properties}} on {{Fault}} by {{Dynamic Rupture Simulations}} and {{Near-Field Observations}}},
  author = {Weng, Huihui and Yang, Hongfeng},
  year = {2018},
  journal = {Journal of Geophysical Research: Solid Earth},
  volume = {123},
  number = {8},
  pages = {6658--6670},
  issn = {2169-9356},
  doi = {10.1029/2017JB015414}
}

@article{Weng2019Dynamics,
  title = {The {{Dynamics}} of {{Elongated Earthquake Ruptures}}},
  author = {Weng, Huihui and Ampuero, Jean-Paul},
  year = {2019},
  journal = {Journal of Geophysical Research: Solid Earth},
  volume = {124},
  number = {8},
  pages = {8584--8610},
  issn = {2169-9356},
  doi = {10.1029/2019JB017684}
}

@article{Wirp2024Dynamic,
  title = {Dynamic {{Rupture Modeling}} of {{Large Earthquake Scenarios}} at the {{Hellenic Arc Toward Physics}}-{{Based Seismic}} and {{Tsunami Hazard Assessment}}},
  author = {Wirp, Sara Aniko and Gabriel, Alice-Agnes and Ulrich, Thomas and Lorito, Stefano},
  year = {2024},
  month = nov,
  journal = {Journal of Geophysical Research: Solid Earth},
  volume = {129},
  number = {11},
  pages = {e2024JB029320},
  issn = {2169-9313, 2169-9356},
  doi = {10.1029/2024JB029320}
}

@article{Wirth2022Occurrence,
  title = {The Occurrence and Hazards of Great Subduction Zone Earthquakes},
  author = {Wirth, Erin A. and Sahakian, Valerie J. and Wallace, Laura M. and Melnick, Daniel},
  year = {2022},
  month = jan,
  journal = {Nature Reviews Earth \& Environment},
  volume = {3},
  number = {2},
  pages = {125--140},
  issn = {2662-138X},
  doi = {10.1038/s43017-021-00245-w}
}

@article{Wollherr2018Fault,
  title = {Off-Fault Plasticity in Three-Dimensional Dynamic Rupture Simulations Using a Modal {{Discontinuous Galerkin}} Method on Unstructured Meshes: Implementation, Verification and Application},
  author = {Wollherr, Stephanie and Gabriel, Alice-Agnes and Uphoff, Carsten},
  year = {2018},
  month = sep,
  journal = {Geophysical Journal International},
  volume = {214},
  number = {3},
  pages = {1556--1584},
  issn = {0956-540X, 1365-246X},
  doi = {10.1093/gji/ggy213}
}

@article{Wong2024Quantitative,
  title = {A {{Quantitative Comparison}} and {{Validation}} of {{Finite}}-{{Fault Models}}: {{The}} 2011 {{Tohoku}}-{{Oki Earthquake}}},
  author = {Wong, Jeremy Wing Ching and Fan, Wenyuan and Gabriel, Alice-Agnes},
  year = {2024},
  month = oct,
  journal = {Journal of Geophysical Research: Solid Earth},
  volume = {129},
  number = {10},
  pages = {e2024JB029212},
  issn = {2169-9313, 2169-9356},
  doi = {10.1029/2024JB029212}
}

@article{Yagi2011Rupture,
  title = {Rupture Process of the 2011 {{Tohoku-oki}} Earthquake and Absolute Elastic Strain Release},
  author = {Yagi, Yuji and Fukahata, Yukitoshi},
  year = {2011},
  month = oct,
  journal = {Geophysical Research Letters},
  volume = {38},
  number = {19},
  pages = {n/a-n/a},
  issn = {00948276},
  doi = {10.1029/2011GL048701}
}

@article{Yagi2012Smooth,
  title = {Smooth and Rapid Slip near the {{Japan Trench}} during the 2011 {{Tohoku-oki}} Earthquake Revealed by a Hybrid Back-Projection Method},
  author = {Yagi, Yuji and Nakao, Atsushi and Kasahara, Amato},
  year = {2012},
  month = nov,
  journal = {Earth and Planetary Science Letters},
  volume = {355--356},
  pages = {94--101},
  issn = {0012821X},
  doi = {10.1016/j.epsl.2012.08.018}
}

@article{Yamazaki2018Self,
  title = {A {{Self}}-{{Consistent Fault Slip Model}} for the 2011 {{Tohoku Earthquake}} and {{Tsunami}}},
  author = {Yamazaki, Yoshiki and Cheung, Kwok Fai and Lay, Thorne},
  year = {2018},
  month = feb,
  journal = {Journal of Geophysical Research: Solid Earth},
  volume = {123},
  number = {2},
  pages = {1435--1458},
  issn = {2169-9313, 2169-9356},
  doi = {10.1002/2017JB014749}
}

@article{Yao2013Compressive,
  title = {Compressive Sensing of Frequency-Dependent Seismic Radiation from Subduction Zone Megathrust Ruptures},
  author = {Yao, Huajian and Shearer, Peter M. and Gerstoft, Peter},
  year = {2013},
  month = mar,
  journal = {Proceedings of the National Academy of Sciences},
  volume = {110},
  number = {12},
  pages = {4512--4517},
  issn = {0027-8424, 1091-6490},
  doi = {10.1073/pnas.1212790110}
}

@article{Yao2020Rupture,
  title = {Rupture {{Dynamics}} of the 2012 {{Nicoya}} {{{\emph{M}}}}{\textsubscript{ {\emph{w}} }} 7.6 {{Earthquake}}: {{Evidence}} for {{Low Strength}} on the {{Megathrust}}},
  author = {Yao, Suli and Yang, Hongfeng},
  year = {2020},
  month = jul,
  journal = {Geophysical Research Letters},
  volume = {47},
  number = {13},
  pages = {e2020GL087508},
  issn = {0094-8276, 1944-8007},
  doi = {10.1029/2020GL087508}
}

@article{Zhang2023Complex,
  title = {Complex Tsunamigenic Near-Trench Seafloor Deformation during the 2011 {{Tohoku}}--{{Oki}} Earthquake},
  author = {Zhang, Kai and Wang, Yanru and Luo, Yu and Zhao, Dineng and Wang, Mingwei and Yang, Fanlin and Wu, Ziyin},
  year = {2023},
  month = jun,
  journal = {Nature Communications},
  volume = {14},
  number = {1},
  pages = {3260},
  issn = {2041-1723},
  doi = {10.1038/s41467-023-38970-z}
}

@article{Zheng1998Conditionsa,
  title = {Conditions under Which Velocity-Weakening Friction Allows a Self-Healing versus a Cracklike Mode of Rupture},
  author = {Zheng, Gutuan and Rice, James R.},
  year = {1998},
  month = dec,
  journal = {Bulletin of the Seismological Society of America},
  volume = {88},
  number = {6},
  pages = {1466--1483},
  issn = {1943-3573, 0037-1106},
  doi = {10.1785/BSSA0880061466}
}

@article{Yue2011Inversion,
  title = {Inversion of High-Rate (1 Sps) {{GPS}} Data for Rupture Process of the 11 {{March}} 2011 {{Tohoku}} Earthquake ({{M}}{\textsubscript{w}} 9.1): {{INVERSION OF HIGH-RATE GPS FOR TOHOKU EQ}}},
  author = {Yue, H. and Lay, T.},
  year = {2011},
  month = apr,
  journal = {Geophysical Research Letters},
  volume = {38},
  number = {7},
  pages = {n/a-n/a},
  issn = {00948276},
  doi = {10.1029/2011GL048700}
}


%% file: custom.bib
@inproceedings{Heinecke_et_al_2014,
	title = {Petascale {High} {Order} {Dynamic} {Rupture} {Earthquake} {Simulations} on {Heterogeneous} {Supercomputers}},
	url = {https://ieeexplore.ieee.org/document/7012188},
	doi = {10.1109/SC.2014.6},
	urldate = {2024-02-06},
	booktitle = {{SC} '14: {Proceedings} of the {International} {Conference} for {High} {Performance} {Computing}, {Networking}, {Storage} and {Analysis}},
	author = {Heinecke, Alexander and Breuer, Alexander and Rettenberger, Sebastian and Bader, Michael and Gabriel, Alice-Agnes and Pelties, Christian and Bode, Arndt and Barth, William and Liao, Xiang-Ke and Vaidyanathan, Karthikeyan and Smelyanskiy, Mikhail and Dubey, Pradeep},
	month = nov,
	year = {2014},
	note = {ISSN: 2167-4337},
	pages = {3--14}
}

@article{Hayek2024,
author = {Hayek, J. N. and Marchandon, M. and Li, D. and Pousse-Beltran, L. and Hollingsworth, J. and Li, T. and Gabriel, A.-A.},
title = {Non-Typical Supershear Rupture: Fault Heterogeneity and Segmentation Govern Unilateral Supershear and Cascading Multi-Fault Rupture in the 2021 7.4 Maduo Earthquake},
journal = {Geophysical Research Letters},
volume = {51},
number = {20},
pages = {e2024GL110128},
doi = {https://doi.org/10.1029/2024GL110128},
year = {2024}
}

@article{Koketsu2009Proposal,
  title={A proposal for a standard procedure of modeling 3-D velocity structures and its application to the Tokyo metropolitan area, Japan},
  author={Koketsu, Kazuki and Miyake, Hiroe and Tanaka, Yasuhisa and others},
  journal={Tectonophysics},
  volume={472},
  number={1-4},
  pages={290--300},
  year={2009},
  publisher={Elsevier}
}

@inproceedings{Koketsu2012Japan,
  title={Japan integrated velocity structure model version 1},
  author={Koketsu, Kazuki and Miyake, Hiroe and Suzuki, Haruhiko},
  booktitle={Proceedings of the 15th world conference on earthquake engineering},
  volume={1},
  pages={4},
  year={2012},
  organization={Lisbon}
}

@article{Fukuyama1998Automated,
  title={Automated seismic moment tensor determination by using on-line broadband seismic waveforms [in Japanese with English abstract]},
  author={Fukuyama, Eiichi},
  journal={J. Seismol. Soc. Jpn.},
  volume={51},
  pages={149},
  year={1998}
}

@misc{Geobco2024,
title={GEBCO 2024 Grid},
author={GEBCO Compilation Group},
year={2024},
doi={doi:10.5285/1c44ce99-0a0d-5f4f-e063-7086abc0ea0f}
}

@article{Heidbach2018World,
  title={The World Stress Map database release 2016: Crustal stress pattern across scales},
  author={Heidbach, Oliver and Rajabi, Mojtaba and Cui, Xiaofeng and Fuchs, Karl and M{\"u}ller, Birgit and Reinecker, John and Reiter, Karsten and Tingay, Mark and Wenzel, Friedemann and Xie, Furen and others},
  journal={Tectonophysics},
  volume={744},
  pages={484--498},
  year={2018},
  publisher={Elsevier},
}

@article{sagiya2004decade,
  title={A decade of GEONET: 1994-2003 The continuous GPS observation in Japan and its impact on earthquake studies},
  author={Sagiya, Takeshi},
  journal={Earth, planets and space},
  volume={56},
  number={8},
  pages={xxix--xli},
  year={2004},
  publisher={Society of Geomagnetism and Earth, Planetary and Space Sciences, The~…}
}

@inproceedings{oeser2006cluster,
  title={Cluster design in the earth sciences tethys},
  author={Oeser, Jens and Bunge, Hans-Peter and Mohr, Marcus},
  booktitle={International conference on high performance computing and communications},
  pages={31--40},
  year={2006},
  organization={Springer}
}

@article{Causse2014Variability,
	title = {Variability of dynamic source parameters inferred from kinematic models of past earthquakes},
	volume = {196},
	issn = {0956-540X},
	url = {https://doi.org/10.1093/gji/ggt478},
	doi = {10.1093/gji/ggt478},
	number = {3},
	urldate = {2024-02-06},
	journal = {Geophysical Journal International},
	author = {Causse, M. and Dalguer, L. A. and Mai, P. M.},
	month = mar,
	year = {2014},
	pages = {1754--1769},
}

@article{Day1998Dynamic,
	author = {Day, Steven M. and Yu, Guang and Wald, David J.},
	doi = {10.1785/BSSA0880020512},
	issn = {0037-1106},
	journal = {Bulletin of the Seismological Society of America},
	month = apr,
	number = {2},
	pages = {512--522},
	title = {Dynamic stress changes during earthquake rupture},
	url = {https://doi.org/10.1785/BSSA0880020512},
	urldate = {2025-02-26},
	volume = {88},
	year = {1998}}

@article{Yang2019Hypocenter,
	author = {Yang, Hongfeng and Yao, Suli and He, Bing and Newman, Andrew V.},
	doi = {https://doi.org/10.1016/j.epsl.2019.05.030},
	issn = {0012-821X},
	journal = {Earth and Planetary Science Letters},
	pages = {10--17},
	title = {Earthquake rupture dependence on hypocentral location along the {Nicoya} {Peninsula} subduction megathrust},
	url = {https://www.sciencedirect.com/science/article/pii/S0012821X19303000},
	volume = {520},
	year = {2019}}

@article{Lambert2020Rupturedependent,
  title = {Rupture-Dependent Breakdown Energy in Fault Models with Thermo-Hydro-Mechanical Processes},
  author = {Lambert, Val{\`e}re and Lapusta, Nadia},
  year = {2020},
  month = nov,
  journal = {Solid Earth},
  volume = {11},
  number = {6},
  pages = {2283--2302},
  issn = {1869-9529},
  doi = {10.5194/se-11-2283-2020}
}

@article{sun2025back,
  title={Back-propagating Earthquakes on a Simple Fault},
  author={Sun, Yudong and Cattania, Camilla},
  journal={Authorea Preprints},
  year={2025},
  publisher={Authorea},
  doi = {10.22541/essoar.173724475.50020741/v1}
}

@article{Andrews1976Rupture,
  title = {Rupture Velocity of Plane Strain Shear Cracks},
  author = {Andrews, D. J.},
  year = {1976},
  month = nov,
  journal = {Journal of Geophysical Research},
  volume = {81},
  number = {32},
  pages = {5679--5687},
  issn = {01480227},
  doi = {10.1029/JB081i032p05679}
}

@article{Hu2019Sustainability,
  title = {The {{Sustainability}} of {{Free}}-{{Surface}}-{{Induced Supershear Rupture}} on {{Strike}}-{{Slip Faults}}},
  author = {Hu, Feng and Oglesby, David D. and Chen, Xiaofei},
  year = {2019},
  month = aug,
  journal = {Geophysical Research Letters},
  volume = {46},
  number = {16},
  pages = {9537--9543},
  issn = {0094-8276, 1944-8007},
  doi = {10.1029/2019GL084318}
}

@article{scholz2014rupture,
  title={The rupture mode of the shallow large-slip surge of the Tohoku-Oki earthquake},
  author={Scholz, Christopher H},
  journal={Bulletin of the Seismological Society of America},
  volume={104},
  number={5},
  pages={2627--2631},
  year={2014},
  publisher={Seismological Society of America}
}

@article{prada2021influence,
  title={The influence of depth-varying elastic properties of the upper plate on megathrust earthquake rupture dynamics and tsunamigenesis},
  author={Prada, Manel and Galvez, Percy and Ampuero, Jean-Paul and Sallar{\`e}s, Valent{\'\i} and S{\'a}nchez-Linares, Carlos and Mac{\'\i}as, Jorge and Peter, Daniel},
  journal={Journal of Geophysical Research: Solid Earth},
  volume={126},
  number={11},
  pages={e2021JB022328},
  year={2021},
  publisher={Wiley Online Library}
}

@article{guatteri2000can,
  title={What can strong-motion data tell us about slip-weakening fault-friction laws?},
  author={Guatteri, Mariagiovanna and Spudich, Paul},
  journal={Bulletin of the Seismological Society of America},
  volume={90},
  number={1},
  pages={98--116},
  year={2000},
  publisher={Seismological Society of America}
}

@article{dunham2005dissipative,
  title={Dissipative interface waves and the transient response of a three-dimensional sliding interface with Coulomb friction},
  author={Dunham, Eric M},
  journal={Journal of the Mechanics and Physics of Solids},
  volume={53},
  number={2},
  pages={327--357},
  year={2005},
  publisher={Elsevier}
}

@article{Oral2022Method,
  title = {A {{Method}} to {{Generate Initial Fault Stresses}} for {{Physics-Based Ground-Motion Prediction Consistent}} with {{Regional Seismicity}}},
  author = {Oral, Elif and Ampuero, Jean Paul and Ruiz, Javier and Asimaki, Domniki},
  year = 2022,
  month = dec,
  journal = {Bulletin of the Seismological Society of America},
  volume = {112},
  number = {6},
  pages = {2812--2827},
  issn = {0037-1106, 1943-3573},
  doi = {10.1785/0120220064}
}

@software{Gabriel2025SeisSol,
  author       = {Gabriel, Alice-Agnes and
                  Kurapati, Vikas and
                  Niu, Zihua and
                  Schliwa, Nico and
                  Schneller, David and
                  Ulrich, Thomas and
                  Dorozhinskii, Ravil and
                  Krenz, Lukas and
                  Uphoff, Carsten and
                  Wolf, Sebastian and
                  Breuer, Alexander and
                  Heinecke, Alexander and
                  Pelties, Christian and
                  Rettenberger, Sebastian and
                  Wollherr, Stephanie and
                  Bader, Michael},
  title        = {SeisSol},
  month        = jun,
  year         = 2025,
  publisher    = {Zenodo},
  version      = {v1.3.2},
  doi          = {10.5281/zenodo.15685917},
  url          = {https://doi.org/10.5281/zenodo.15685917},
  swhid        = {swh:1:dir:5b7e9c3e2390364526c34903e00a2c425b8845db
                   ;origin=https://doi.org/10.5281/zenodo.4672483;vis
                   it=swh:1:snp:5df1323004d8ab8201bf82b8ee556660f67c3
                   001;anchor=swh:1:rel:0dd37599d1f5aaeb3034fa4c4bcc1
                   d88b35e37e8;path=SeisSol-SeisSol-20334d0
                  },
}

@article{schmedes2010correlation,
  title={Correlation of earthquake source parameters inferred from dynamic rupture simulations},
  author={Schmedes, Jan and Archuleta, Ralph J and Lavall{\'e}e, Daniel},
  journal={Journal of Geophysical Research: Solid Earth},
  volume={115},
  number={B3},
  year={2010},
  publisher={Wiley Online Library}
}
